\documentclass[a4paper,12pt,numbers,sort&compress]{article}

\pdfoutput=1

\usepackage[paper=letterpaper,margin=1in]{geometry}

\usepackage{amsmath,amssymb,amsfonts,graphicx,cite,setspace,bigstrut,longtable,array,hyperref,url,breqn,float}

\begin{document}

\vskip 0.25in

\newcommand{\todo}[1]{{\bf ?????!!!! #1
?????!!!!}\marginpar{$\Longleftarrow$}}

\newcommand{\nn}{\nonumber}
\newcommand{\tr}{\mathop{\rm Tr}}
\newcommand{\comment}[1]{}

\newcommand{\cM}{{\cal M}}
\newcommand{\cW}{{\cal W}}
\newcommand{\cN}{{\cal N}}
\newcommand{\cH}{{\cal H}}
\newcommand{\cK}{{\cal K}}
\newcommand{\cZ}{{\cal Z}}
\newcommand{\cO}{{\cal O}}
\newcommand{\cB}{{\cal B}}
\newcommand{\cC}{{\cal C}}
\newcommand{\cD}{{\cal D}}
\newcommand{\cE}{{\cal E}}
\newcommand{\cF}{{\cal F}}
\newcommand{\cX}{{\cal X}}
\newcommand{\IA}{\mathbb{A}}
\newcommand{\IP}{\mathbb{P}}
\newcommand{\IQ}{\mathbb{Q}}
\newcommand{\IH}{\mathbb{H}}
\newcommand{\IR}{\mathbb{R}}
\newcommand{\IC}{\mathbb{C}}
\newcommand{\IF}{\mathbb{F}}
\newcommand{\IV}{\mathbb{V}}
\newcommand{\II}{\mathbb{I}}
\newcommand{\IZ}{\mathbb{Z}}
\newcommand{\IN}{\mathbb{N}}
\newcommand{\re}{{\rm Re}}
\newcommand{\im}{{\rm Im}}
\newcommand{\sym}{{\rm Sym}}

\newcommand{\tmat}[1]{{\tiny \left(\begin{matrix} #1 \end{matrix}\right)}}
\newcommand{\mat}[1]{\left(\begin{matrix} #1 \end{matrix}\right)}
\newcommand{\diff}[2]{\frac{\partial #1}{\partial #2}}
\newcommand{\gen}[1]{\langle #1 \rangle}
\newcommand{\ket}[1]{| #1 \rangle}
\newcommand{\jacobi}[2]{\left(\frac{#1}{#2}\right)}

\newcommand{\drawsquare}[2]{\hbox{%
\rule{#2pt}{#1pt}\hskip-#2pt
\rule{#1pt}{#2pt}\hskip-#1pt
\rule[#1pt]{#1pt}{#2pt}}\rule[#1pt]{#2pt}{#2pt}\hskip-#2pt
horizontal
\rule{#2pt}{#1pt}}
\newcommand{\fund}{\raisebox{-.5pt}{\drawsquare{6.5}{0.4}}}
\newcommand{\antifund}{\overline{\fund}}

\newtheorem{theorem}{\bf THEOREM}
\def\thetheorem{\thesection.\arabic{theorem}}
\newtheorem{proposition}{\bf PROPOSITION}
\def\thetheorem{\thesection.\arabic{proposition}}
\newtheorem{observation}{\bf OBSERVATION}
\def\thetheorem{\thesection.\arabic{observation}}

\def\theequation{\thesection.\arabic{equation}}
\newcommand{\setall}{\setcounter{equation}{0}
         \setcounter{theorem}{0}}
\newcommand{\setequation}{\setcounter{equation}{0}}
\renewcommand{\thefootnote}{\fnsymbol{footnote}}

\setcounter{secnumdepth}{4}
\setcounter{tocdepth}{4}

~\\
\vskip 1cm

\begin{center}
{\Large \bf Gauge Theories, Tessellations \& Riemann Surfaces}
\end{center}
\medskip

\vspace{.4cm}
\centerline{
{\large Yang-Hui He}$^{1,2,3}$ \&
{\large Mark van Loon}$^3$
\footnote{hey@maths.ox.ac.uk; \quad mark.vanloon@merton.ox.ac.uk }
}
\vspace*{3.0ex}

\begin{center}
{\it
{\small
{
${}^{1}$ Department of Mathematics, City University, London, EC1V 0HB,
UK and \\
${}^{2}$ School of Physics, NanKai University, Tianjin, 300071,
P.R.~China and \\
${}^{3}$ Merton College, University of Oxford, OX1 4JD, UK\\
}
}}
\end{center}

\vspace*{4.0ex}
\centerline{\textbf{Abstract}} \bigskip
We study and classify regular and semi-regular tessellations of Riemann surfaces of various genera and investigate their corresponding supersymmetric gauge theories.
These tessellations are generalizations of brane tilings, or bipartite graphs on the torus as well as the Platonic and Archimedean solids on the sphere. On higher genus they give rise to intricate patterns.
Special attention will be paid to the master space and the moduli space of vacua of the gauge theory and to how their geometry is determined by the tessellations.

\newpage

\tableofcontents

\section{Introduction}
\comment{
The Standard Model of Particle Physics has been heralded as one of the greatest achievements of 20th century physics. It is a non-abelian gauge theory with symmetry group $U(1)\times SU(2) \times SU(3)$ and has provided us with a unified description of the strong and electroweak forces. However, its main shortcoming is that it does not incorporate gravity, as described by Einstein's theory of General Relativity, and hence cannot be a complete theory, which needs to go beyond the Standard Model and reproduce it at the appropriate energy scale. \\
String Theory is one such theory. It proposes that point-like particles are actually 1-dimensional strings, naturally living in 10 dimensions. Hence we see the need to reduce the dimensionality to the 4 effective dimensions of familiar spacetime.  The conventional approach is to achieve this through \emph{compactification}, with the 6 'extra' dimensions (i.e. 3 complex dimensions) manifesting itself as $M^{(6)}$, a smooth space that locally resembles a \emph{Calabi-Yau threefold}. \cite{He:2004rn} Noting that the only smooth Calabi-Yau threefold is $\mathbb{C}^3$, we see that we are naturally driven to study singular Calabi-Yau spaces. It is exactly the geometry of these spaces that allow for a reduction of the expected $U(n)$ gauge group in string theory to the $U(1)\times SU(2) \times SU(3)$ group of the Standard Model.\\

In recent years, it has been found that $\mathcal{N} = 1$ quiver gauge theories can be elegantly described in \emph{dimer models}, which offer an elegant way of analysing these quiver gauge theories. Conversely, any such tiling gives rise to a quiver gauge theory. In a recent paper, quiver gauge theories arising from tilings not on a torus, but on the genus $g=2$ Riemann surface are studied \cite{Cremonesi:2013aba}. 
We find all the quiver gauge theories arising from regular tilings of the Riemann surfaces of genus $g=0,1,2,3$, from all semi-regular tilings of the sphere and torus ($g=0$ and $g=1$ respectively) and some semi-regular on Riemann surfaces of genus $g=2,3$. \\
\\
}

Recently, there has been a host of activity in exploring a remarkable {\it bipartite structure} of supersymmetric gauge theories.
This has ranged from a relatively well-established programme of using doubly periodic brane-tilings, or equivalently, dimer models on the torus, to understand the four-dimensional $\cN=1$ quiver gauge theories of D3-branes probing toric Calabi-Yau spaces \cite{Feng:2000mi,Hanany:2005ve,Franco:2005sm,Jejjala:2010vb,Hanany:2011ra}, to matrix models and quiver calculators as a BPS state-counting mechanism \cite{Koch:2010zza,Pasukonis:2013ts}; from a systematic outlook of bipartite field theories (BFT) \cite{Franco:2012mm,Franco:2012wv,Franco:2013ana,Heckman:2012jh,Xie:2012mr} to relations to Grothendieck's {\it dessin d'enfant} and subsequent connections to number theory \cite{Ashok:2006br,He:2012kw,He:2012js,He:2012jn}; as well as to the vast and exciting subject of encoding scattering amplitudes in $\cN=4$ super-Yang-Mills theory in terms of planar bipartite graphs and cells in the positive Grassmannian \cite{ArkaniHamed:2012nw,Golden:2013xva,Cachazo:2012pz,Amariti:2013ija,Franco:2013nwa}.
Amongst these various facets there has been an ever-steady emergence of new connections and understanding.

At the crux of all of the above is the question of representing the gauge theory information in terms of bipartite graphs on Riemann surfaces, which is an important problem in and of itself in combinatorial geometry.
Along this vein there has been, with the aid of modern computing and algorithmic geometry \cite{book}, a persistent series of classification results in cataloguing these graphs in the context of gauge theories \cite{Hanany:2008fj,Hanany:2008gx,Davey:2009bp,Hanany:2011iw,Hanany:2012hi,Hanany:2012vc,Cremonesi:2013aba,He:2012jn,He:2013eqa,toappear}.

Motivated by this taxonomy of bipartite structures in gauge theory, our vision is two-fold: (1) to go beyond planar and doubly-periodic tilings and (2) to explore the terra incognita of non-bipartite field theories.
In the first direction, a recent work \cite{Cremonesi:2013aba} has nicely addressed the situation of genus 2 Riemann surfaces.
We will focus on a particular case of so-called {\it (semi-)regular} tilings on aribitrary Riemann surfaces and proceed, in incremental genus, to explicitly construct the relevant gauge theories.
Such tilings are fundamental to the study of {\it tessellations}, the planar case of which dates back to the geometric patterns known to early civilizations.
Throughout we will use the terms ``(brane) tilings'' and ``tessellations'' interchangeably. In particular, we will focus on the case where the tilings admit a dimer model and touch upon the cases beyond such bipartite colouring of the nodes.

In the second direction, very little work has been due to the augmenting complexity and the lack of physical intuition.
In the context of gauge theories, the toric condition restricts the superpotential to assume a particular form, consisting of plus/minus terms only, and grouped in pairs. This is the string-theoretic (AdS/CFT) origin of the bipartite graphs. There are, of course, geometrically engineered gauge theories which transcend this restriction and do no afford tiling/dimer descriptions.
Very quickly, in leaving the realm of toric Calabi-Yau spaces, we lose control of (coupling) parametres in the superpotential.
Constrained by (semi-)regularity, we will take the first steps in probing some of these theories and compute their moduli space of vacua.

The outline of the paper is as follows. In section \ref{subsec:dimermodels} we give a brief introduction to the use of dimer models in describing quiver gauge theories. Then in section \ref{subsec:modulispace} we explain how to use these models to recover the mesonic moduli space of the theory. Section \ref{sec:methodology} explains the exact methods we use to analyse our tilings. The methodology used to classify the regular and semi-regular tilings is described in section \ref{sec:Classification}. Our analysis of the tilings is found in sections \ref{sec:RegTilingResults} and \ref{sec:SemiRegTilingResults}.
The results are summarised in \ref{sec:conclusions}, which also provides directions of further research.
A brief foray into the study of tilings of Riemann surfaces which do not have bipartite structure is discussed in section \ref{sec:nonbipartite}.

\section{Bipartite Tessellation of Riemann Surfaces} 
\label{sec:Background}\setall

Let us begin by collecting some rudiments of the requisite mathematics and physics.
We will introduce the bipartite tiling of Riemann surfaces, emphasizing the approach from {\it dessins d'enfants} and permutation triples, and their physical realization of four-dimensional supersymmetric gauge theory.
The vacuum moduli space of the gauge theory will be an associated Calabi-Yau variety.
In a string theoretic realization of the gauge theory, geometrically engineered by configurations of brane tilings, or equivalently, as the dual world-volume theory of a D3-brane, the Calabi-Yau geometry is precisely the one which the branes probe.

\subsection{Dimer Models: Bipartite Graphs on Riemann Surfaces} 
\label{subsec:dimermodels}
Our dimer model consists of a balanced bipartite graph drawn on a Riemann surface $\Sigma$, i.e. a finite graph embedded into $\Sigma$ with an equal number of nodes coloured black and white, and with every black node only connected to white nodes and vice versa (cf.~e.g., \cite{He:2012js} for a rapid introduction).
We will use the dimer to encode a four-dimensional $\cN=1$ quiver gauge theory as follows:
the edges represent bi-fundamental (adjoint) fields $\Phi_i$, while each face with $N$ sides represents a $U(1)$ gauge group with $N$ fields charged under it. 
The direction of the bi-fundamental is determined by an overall choice of orientation of black-to-white.

A dimer with faces $F_1,\dots,F_k$ thus gives a total gauge symmetry of $U(1)^k$. Note each field is charged exactly twice as it is the border between two faces. The superpotential can also be recovered from the dimer as follows: to each vertex we assign a monomial equal to the product of the fields associated to the edges incident to the vertex. We circle the black vertices clockwise and the white vertices anti-clockwise. The superpotential is then the (weighted) sum of these monomials, where each monomial from a black node appears as a positive term and each one from a white node appears as a negative term \cite{Hanany:2005ve,Franco:2005sm}. 
As a result, each superpotential term appears exactly twice in the superpotential, once with a positive and once with a negative sign, hence capturing the toric nature of the theory in what has become known as the ``toric condition'' \cite{Feng:2000mi}.

The quiver diagram, which is a directed graph representing the gauge theory, with nodes representing the gauge groups and edges representing the bifundamental fields, is then found as the dual graph of the dimer. 
The direction is given by the orientation of the Riemann surface on which the dimer is drawn: the arrows are pointing in such a way that in the original tiling, the black node is on the right and the white node on the left. Note this is contrary to the convention in \cite{Hanany:2005ve}. This does not matter however: for any tiling, we could interchange our white and black nodes to use the convention in \cite{Hanany:2005ve}; this interchange merely changes the signs of the terms in the superpotential and hence does not affect the gauge theory.\\
The gauge invariants of the theory are then found as closed loops in the quiver \cite{Benvenuti:2006qr}.

The dimer model can also be represented by \emph{permutation triples} and \emph{Belyi pairs} \cite{Jejjala:2010vb,Hanany:2011ra}. 
The permutation triple $\{ \sigma_B, \sigma_W, \sigma_{\infty} \}$ is such that each cycle of $\sigma_B$ gives the clockwise ordering of the edges around a certain black node. The cycles of $\sigma_W$ correspond to clockwise orderings around white nodes. Note that the {\it same} direction is chosen, unlike the case of reading out the plus/minus terms in the superpotential.
The third permutation $\sigma_{\infty}$ is then found by imposing the condition, with multiplication as permutations in the symmetric group,
\begin{equation}
\sigma_B \sigma_W \sigma_{\infty} = {\rm Id} \ .
\end{equation}
An interesting thing to note here is that the cycle decomposition of $\sigma_{\infty}$ gives information about the faces of the bipartite tiling: there is a 1-1 correspondence between a $p-$cycle in $\sigma_{\infty}$ and a $2p-$gon in our tiling. In fact, this correspondence is such that each cycle in $\sigma_{\infty}$ corresponds to a set of outgoing arrows at a node in the quiver and giving the clockwise orientation with which they appear in the tiling.

Finally, we can realise the dimer concretely by having an algebraic model for the Riemann surface $\Sigma$, together with a rational map $\pi$ to $\IP^1$, ramified only at 3 points (say $0,1$ and $\infty$); this is the {\it Belyi pair}:
\begin{equation}
(\Sigma_g, \quad \pi: \Sigma_g \rightarrow \IP^1) \ .
\end{equation}
In the above and the ensuing, we write the genus $g$ of the Riemann surface as a subscript.
For the details on the construction of Belyi pairs, we refer the interested reader to \cite{Jejjala:2010vb,Hanany:2011ra}.
The key properties we need are the following.

We are mapping a Riemann surface of genus $g$ to $\IP^1$, of genus $G=0$, using a degree $d$ rational map, so it should satisfy the Riemann-Hurwitz relation:
\begin{equation}
2g - 2 = d(2G-2) + B \ , \quad
B = \sum_{i\in \{B,W,\infty \} } (d-C_{\sigma_i})
\end{equation}
where $B$ is the branching number and $C_{\sigma_i}$ represents the number of cycles in permutation $\sigma_i$.
Whence, we find that
\begin{equation}
d - n = 2 - 2g \ ,
\end{equation}
where $n$ is the number of ramification points. All ramification structures given in the paper can be checked to satisfy this condition.

\subsection{Master Space and Moduli Space} \label{subsec:modulispace}
An object of crucial importance in the study of $\mathcal{N} = 1$ supersymmetric gauge theories is the moduli space of vacua $\mathcal{M}$, given by the vanishing of scalar potential of the field theory \cite{argyres}.
It is the space of zeroes to the \emph{F-terms}:
\begin{equation}
\frac{\partial W(\phi)}{\partial \phi_i} = 0
\end{equation}
describing the extremisation of the \emph{superpotential} of the theory, and the \emph{D-terms}:
\begin{equation}
D^A = \sum_i \phi_{i0}^{\dagger} T^A \phi_{i0} = 0
\end{equation}
describing the orbits of the gauge invariant operators of the theory.

The space of F-flatness (solutions to the F-terms) is called the Master Space \cite{Forcella:2008bb,Forcella:2008eh}, denoted $\mathcal{F}^{\flat}$. The moduli space $\mathcal{M}$ is then the symplectic quotient:
\begin{equation}
\mathcal{M} \simeq \mathcal{F}^{\flat} // G_{D^{\flat}}
\end{equation}
where $G_{D^{\flat}}$ describes the D-flatness conditions.

It is worth noting that the vacuum moduli space is an affine variety \cite{Luty:1995sd,Gray:2006jb}, i.e., it is a submanifold of $\mathbb{C}^n$, the coordinates of whose points vanish exactly on some (finite) set of polynomials $\{ f_i(z_1,\dots,z_n)\}$.
The following theorem by Luty and Taylor proves useful to us \cite{Luty:1995sd}:
\begin{theorem}
Given a group $G_{D^{\flat}}$ acting on a variety $A$, there is a one-to-one correspondence between $A//G_{D^{\flat}}$ and the set of points in the affine variety $A^G$ defined by the ring $R_G$ of $G$-invariant elements in $R=R(A)$, where $R(A)$ is the ring of polynomials defining the variety $A$.
\end{theorem}

For us, this effectively means we can find the moduli space $\mathcal{M}$ as a quotient polynomial ring as follows \cite{Gray:2006jb}:
\begin{itemize}
\item First we define the polynomial ring $S = \mathbb{C}[\phi_1,\dots,\phi_n]$ where the $\{ \phi_1, \dots,\phi_n \}$ are the fields. We then wish to impose the F-term constraints $\partial_i W = 0$. This can be achieved by considering the ideal $I_1 = \left< \partial_i W \right>_{i=1,\dots,n}$. Then by definition, the quotient ring $\mathcal{F} = \mathbb{C}[\phi_1,\dots,\phi_n] / \left< \partial_i W \right>_{i=1,\dots,n} $ is a polynomial ring in which exactly all F-flatness is satisfied.
\item The D-term conditions are captured exactly by the holomorphic gauge invariants. Generally there is a large number of gauge invariants, carrying a certain amount of redundancy. We consider a minimal generating set $D = \{ r_j(\phi_i) \}_{j=1,\dots,k}$. As the $r_j$ are polynomials in the $\phi_i$, we can consider the set $D$ as a map from $S = \mathbb{C}[\phi_1,\dots,\phi_n]$ to $R=\mathbb{C}[r_1,\dots,r_k]$. To satisfy F-term constraints, we simply restrict the map to $\mathcal{F}$:
\begin{equation}
D: \mathcal{F} \rightarrow \mathbb{C}[r_1,\dots,r_k] \ .
\end{equation}
The moduli space $\mathcal{M}$ is then the image of this map:
\begin{equation}
\mathcal{M} \simeq \text{Im} \left( \mathcal{F} \xrightarrow{D} R\right) \ .
\end{equation}
\end{itemize}

One can re-phrase this as an elimination problem \cite{Hauenstein:2012xs} which can then be addressed using parallelisable algorithms.
Alternatively, we can also use Gr\"obner-basis techniques to study the affine variety \cite{Gray:2006jb}, computing such quantities as dimension, degree and Hilbert series of the moduli space of vacua \cite{Benvenuti:2006qr}.


\subsection{Classification of (Semi-)Regular Tessellations} \label{sec:Classification}
We now move on to describe in detail the protagonist of our concerns.
In this section, we provide a complete classification of all regular tilings on Riemann surfaces of genus $g=0,1,2,3$. We first define a regular and semi-regular tiling and show our methods for classifying them. We then show the results of our computational analysis of the gauge theory arising from the tiling.

As there are several inequivalent definitions of (semi-)regular tessellations, and to assist the reader not familiar with the terminology used, we first give an overview of some important definitions and terms.
We then discuss properties of (semi-)regular tessellations that will aid us in their classification, after which we restrict ourselves to the easier case of the regular tessellations. We discuss the construction we used to attempt a complete classification of these.
For the interested reader, we have also included a proof (unfortunately non-constructive) that for any Riemann surface $\Sigma$ of a given genus $g$, the number of semi-regular maps that can be embedded in it is finite.

\subsubsection{Terminology and Definitions}

A tessellation of a Riemann surface $\Sigma$, being a graph embedded onto $\Sigma$, consists of vertices, edges and faces and has an associated symmetry group we will denote by $G$.
We define a tessellation to be \emph{semi-regular} if its symmetry group $G$ acts transitively on its vertices.
To define \emph{regular} tessellations, we first define a \emph{flag} of the tiling to be a triple of vertex, incident edge and face of the tessellation. A tessellation is then \emph{regular} if the symmetry group $G$ acts transitively on the flags. Note this implies that $G$ also acts transitively on the the vertices, so any result holding for semi-regular maps holds for regular maps.

Given the vertex transitivity condition of semi-regular tessellations, we note that, given any surface to embed in, we can classify tessellations according to the structure of the edges around each vertex. Following notation from \cite{walsh72}, we will write the cyclic sequence $\mathbf{x} = (p_1,p_2,\dots,p_q)$ for a semi-regular tessellation where every vertex is of valency $q$ and is surrounded by faces that are $p_i$-gons, appearing in the cyclic order given by $\mathbf{x}$. Note physics imposes on us the constraints $q, p_i \geq 3$.

In the case of a regular tiling, we see that by flag-transitivity, we must have $p_1 = p_2 = \dots = p_q =: p$, so that $\mathbf{x} = (p,p,\dots,p)$. In this case, it is more useful to classify the type of tessellation by its \emph{Schl{\"a}fli symbol} $\{p,q\}$, where $q$ is the valency of the vertices and all the faces are $p$-gons.

A further useful definition is that of a \emph{dart}: a half-edge of the tessellation, of which there are $2E$.
We consider in our symmetry group $G$ the subgroup $H$ of rotational symmetries. We note that any element $h \in H$ is completely determined by its action on one of the darts of the tessellation, hence we see $H$ is a group of order $|H| = 2E$, a fact that will come in useful later in classifying all regular tilings.

\subsubsection{Classification}
We consider a regular tessellation of type $\mathbf{x} = (p_1,p_2,\dots,p_q)$.
Define $m_p$ to be the multiplicity of $p$ in $\mathbf{x}$, i.e. $m_p$ is such that every vertex has $m_p$ incident $p$-gons. Then we must have the relation:
\begin{equation} \label{eq:FpMultiplicity}
F_p = \frac{m_p}{p}V \ ,
\end{equation}
where $F_p$ is the number of $p$-gons in our tiling. This rearranges to give
\begin{equation} 
F = \sum_p F_p = V \sum_p \frac{m_p}{p} = V \sum_{i=1}^q \frac{1}{p_i}  \ .
\end{equation}
By considering the number of edges meeting at each vertex, and by noting genus $g$ Riemann surfaces are orientable and have Euler characteristic $\chi = 2-2g$, we have the relations:
\begin{equation} \label{eq:SemiRegTilingRelations}
2E = qV,\qquad\qquad V-E+F = 2-2g \ .
\end{equation}
Combining these results, we get the relation:
\begin{equation} \label{eq:SemiRegTilingEqn} \sum_{i=1}^q \frac{1}{p_i} - \frac{q}{2} + 1 +\frac{2g-2}{V} = 0 \ .
\end{equation}

We now consider two cases.

\begin{figure}
\begin{center}
\includegraphics[width=18cm]{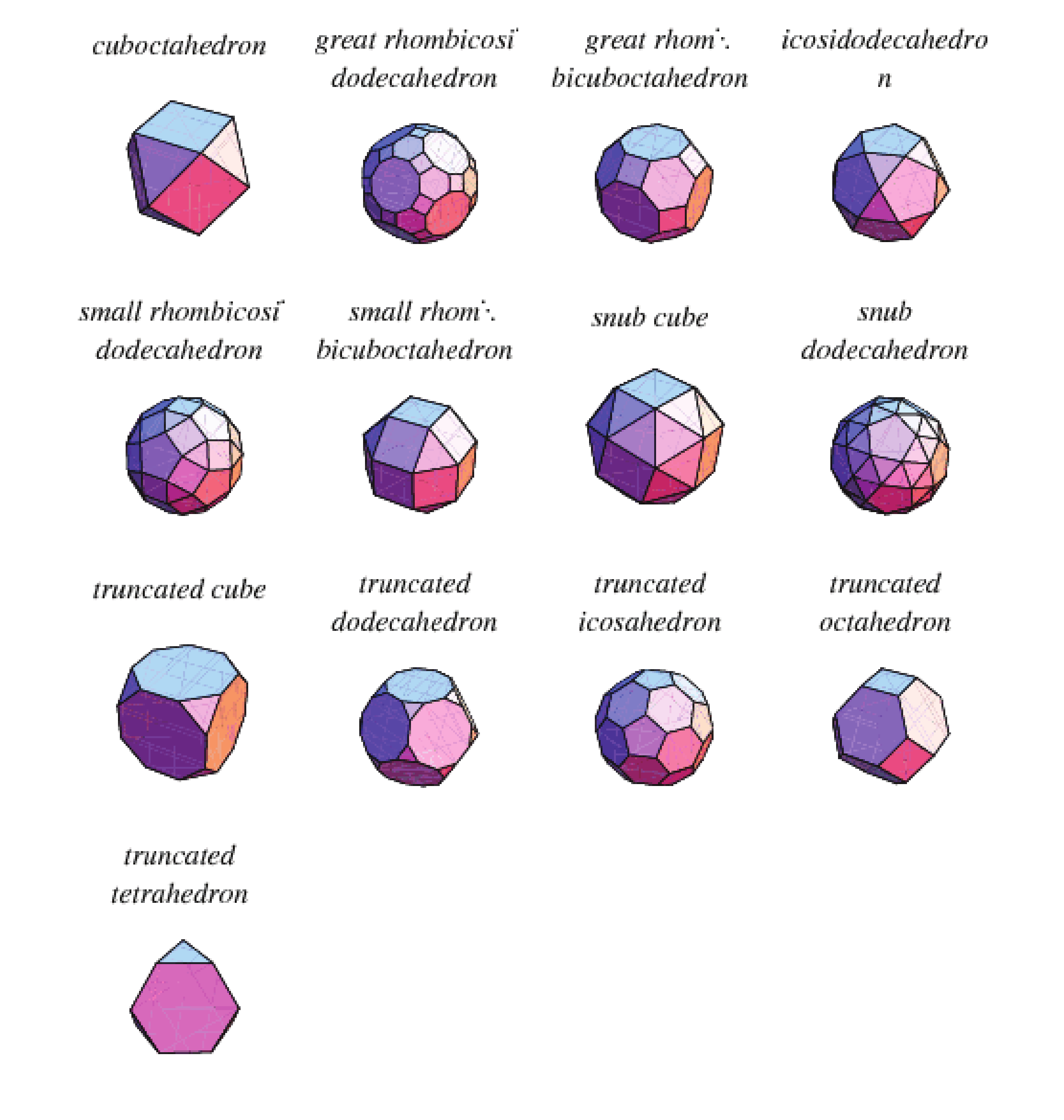}
\end{center}
\sf{\caption{The Archimedean Solids, corresponding to the semi-regular tessellations of the plane.}
\label{f:archimedean}}
\end{figure}

\paragraph*{Case $\mathbf{g=0}$: }  We get that $\frac{1}{q} \left( 1+ \sum\limits_{i=1}^q \frac{1}{p_i}\right) = \frac{1}{2} +\frac{2}{V}\geq \frac{1}{2}$. This condition gives us exactly five regular tessellations of the sphere, corresponding to the Platonic solids, and thirteen other semi-regular tessellations, corresponding to the well-known Archimedean solids, pictured in Figure \ref{f:archimedean} (cf.~\cite{wolframg=0} and detailed discussions in \cite{He:2013eqa}).

\paragraph*{Case $\mathbf{g\geq1}$: } Note this holds if and only if $1-g \leq 0$. Then we have that
\begin{equation} 
\sum_{i=1}^q \frac{1}{p_i} - \frac{q}{2} \leq \sum_{i=1}^q \frac{1}{3} - \frac{q}{2} \leq -\frac{q}{6} \ ,
\end{equation}
where we have used the fact that $p_i \geq 3$ for all $i$.
Hence we see that 
\begin{equation} \label{IntermediateBoundEqn}
-\frac{q}{6} \geq \sum_{i=1}^q \frac{1}{p_i} - \frac{q}{2} = -1 +\frac{2-2g}{V}\geq -1 + 1 - g = -g \ ,
\end{equation}
in the case that $V \geq 2$, which leads to the bound:
\begin{equation} \label{qbound}
 q\leq 6g
 \end{equation}
for (semi-)regular tessellations.
Had we restricted ourselves to bipartite tessellations, we would need $p_i \geq 4$ and we would get the bound
\begin{equation} \label{qboundbipartite}
 q\leq 4g \ .
 \end{equation}
In the case that $V=1$, we see that equation \ref{IntermediateBoundEqn} gives us $-\frac{q}{6} \geq 1-2g$, or bound $q \leq 12g - 6$.

\subsubsection{Regular Tessellations} \label{sec:RegTilingClass}

We first limit our discussion to that of regular tessellations, using the same notation as before. Let $\Sigma_g$ be the Riemann surface of genus $g$. Consider a regular map with Schl{\"a}fli symbol $\{p,q\}$. Then by considering each edge as seperating two faces, we get the relation $pF = 2E$. Combining this with equation \ref{eq:SemiRegTilingRelations}, we get the following relations:
\begin{equation} \label{eq:RegTilingRelations}
qV = pF = 2E, \qquad \qquad V- E + F = 2-2g \ ,
\end{equation}
which rearranges to:
\begin{equation} \label{eq:RegTilingEqn}
\frac{1}{p} + \frac{1}{q} + \frac{g-1}{E} = \frac{1}{2} \ .
\end{equation}

Note this is merely a necessary, and not sufficient condition. However, for any fixed $g\in \mathbb{N}$, we will see there is only a finite number of solutions $(p,q,E)\in \mathbb{N}^3$ and hence we can consider these on a case-by-case basis.

From a physical point of view,  we must also impose for consistency that $p,q\geq3$.
It is easy to see from \ref{eq:RegTilingEqn} that there is something interesting going in when $g=1$. Hence we first consider the case $g\neq 1$, leaving the $g=1$ case to a separate section.
In our analysis, it turns out to be expedient to first consider seperately the case $g=0$.

\begin{figure}[h!!!]
\begin{center}
\includegraphics[width=10cm]{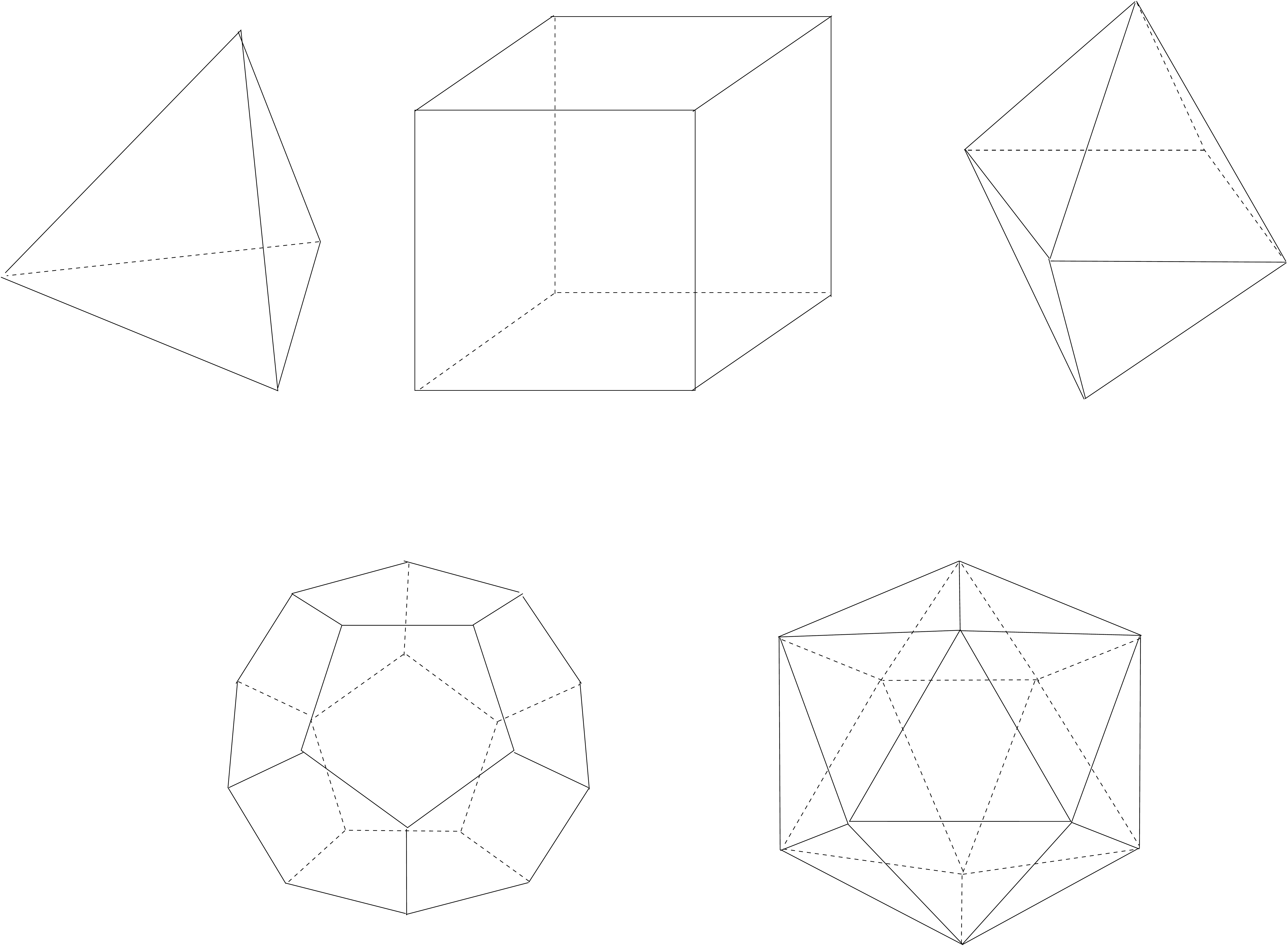}
\end{center}
\sf{\caption{The five Platonic Solids, corresponding to the regular tessellations of the plane.}
\label{f:platonic}}
\end{figure}

\paragraph*{Case $\mathbf{g=0}$: }
we see that $ \frac{1}{p} + \frac{1}{q}  = \frac{1}{2} + \frac{1}{E} > \frac{1}{2}$. As $p,q\in\mathbb{N}$, this means that the only solutions in the required range are 
\begin{equation} \label{eq:genus0solutions}
\{p,q\} = \{3,3\},\{3,4\},\{4,3\},\{3,5\},\{5,3\} \ .
\end{equation}
Note that by \ref{eq:RegTilingEqn}, each solution fixes the number of edges $E$.
These solutions describe the famous Platonic solids: they are (respectively) the tetrahedron, the octahedron, the cube, the icosahedron and dodecahedron, pictured in Figure \ref{f:platonic}.

\paragraph*{Case $\mathbf{g>1}$: }
For the case of arbitrary genus $g\geq 2$, we must solve over the natural numbers for:
\begin{equation} \label{regtilingeqn}
\frac{1}{p} + \frac{1}{q} = \frac{1}{2} - \frac{g-1}{E} \ .
\end{equation}
We note that $\frac{1}{p},\frac{1}{q} \in \{ \frac{1}{3},\frac{1}{4},\frac{1}{5},\dots \}$ and that $\frac{1}{p} + \frac{1}{q} < \frac{1}{2}$, whence
\begin{equation}
\frac{1}{p} + \frac{1}{q} \leq \max \left( \{ \frac{1}{n} + \frac{1}{m} : n,m \in \mathbb{N} \} \cap (0,\frac{1}{2}) \right)= \frac{1}{3} + \frac{1}{7}
\ . 
\end{equation}
Thus we see that $\frac{g-1}{E} \geq \frac{1}{2} - (\frac{1}{3} + \frac{1}{7}) = \frac{1}{42}$, so we get an upper bound: 
\begin{equation} 
E \leq 42(g-1) \ .
\end{equation}
Now we note that for a fixed $E$, we have $\frac{1}{2} - \frac{g-1}{E} \in \left( \frac{1}{n+1},\frac{1}{n} \right]$ for some $ n\in\mathbb{N}$. Hence we need $p,q \geq n+1$.
We saw before that for $g\geq 1$ we have bound $q\leq 4g$ for (semi-)regular tilings with bipartite structure with $V>1$. Noting that any regular tiling of type $\{p,q\}$ has a regular tiling that is realised as its dual (i.e. by interchanging vertices and faces) and is of type $\{q,p\}$, we see by symmetry that for regular tilings we must also have bound $p\leq 6g$ whenever $F>1$.
In the case that $F=1$, we can use the bound previously obtained for $q$ when $V=1$ to see that $p \leq 12g - 6$.
So we see there are only finitely many solutions in the positive integers for any genus $g$.

We used the bounds thus obtained to generate, in Matlab or Mathematica, for example, all possible solutions for $\{p,q\}$. Note that the existence of such a solution does not automatically imply a tiling with this $p$ and $q$ actually exists. A complete overview of the solutions for each genus and details regarding their existence can be found in the ensuing subsections.

An interesting thing to note about the regular tilings is that for $g\geq 2$ we can see certain ``families" of tilings appearing, with very similar looking tilings and quivers, such as the family with $V=2,E=4g,F=2g,p=4,q=4g$, the family with $V=2g,E=4g,F=2,p=4g,q=4$, or more obviously, the families with $V=1,E=2g, F=1,p=4g,q=2g$ or $V=2,E=2g+2,F=2,p=2g+2,q=2g+2$.

\paragraph*{Case $\mathbf{g=1}$: }
In this special case, equation \ref{eq:RegTilingEqn} becomes $ \frac{1}{p} + \frac{1}{q}  = \frac{1}{2}$, and again we can proceed via simple elimination, getting solutions $\{p,q\} = \{3,6\},\{4,4\},\{6,3\}$.
Alternatively, we can note this is the torus, whose fundamental polygon is doubly periodic (i.e. a parallelogram) with opposite edges identified. Hence we see that (both for regular and semi-regular tilings) these tessellations are in a one-to-one correspondence with regular and semi-regular, respectively, tessellations of the plane by polygons. Complete classifications of these already exist in the literature \cite{wolframg=1}.

We note that, unlike for other genera, the number of edges $E$ is not fixed by equation \ref{eq:RegTilingEqn} for $g=1$. However, note that each regular tiling corresponds to a regular tessellation of the plane with a fundamental domain imposed. Hence each choice of fundamental domain gives a new gauge theory, but having $m\times n$ copies of the fundamental domain simply corresponds to an orbifold action by $\mathbb{Z}_m \times \mathbb{Z}_n$ \cite{Franco:2005sm}. We do not consider these cases here.

The image below shows the three regular tilings of the plane, by triangles, squares and hexagons respectively:
\begin{center}
\includegraphics[width=15cm]{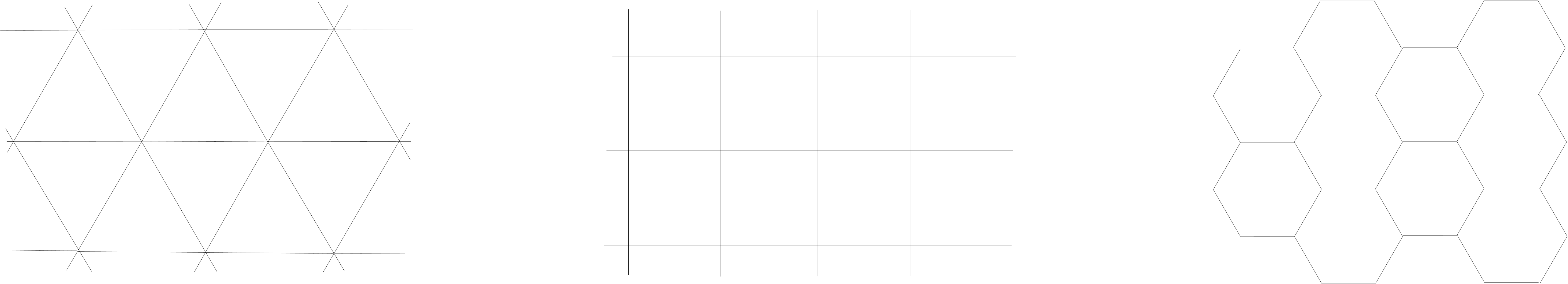}
\end{center}

To find a fundamental domain of our tessellation we need to specify a parallelogram with a base point $\mathbf{a}$ and sides given by vectors $\mathbf{x}$  $\mathbf{y}$, where $\mathbf{x}$ and $\mathbf{y}$ are such that our tessellation is invariant under translation by them.
However, specifying $\mathbf{x}$ and $\mathbf{y}$ is  actually sufficient. Imagine having a parallelogram superimposed upon our lattice. Identify this parallelogram as the ordered triple $( \mathbf{a},\mathbf{x},\mathbf{y})$, where $\mathbf{a}$ is the `base point' of the parallelogram and $\mathbf{x},\mathbf{y}$ giving the sides (i.e. our parallelogram has vertices $\mathbf{a}, \mathbf{a + x}, \mathbf{a+y}, \mathbf{a+x+y}$).

Now consider a second parallelogram $( \mathbf{b},\mathbf{x},\mathbf{y})$. The two parallelograms are easily seen to be homotopic by ``sliding" one onto the other. We also note that at every step in this sliding, the diagram - as living on the torus - is preserved, because whenever a vertex, edge or face `disappears' on one side, it reappears at the opposite side (by the translational symmetry of our lattice under $\mathbf{x},\mathbf{y}$).

We find any two vectors $\mathbf{g},\mathbf{h}$ that give us a primitive fundamental domain, where primitive means there is not fundamental domain contained strictly within it (see e.g. figure below). We can then form new vectors $\mathbf{z}(m,n) := m\mathbf{g} + n\mathbf{h}$ (where $m,n\in\mathbb{Z}$) that are again translational symmetries of our lattice.

Conversely, any translational symmetry must be of this form.
To see this, consider a tessellation in which $\mathbf{\mathbf{g}},\mathbf{h}$ are such that they give a primitive fundamental domain. Suppose $\mathbf{x} := \vec{A B}$ is a symmetry of the lattice. As $B$ is in some parallelogram, we can write $\mathbf{x} = a\mathbf{g} + b\mathbf{h} + \mathbf{r_x}$ where $a,b\in\mathbb{Z}$ and $\mathbf{r_x}$ is contained in the parallelogram $\left( (0,0),\mathbf{g},\mathbf{h} \right)$. Then note $\mathbf{r_x} = \mathbf{x} - a\mathbf{g} - b\mathbf{h}$ is also a translational symmetry (as $\mathbf{x},\mathbf{g},\mathbf{h}$ are). Consider then for example the fundamental domain given by vectors $\mathbf{g}$ and $\mathbf{r_x}$. Note that this gives parallelograms that are strictly contained in those given by vectors $\mathbf{g},\mathbf{h}$, which is a contradiction.

So once we have a primitive fundamental domain given by two vectors $\mathbf{g}, \mathbf{h}$, we can form any other fundamental domain by making two vectors $\mathbf{x} = \mathbf{z}(a_1,b_1) =a_1 \mathbf{g} + b_1 \mathbf{h}$ and $\mathbf{y} = \mathbf{z}(a_2,b_2)= a_2 \mathbf{g} + b_2 \mathbf{h}$ where we require that $\mathbf{x_1}$ and $\mathbf{x_2}$ are not linearly dependent, if and only if $\frac{a_1}{b_1} \neq \frac{a_2}{b_2}$, if and only if
$J = \det \left( \begin{array}{cc}
a_1 &a_2 \\
b_1 & b_2
\end{array} \right) \neq 0$.

Now note that the Jacobian $J$ is in fact the scaling factor of the area, i.e.
\begin{equation} \label{eq:JacobianArea}
J = \frac{A_{new}}{A_{old}} \ ,
\end{equation}
where $A_{new}, A_{old}$ denote the area of our new, respectively old, fundamental domain.
The mapping $f: \{ \mathbf{g,h} \} \rightarrow \{ \mathbf{x,y} \}$ with associated Jacobian $J_f\neq 0$ is invertible with inverse $f^{-1}$ which has Jacobian $1/J_f$. 

We see that hence we can classify all fundamental domains by their area and define a minimal fundamental domain to be a pair of vector symmetries of the planar tiling $\{\mathbf{g,h}\}$ such that any mapping $f: \{ \mathbf{g,h} \} \rightarrow \{ \mathbf{x,y} \}$ to another such pair has an associated Jacobian $J_f \geq 1$. Note we have equality if and only if the target fundamental domain is also minimal. Equivalently, this means that the area of the fundamental domain is minimal in the set $\{ A(\rho) | \rho \text{ is a fundamental domain} \}$.
Because the fundamental domain $\{ \mathbf{g,h} \}$ is minimal, this means that any mapping $\pi$ onto another fundamental domain $\{ \mathbf{x} = a_1 \mathbf{g} + b_1 \mathbf{h} ,\mathbf{y}= a_2 \mathbf{g} + b_2 \mathbf{h} \}$ must satisfy $a_1,a_2,b_1,b_2 \in \mathbb{Z}$ and hence:
\begin{equation} \label{eq:intJacobian}
J_{\pi} = |a_1b_2 - a_2b_1| \in \mathbb{Z} \ .
\end{equation}

We now show there is a direct correspondence between the area of a fundamental domain and the number of vertices it contains.
To see this, we will apply Pick's theorem, which states that any polygon on a square lattice (where each square has unit area), with its vertices all lattice points, has an area as follows:
\begin{equation} \label{eq:pick}
A = i + \frac{b}{2} - 1 \ ,
\end{equation}
where $i$ is the number of lattice points interior to the polygon and $b$ the number of lattice points on the boundary.

To see this applies, we consider a minimal fundamental domain of a tiling, which - without loss of generality - we can take to have corners coinciding with the vertices of the tiling. By using an appropriate shear mapping (which preserves the area of polygons) and imposing a suitable distance measure, we can then take the fundamental domain to be a square with its sides parallel to the $x$-axis and $y$-axis respectively.
To any fundamental domain, whose corners have to be lattice points, we can apply Pick's theorem to see that 
\begin{equation}
A = V \ .
\end{equation}

This happens as we can split the boundary vertices into two sets: the four vertices on the corners, which, after identification of opposite sides, only contribute to one vertex in our toric tiling, and the other boundary vertices $\tilde{b}$, which contribute in pairs to one vertex in the toric tiling, hence negating the fact that equation \ref{eq:pick} has a factor of $\frac{1}{2}$ in it.
Combining this result with equations \ref{eq:JacobianArea} and \ref{eq:intJacobian}, we see that the number of vertices $V(new)$ in \emph{any} fundamental domain is an integer multiple $n$ of the number of vertices $V(old)$ in the minimal domain.
For any (semi-)regular tiling, we can then use the relation $2E = qV$, to see $\frac{E(new)}{E(old)} = n = \frac{V(new)}{V(old)}$.

If we look now at the number $F_p$ of $p$-gons, we have according to equation \ref{eq:FpMultiplicity} that $F_p = V \frac{m_p}{p}$, where $m_p$ is the multiplicity of $p$-gons around each vertex, so that again: $\frac{F_p(new)}{F_p(old)} = n = \frac{V(new)}{V(old)}$.
Hence we see that in fact, any fundamental domain gives a gauge theory that is an orbifold of the gauge theory given by a minimal fundamental domain.

So we need only consider each solution for $\{p,q\}$ once. Also note that we cannot impose a bipartite structure on the tessellation by triangles, so we are simply left with two regular tessellations, corresponding to the $\mathbb{C}^3$ and conifold theories.

\subsubsection{Semi-Regular Tessellations} \label{sec:SemiRegTilingClass}
\begin{figure}[H]
\begin{center}
\includegraphics[width=10cm]{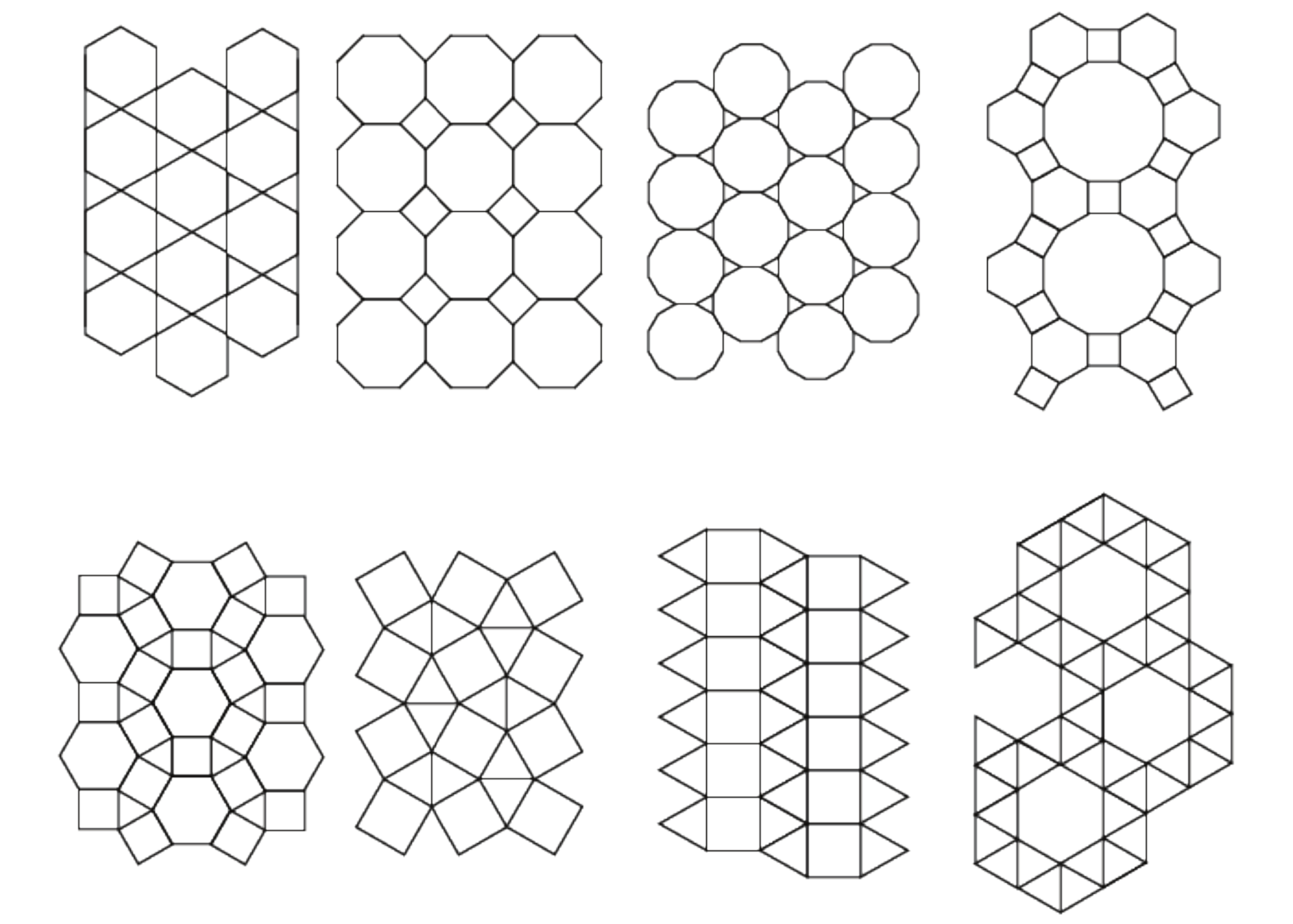}
\end{center}
\sf{\caption{The semi-regular tessellations of the plane. Picture obtained from \url{http://mathworld.wolfram.com/Tessellation.html}}
\label{f:semiregplanetilings}}
\end{figure}

In this subsection, we give a proof that the number of semi-regular tessellations of a Riemann surface $\Sigma_g$ is finite. This proof is not necessary for the remainder of the paper and hence can be skipped; it is included for the interested reader.\\ 
The proof is not constructive and does not aid in actually finding these tilings; fortunately however, for $g=0$ and $g=1$ complete classifications already exist \cite{wolframg=0,wolframg=1}. They correspond to the Archimedean solids and semi-regular tessellations of the planes respectively, the latter of which are pictured in Figure \ref{f:semiregplanetilings}.

To see the number of tessellations of $\Sigma_g$ is finite, we first recall that $q$ is bounded.
We now show that $E$ - and hence $V$ - is bounded for $g > 1$. Note we need not consider $g=0$ or $g=1$, as we already know the number of tessellations is finite.
We consider equation \ref{eq:SemiRegTilingEqn} and rearrange to get: 
\begin{equation} 
\frac{g-1}{E} = \frac{1}{2} - \frac{1}{q} - \frac{1}{q}\sum_{i=1}^q \frac{1}{p_i} \geq  \frac{1}{3}\left( \frac{1}{2} - \sum_{i=1}^q \frac{1}{p_i} \right) \ ,
\end{equation}
where we use the fact that $q \geq 3$.
We now show that either $ \sum\limits_{i=1}^q \frac{1}{p_i} = \frac{1}{2}$, which happens if and only if $g=1$, or there exists $\epsilon_0 > 0$ such that $ \sum\limits_{i=1}^q \frac{1}{p_i} < \frac{1}{2} - 3 \epsilon_0$, whence it follows that for $g>1$:
\begin{equation} \label{Eboundeqn} E \leq \frac{1}{\epsilon_0} (g-1) \ .
\end{equation}
To see this, we first fix $ 3\leq q \leq \max \{ 6g, 12g - 6 \}$. Suppose for a contradiction that given any $\epsilon > 0$, there are $(p_1,\dots,p_q)$ such that 
\begin{equation} \label{p_ieqn} 0 < \frac{1}{2} - \sum_{i=1}^q \frac{1}{p_i} \leq \epsilon \ .
\end{equation}
Without loss of generality we can impose that $p_1 \leq p_2 \leq \dots \leq p_{q-1} \leq p_q$.
We first consider $p_1$. Note that $\sum_{i=1}^q \frac{1}{p_i} \geq \sum_{i=1}^q \frac{1}{p_1} = \frac{q}{p_1}$. Hence equation \ref{p_ieqn} implies that $\frac{1}{p_1} \geq \frac{1}{q} S^{(\epsilon)}_1$ where $S^{(\epsilon)}_1 = \frac{1}{2} - \epsilon$, or equivalently:
\begin{equation} 
p_1 \leq \frac{q}{S^{(\epsilon)}_1} \ .
\end{equation}
So we see that $p_1$ can only take the finite number of values $\{ 3,4,\dots,\lfloor \frac{q}{S^{(\epsilon)}_1} \rfloor \}$.

Fix any such value of $p_1$. We then want to choose $(p_2,\dots,p_q)$ such that $\sum\limits_{i=2}^q \frac{1}{p_i} \geq \frac{1}{2} - \frac{1}{p_1} - \epsilon$. Now we define for any fixed $(p_1,\dots,p_{n-1})$:
\begin{equation} 
S_n^{(\epsilon,p_1,\dots,p_{n-1})} = \frac{1}{2} - \sum_{i=1}^{n-1} \frac{1}{p_i} - \epsilon  \ .
\end{equation}
Again, we take note of the ordering on the $p_i$ and conclude that $\frac{1}{p_2} \geq \frac{1}{q-1}  S_2^{(\epsilon,p_1)}$, or equivalently:
\begin{equation} 
p_2 \leq \frac{q-1}{S^{(\epsilon,p_1)}_2} \ . 
\end{equation}
Now fixing $p_2$ to be any of the finite number of values it can take, we proceed to inductively get the bound:
\begin{equation} p_n \leq \frac{q-n+1}{S^{(\epsilon,p_1,\dots,p_{n-1})}_n} 
\end{equation}
for $n\leq q-1$.

Now we consider again equation \ref{p_ieqn} and note we get, upon rearranging: 
\begin{equation} \frac{1}{p_q} \geq \frac{1}{2} - \sum_{i=1}^{q-1} \frac{1}{p_i} - \epsilon \end{equation}
We have two cases to consider:
\begin{itemize}
\item If $\frac{1}{2} - \sum\limits_{i=1}^{q-1} \frac{1}{p_i} \neq \frac{1}{N}$ for some $n \in \IN$, we can pick $\epsilon >0$ small enough such that no $p_q \in \IN$ satisfies this equation.
\item If $\frac{1}{2} - \sum\limits_{i=1}^{q-1} \frac{1}{p_i} = \frac{1}{N}$ for some $n \in \IN$, we can pick $\epsilon >0$ small enough such that the only solution is such that $\frac{1}{p_q} = \frac{1}{2} - \sum\limits_{i=1}^{q-1} \frac{1}{p_i}$, which is a contradiction of equation \ref{p_ieqn}.
\end{itemize}
Either way, by the finiteness of the possibilities of $(p_1,\dots,p_q)$, we see that it is not possible to find, given any $\epsilon > 0$, $(p_1,\dots,p_q)$ such that \ref{p_ieqn} holds. Hence there must exist some $\epsilon_0$ such that 
\begin{equation} 
\sum_{i=1}^q \frac{1}{p_i} < \frac{1}{2} - 3 \epsilon_0 \ .
\end{equation}
We have hence shown that \ref{Eboundeqn} holds.
Furthermore, to see all the $p_i$ are bounded, we simply note that each edge of the tiling can only appear as an edge of any face twice. Hence we see that necessarily $p_i \leq 2E$ for all $p_i$.

Finally, we see that on any Riemann surface, there are only a finite number of solutions to equation \ref{eq:SemiRegTilingEqn}, a necessary equation for a semi-regular tiling of type $(p_1,\dots,p_q)$ to exist. \\
We will see in section \ref{sec:regulartilings} that it is possible to have multiple (semi-)regular tilings of the same type. However, to see this number is finite, we simply note that for any two such tilings $T_1, T_2$ there is a bijection $f: V_1 \times E_1 \times F_1 \rightarrow V_2 \times E_2 \times F_2$. As these sets are all finite by the above, we note there can only be finitely many such $f$. So there are only finitely many semi-regular maps of a certain type on any given Riemann surface.

We conclude that on any Riemann surface of a given genus $g$, there are only finitely many semi-regular tilings.

\subsection{Outline of Computation} 
\label{sec:methodology}
The course of our action is therefore clear.
\begin{itemize}
\item We will study tessellations of Riemann surfaces, focusing on the regular and semi-regular classes. Having obtained the finiteness results from the above discussions, we will explicitly find regular and semi-regular tessellations in sections \ref{sec:regulartilings} and \ref{sec:semiregulartilings} respectively for various genera.
\item Having found a tiling, we then impose a bipartite structure on it, noting that this is unique up to interchange of black and white vertices (which would merely flip all signs in later calculations, hence not affecting the gauge theory). This may not always be possible and we will leave the impossible cases to a discussion in section \ref{sec:nonbipartite}.
\item We then establish the quiver gauge theory associated to the tiling. As described in section \ref{subsec:dimermodels}, we assign a superfield $\Phi_i$ and a $U(1)$ gauge group to each face. As described earlier, we then find the superpotential $W$ by considering each vertex as generating a monomial.
We also draw the corresponding quiver diagram as the dual of the regular tiling.
\item Given the gauge data in terms of the matter content and the superpotential, we can proceed to compute the relevant moduli space.
First, we list the partials $ \{ \frac{\partial W}{\partial \Phi_j}\}_{j=1,\dots,n} $ of the superpotential. We then use a computer algebra system, for example Macaulay2 \cite{mac2}, to generate the master space $\mathcal{F}^{\flat}$, which is realised as the quotient:
\begin{equation}
\mathbb{C}[\Phi_i]_{i=1,\dots,n} / \left<\partial_j W\right>_{j=1,\dots,n} \ .
\end{equation}
We would then like to study the {\it branches} of the vacua.
This is done by finding the {\it primary decomposition} of the ideal $I=\left< \partial_j W \right>_{j=1,\dots,n}$ and listing its components. 
Generally, all but one of these will be low-dimensional, linear pieces. The most interesting is the top-dimensional ideal, the \emph{coherent component}, usually written $^{\text{Irr}}\mathcal{F}^{\flat}$, which is a Calabi-Yau manifold of the same dimension and degree as the master space itself \cite{Forcella:2008bb,Forcella:2008eh}.
\item
We compute the Hilbert series of these ideals. 
This is the central object to the {\it Plethystic Programme} of studying operator enumeration of supersymmetric gauge theories \cite{Benvenuti:2006qr,Feng:2007ur}.
For a (quotient) polynomial ring $X = \bigoplus_{i\geq0} X_i$, the Hilbert series is given by 
\begin{equation}
H(t) = \sum_{i=0}^{\infty} \text{dim}(X_i)t^i = P(t) / (1-t)^{\text{dim } X} 
\end{equation}
for some polynomial $P(t)$ with integer coefficients, where dim$(X_i)$ represents the number of independent polynomials of a degree $i$ on the variety.
\item From the Master space, we can proceed to compute the full (mesonic) moduli space.
We consider the gauge invariants $\{ r_j \}_{j=1,\dots,k}$, corresponding to closed loops of minimal length in the quiver, i.e., those that do strictly contain another closed loop. We bypass the mapping as described in section \ref{subsec:modulispace} by considering ring $S = \mathbb{C}[\Phi_i,y_j]_{i=1,\dots,n;\, j=1,\dots,k} $ with ideal $J=\left< \partial_i \Phi, y_j\right>_{i=1,\dots,n;\, j=1,\dots,k}$. We subsequently eliminate the $\Phi$ variables and substitute the ideal $J$ into the new ring $R = \mathbb{C}[y_j]_{j=1,\dots,k}$ to get the (mesonic) moduli space as:
\begin{equation}
\mathbb{C}[y_j]_{j=1,\dots,k} / \left< \partial_i \Phi, y_j\right>_{i=1,\dots,n;\, j=1,\dots,k} \ .
\end{equation}
We find that the dimension of $\mathcal{M}_{mes}$ is equal to $1+2g$, agreeing with well-known $g=1$ gauge theories (i.e., Calabi-Yau threefolds) and the $g=2$ theories in \cite{Cremonesi:2013aba}. The moduli space is a Calabi-Yau manifold if the tiling is consistent \cite{Cremonesi:2013aba}.


Due to the complexity of Groebner basis calculations, based on the Buchberger algorithm, which has large space complexity and doubly exponential time complexity in the input size \cite{buch1,buch2}, we were not able to compute the full information for the moduli spaces for tilings with too many edges or gauge invariants.
\item
We also provide a permutation triple with its associated \emph{ramification structure}, which gives information about the cycle decomposition, in accordance with the procedure outlined in \cite{Jejjala:2010vb}, to aid in any future construction of a Belyi pair from the genus $g$ Riemann surface to the sphere.
\end{itemize}

\section{Regular Tessellations} \label{sec:regulartilings}\setall
We follow the prescription of the above subsection and proceed to study the tessellations in detail.
We begin with the regular cases and then move on to the semi-regular ones in the next section.
We find it expedient to summarize, for the reader's ease, the results of the geometrical properties first before addressing the individual examples.

\subsection{Summary of results} \label{sec:RegTilingResults}

The table below shows a summary of the results, with more detailed results in the relevant subsections.
The first column is the genus $g$ of the Riemann surface and the Schl{\"a}fli symbols $\{p,q\}$ (recall $q$ is the valency of the vertices and all the faces are $p$-gons).
The second column lists the dimensions and the degrees of the master space ${\cal F}^{\flat}$ and the full mesonic moduli space ${\cM_{mes}}$.
The third column presents the Hilbert spaces of the affine varieties ${\cal F}^{\flat}$, its coherent (highest dimensional irreducible) component $^{\text{Irr}}\mathcal{F}^{\flat}$ and ${\cM_{mes}}$.
Due to the computational complexity of our calculations, as described in section \ref{sec:methodology}, the expensiveness of the primary decomposition and Gr\"obner basis calculations prevented us in giving all results with our present computing resources; these will be marked ``n/a''.
We show as much information as possible, to make any future calculations easier.

\begin{longtable}{cc|c|rl}
$g$ & \{$p$,$q$\} & (dim,deg)$_{\mathcal{F}^{\flat}}$	&	& Hilbert series  \\	&	&	&						& \\
				&											&	(dim,deg)$_{\cM_{mes}}$							&						& \\
\cline{1-5}
$0$ 		& \{$ 4$,$3$\} 			& $(6$,$14)$ 			& $\mathcal{F}^{\flat}:$ 					& $(1-t)^{-6}(1 + 6t  + 9t^2 -5t^3 +3t^4)$  \\
				&											&									& $^{\text{Irr}}\mathcal{F}^{\flat}:$ 		& $(1-t)^{-6}( 1+6t+6t^2+t^3)$ \\
				& 							 			& $(1$,$1)$				&	$ \cM_{mes}:$ 									& $(1-t^3)^{-1}$  \\
\cline{1-5}
$1$ 		&	\{$6$,$3$\}					& $(3$,$1)$				&$\mathcal{F}^{\flat}:$ 					& $(1-t)^{-3}$  \\
				&											&									&$^{\text{Irr}}\mathcal{F}^{\flat}:$ 		& $(1-t)^{-3}$ \\
				&											& $(3$,$1)$				&$ \cM_{mes}:$										& $(1-t)^{-3}$  \\
				\cline{2-5}
				&	\{$4$,$4$\}					&	$(4$,$1)$				&$\mathcal{F}^{\flat}:$ 					& $(1-t)^{-4}$ \\
				&											&									&$^{\text{Irr}}\mathcal{F}^{\flat}:$ 		& $(1-t)^{-4}$ \\
				&											&	$(3$,$2)$				&$ \cM_{mes}:$ 										& $(1-t^2)^{-3}(1+t^2)$ \\
\cline{1-5}
$2$			& \{$4$,$6$\} 				& $(10$,$9)$			&$\mathcal{F}^{\flat}:$ 					& $(1-t)^{-10}(1+2t+3t^2+4t^3+5t^4-6t^5-8t^6+8t^7)$  \\
				&											&									&$^{\text{Irr}}\mathcal{F}^{\flat}:$ 		& $(1-t)^{-10}(1+2t + 3t^2  + 2t^3 + t^4)$\\
				&											&	$(5$,$216)$			&$ \cM_{mes}:$ 										& $(1-t^6)^{-5}(1+44t^6 + 126t^{12}+44t^{18}+t^{24})  $ \\
				\cline{2-5}
				& \{$4$,$8$\} 				& $(8$,$1)$				&$\mathcal{F}^{\flat}:$ 					& $(1-t)^{-8}$  \\
				&											&									&$^{\text{Irr}}\mathcal{F}^{\flat}:$ 		& $(1-t)^{-8}$ \\
				& 										& $(5$,$24)$			&$ \cM_{mes}:$ 										& $(1-t^4)^{-5}(1+11t^4 + 11t^8+t^{12})$  \\
				\cline{2-5}
				& \{$6$,$4$\} 				& $(8$,$16)$			&$\mathcal{F}^{\flat}:$ 					& $(1-t)^{-8}(1 + 4t + 10t^2 + 8t^3 -6t^4 - 20t^5 $\\
				&											&									&																	& $+ 28t^6 - 12t^7 + 3t^8)$  \\
				&											&									&$^{\text{Irr}}\mathcal{F}^{\flat}:$ 		& $(1-t)^{-8}(1+4t+6t^4+4t^3+t^4)$ \\
				& 										& $(5$,$216)$			&$ \cM_{mes}:$										& $(1-t^4)^{-5}(1+47t^4+114t^8+62t^{12}-11t^{16}+3t^{20})$  \\
				\cline{2-5}
				& \{$6$,$6$\} 				& $(6$,$1)$				&$\mathcal{F}^{\flat}:$ 					& $(1-t)^{-6}$  \\
				&											&									&$^{\text{Irr}}\mathcal{F}^{\flat}:$ 		& $(1-t)^{-6}$ \\
				& 										& $(5$,$6)$				&$ \cM_{mes}:$ 										& $(1-t^2)^{-5}(1+4t^2+t^4)$  \\
				\cline{2-5}
				& \{$8$,$3$\} 				& $(10$,$594)$		&$\mathcal{F}^{\flat}:$ 					& $(1-t)^{-10}(1 + 14 t + 81 t^2 + 233 t^3 + 268 t^4 $\\
				&											&									&																	& $- 45 t^5 - 63 t^6 + 105 t^7)$ \\
				&											&									&$^{\text{Irr}}\mathcal{F}^{\flat}:$ 		& $n$/$a$ \\
				& 										& $(5$,$96)$			&$ \cM_{mes}:$ 										& $(1-t^3)^{-5}(1+20t^3+54t^6+20t^9+t^{12})$ \\	
				\cline{2-5}
				& \{$8$,$4$\} 				& $(6$,$4)$				&$\mathcal{F}^{\flat}:$ 					& $(1-t)^{-6}( 1 + 2t + 3t^2 - 4t^3 + 2 t^4)$  \\
				&											&									&$^{\text{Irr}}\mathcal{F}^{\flat}:$ 		& $(1-t)^{-6}(1+2t+t^2)$\\
				& 										& $(5$,$6)$				&$ \cM_{mes}:$ 										& $(1-t^2)^{-5}(1+11t^2+11t^4+t^6)$  \\
				\cline{2-5}
				& \{$10$,$5$\} 				& $(5$,$1)$				&$\mathcal{F}^{\flat}:$ 					& $(1-t)^{-5}$ \\
				&											&									&$^{\text{Irr}}\mathcal{F}^{\flat}:$ 		& $(1-t)^{-5}$ \\
				& 										& $(5$,$1)$				&$ \cM_{mes}:$ 										& $(1-t)^{-5}$ \\
\cline{1-5}
$3$			&	\{$4$,$6$\}					&	$(18$,$896)$		&$\mathcal{F}^{\flat}:$ 					& $(1-t)^{-18}(1 + 6t +21t^2 + 56t^3 + 126t^4 + 228t^5 $ \\
				&											&									&																	& $+ 335t^6 + 390t^7+ 300t^8 - 70t^9 - 543t^{10} $ \\
				&											&									&																	& $ -660t^{11} -187t^{12} +282t^{12}+ 1329t^{14} - 1340 t^{15} $ \\
				&											&									&																	& $ + 894t^{16} - 384t^{17} + 139t^{18} - 30t^{19} + 3t^{20})$\\
				&											&									&$^{\text{Irr}}\mathcal{F}^{\flat}:$ 		& $n/a$ \\
				& 										& $n/a$						&$ \cM_{mes}:$ 										& $n/a$ \\				
				\cline{2-5}
				& \{$4$,$8$\}$_A$			& $(14$,$16)$			&$\mathcal{F}^{\flat}:$ 					& $(1-t)^{-16}(1 + 2t+3t^2 + 4t^3 + 5t^4 + 6t^5 + 7t^6$ \\
				&											&									&																	& $- 8t^7 - 10t^8 - 12t^9 +18t^{10})$  \\
				&											&									&$^{\text{Irr}}\mathcal{F}^{\flat}:$ 		& $n$/$a$\\
				& 										& $(7$,$88)$			&$ \cM_{mes}:$ 										& $(1-t^4)^{-7}(1 +9t^4 + 37t^8 + 29t^{12} + 32t^{20} $ \\
				&											&									&																	&	$- 35t^{24}+ 21t^{28} - 7t^{32} + t^{36})$ \\
				\cline{2-5}
				& \{$4$,$8$\}$_B$			& $(14$,$16)$			&$\mathcal{F}^{\flat}:$ 					& $(1-t)^{-16}(1 + 2t+3t^2 + 4t^3 + 5t^4 + 6t^5 + 7t^6$ \\
				&											&									&																	& $- 8t^7 - 10t^8 - 12t^9 +18t^{10})$  \\
				&											&									&$^{\text{Irr}}\mathcal{F}^{\flat}:$ 		& $n$/$a$\\
				& 										& $n/a			$			&$ \cM_{mes}:$ 										& $n/a$ \\
				\cline{2-5} 
				& \{$4$,$12$\} 				& $(12$,$1)$			&$\mathcal{F}^{\flat}:$ 					& $(1-t)^{-12}$\\
				& 										&									&$^{\text{Irr}}\mathcal{F}^{\flat}:$ 		& $(1-t)^{-12}$ \\
				& 										& $(7$,$720)$			&$ \cM_{mes}:$ 										& $(1-t^6)^{-7}(1+57t^6 + 302t^{12}+302t^{18}+57t^{24}+t^{30})$\\
				\cline{2-5}
				& \{$6$,$4$\}					& $(14$,$2048)$		&$\mathcal{F}^{\flat}:$ 					&$(1-t)^{-14}(1 + 10t + 55t^2 + 196t^3 + 488t^4 $ \\
				&											&									&																	&	$+ 812t^5 + 716t^6 -284t^7 - 484t^8 + 212t^9 + 500t^{10} $\\
				&											&									&																	&$-276t^{11} + 117t^{12} - 18t^{13} + 3t^{14})$\\
				& 										&									&$^{\text{Irr}}\mathcal{F}^{\flat}:$ 		& $n$/$a$ \\
				& 										& $n$/$a$				 	&$ \cM_{mes}:$ 										& $n$/$a$\\				
				\cline{2-5}
				& \{$8$,$4$\}$_A$			& $(10$,$96)$			&$\mathcal{F}^{\flat}:$ 					& $(1-t)^{-10}(1 + 6t + 21t^2 + 40t^3 + 39t^4 -30t^5 + 19t^6)$\\
				& 										&									&$^{\text{Irr}}\mathcal{F}^{\flat}:$ 		& $n$/$a$ \\
				& 										& $n$/$a$				 	&$ \cM_{mes}:$ 										& $n$/$a$\\
				\cline{2-5}
				& \{$8$,$4$\}$_B$ 		& $(10$,$64)$			&$\mathcal{F}^{\flat}:$ 					& $(1-t)^{-10}(1 + 6t +21t^2 + 40t^3 + 39t^4 - 30t^5$ \\
				&											&									&																	& $- 99t^6 + 44t^7 + 106t^8 - 96t^9 + 32t^{10})$ \\
				&											&									&$^{\text{Irr}}\mathcal{F}^{\flat}:$ 		& $n$/$a$ \\
				& 										& $n$/$a$					&$ \cM_{mes}:$ 										& $n$/$a$ \\
				\cline{2-5}
				& \{$8$,$8$\}			 		& $(8$,$1)$				&$\mathcal{F}^{\flat}:$ 					& $(1-t)^{-8}$ \\
				&											&									&$^{\text{Irr}}\mathcal{F}^{\flat}:$ 		& $(1-t)^{-8}$ \\
				& 										& $(7$,$20)$			&$ \cM_{mes}:$ 										& $(1-t)^{-7}(1+9t^2+9t^4+t^6)$\\				
				\cline{2-5}
				& \{$12$,$4$\}				& $(8$,$16)$			&$\mathcal{F}^{\flat}:$ 					& $(1-t)^{-8}(1+4t+10t^2+8t^3-6t^4-20t^5$ \\
				&											&									&																	& $+28t^6-12t^7+3t^8)$ \\
				&											&									&$^{\text{Irr}}\mathcal{F}^{\flat}:$ 		& $(1-t)^{-8}(1+4t+6t^2+4t^3+t^4)$ \\
				& 										& $(7$,$320)$			&$ \cM_{mes}:$ 										& $(1-t^2)^{-7}(1+29t^2+145t^4+109t^6+23t^8$ \\
				&											&									&																	& $+19t^{10}-9t^{12}+3t^{14})$\\
				\cline{2-5}
				& \{$14$,$7$\}				& $(7$,$1)$				&$\mathcal{F}^{\flat}:$ 					& $1$ \\
				&											&									&$^{\text{Irr}}\mathcal{F}^{\flat}:$ 		& $1$ \\
				& 										& $(7$,$1)$				&$ \cM_{mes}:$ 										& $(1-t)^{-7}$ \\
\end{longtable}

\subsection{Detailed results} \label{sec:DetailedResults}
Having seen the geometries of the various moduli spaces at a glance, we now study the physics of the associated gauge theories in detail, genus by genus, and Schl{\"a}fli symbol by Schl{\"a}fli symbol.

Since the same Riemann surface can be represented in the plane in many different ways, we have carte blanche to choose whichever representation works best in portraying the particular tiling we are interested in. Hence there will many different looking representations of the same surface.\\
We represent the \emph{surfaces} in which our tilings are embedded in brown, whereas the \emph{edges} of the tiling are represented by black lines. If the surface is represented by identification of sides of a polygon, the arrows on the sides show both which sides match up and the direction in which they do so. If the surface is represented as circles which are identified with each other, then lowercase letters around these circles show which circles match up and in which direction they do so.
In the $g=2$ case, because we can conformally map things to the hyperbolic plane, the tilings are perhaps better represented inside polygons, as was done in \cite{Cremonesi:2013aba}, which is no longer guaranteed for general genus.

Also, to avoid too many over-crossings in the planar representation of the quiver diagrams, we will often represent the quiver diagram as a 3D-drawing, so as to better illustrate its symmetries. The topology of the graph should be clear from the context.

\subsubsection{Genus 0}
For the Riemann sphere (genus 0), we choose as representation a disk with its entire boundary identified to one point.
We found the possible Schl{\"a}fli symbols for regular tilings in equation \eqref{eq:genus0solutions}, which correspond to the five platonic solids. Note the cube with $\{ p,q \} = \{ 4,3 \} $ is the only regular tiling that allows us to impose a bipartite structure on it.
This has been studied in the context of modular groups and dessins in \cite{He:2012kw,He:2013eqa}.

\paragraph*{\fbox{$\mathbf{\{p,q\} = \{4,3\}}$}}
This is the cube, with $V=8,E=12,F=4$.
\begin{center}
\includegraphics[width=15cm]{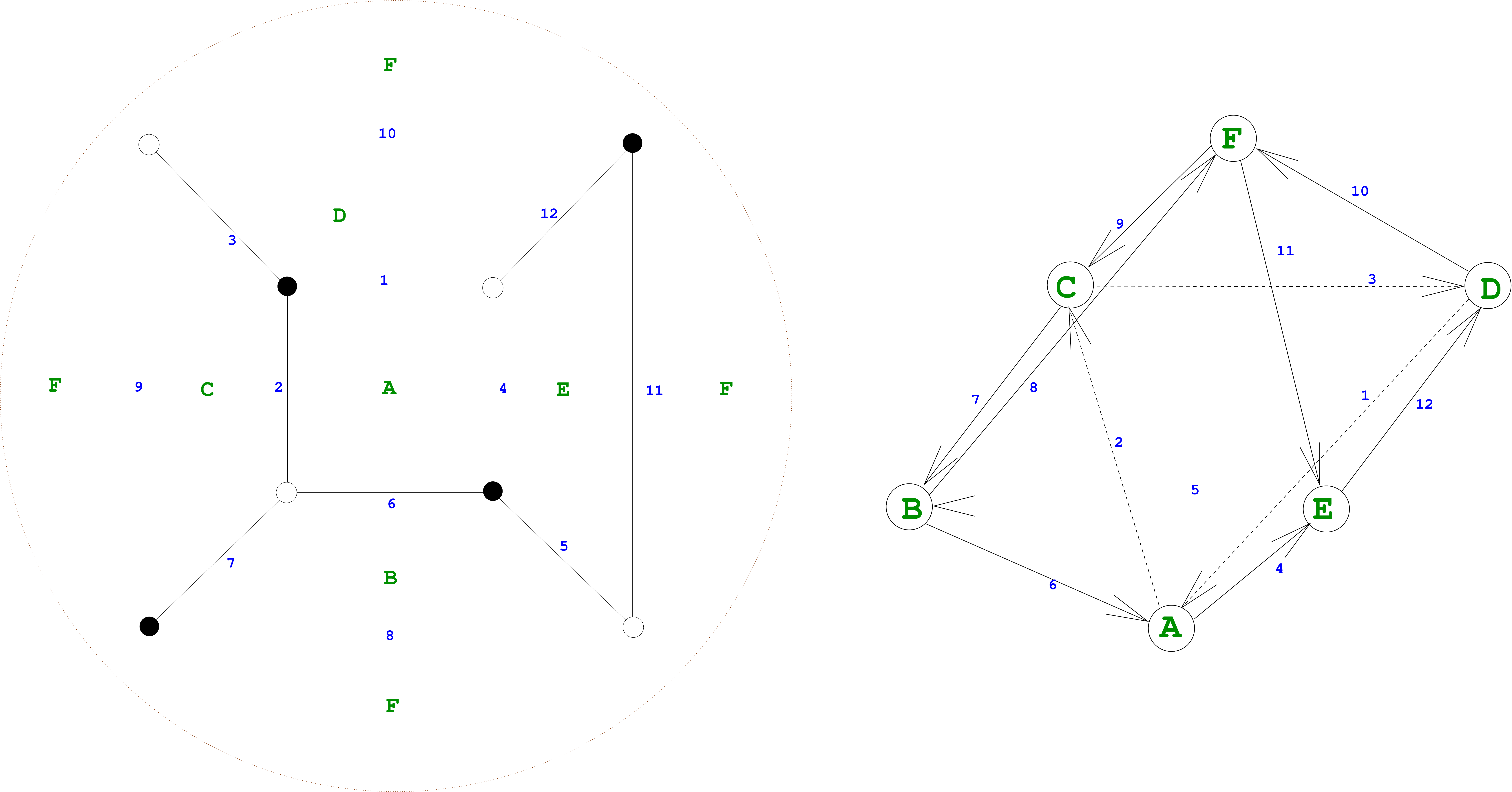}
\end{center}
We have superpotential
\begin{align}
W = \tr (& \Phi_1\Phi_2\Phi_3 + \Phi_4\Phi_5\Phi_6 + \Phi_7\Phi_8\Phi_9 + \Phi_{10}\Phi_{11}\Phi_{12} \nonumber \\
&- \Phi_1\Phi_4\Phi_{12} - \Phi_2\Phi_7\Phi_6 - \Phi_3\Phi_{10}\Phi_9 - \Phi_5\Phi_8\Phi_{11} )
\end{align}
and 8 gauge invariants, which are exactly the superpotential terms:
\begin{align}
\nonumber
r_1 = \Phi_1\Phi_2\Phi_3 , \ , \quad
r_2 = \Phi_1\Phi_4\Phi_{12} , \ , \quad
r_3 = \Phi_2\Phi_7\Phi_6 , \ , \quad
r_4 = \Phi_3\Phi_{10}\Phi_9, \\
r_5 =  \Phi_4\Phi_5\Phi_6 \ , \quad
r_6 = \Phi_5\Phi_8\Phi_{11} \ , \quad
r_7 = \Phi_7\Phi_8\Phi_9 \ , \quad
r_8 = \Phi_{10}\Phi_{11}\Phi_{12} \ .
\end{align}

To find the master space $\cF^{\flat}$, we define polynomial ring $S = \mathbb{C}[\Phi_1,\dots,\Phi_{12}]$ and ideal $I_1 = \left< \partial_i W \right>_{i=1,\dots,12}$. We then generate master space $R = S /I_1$, and using singular we find that $I_1$ has dimension 6, degree 14 and Hilbert series
\begin{equation} 
H(t, \cF^{\flat}_{g=0, \ (p,q)=(4,3)}) = 
\frac{1 + 6t  + 9t^2 -5t^3 +3t^4 }{(1-t)^6} \ .
\end{equation}

Using primary decomposition with \cite{mac2}, we get that the variety given by $I_1$ is the union of those given by ideals of which:
(a) 3 are trivial with degree 1, dimension 4 and Hilbert series $\frac{1}{(1-t)^4}$; these are just copies of $\IC^4$;
(b) the \emph{coherent component} is of degree 14, dimension 6 and has Hilbert series
\begin{equation} 
H(t, \ ^{\text{Irr}}\cF^{\flat}_{g=0, \ (p,q)=(4,3)}) = 
\frac{1+6t+6t^2+t^3}{(1-t)^6} \ .
\end{equation}
This Hilbert series has a palindromic numerator, and thus by Stanley's theorem \cite{Forcella:2008bb} the coherent component is an affine Calabi-Yau space, of complex dimension 6.

To find the full (mesonic) vacuum moduli space, we consider the ring $R = \mathbb{C} \left[ \Phi_1,\dots\Phi_{12},y_1,\dots,y_8 \right]$ and ideal $I_2 = \left< \partial_i W,y_j - r_j \right>_{i=1,\dots,12;j=1,\dots,8}$.
We then eliminate all the $\Phi$s and substitute the resulting ideal into ring $R' = \mathbb{C}\left[ y_1,\dots,y_8\right]$ to get ideal $V$ representing the vacuum moduli space. Using \cite{mac2}, we see that V has dimension 1, degree $1$ and after assigning weights to each $y_j$ equal to the degree of the monomial they represent, we get Hilbert series
\begin{equation}
H(t, \cM_{g=0, \ (p,q)=(4,3)}) = \frac{1}{1-t^3} \ .
\end{equation}
This shows that the full (mesonic) moduli space is nothing but $\IC^3$ here.

\subsubsection{Genus 1}
The genus $g=1$ situation is familiar to us as {\it brane tilings} which are dimer models on the doubly periodic plane and constitutes all the so-called toric AdS$_5$/CFT$_4$ theories \cite{He:2012js}.
Therefore, the gauge theories below will be familiar to us.

The tilings are represented as a (brown) fundamental domain imposed on an infinite tiling of the plane.

\paragraph*{\fbox{$\mathbf{\{p,q\} = \{6,3\}}$}}
This is the well-known, $\cN=4$ SYM, or, the $\mathbb{C}^3$ theory with dimer model and quiver as follows:
\begin{center}
\includegraphics[width=12cm]{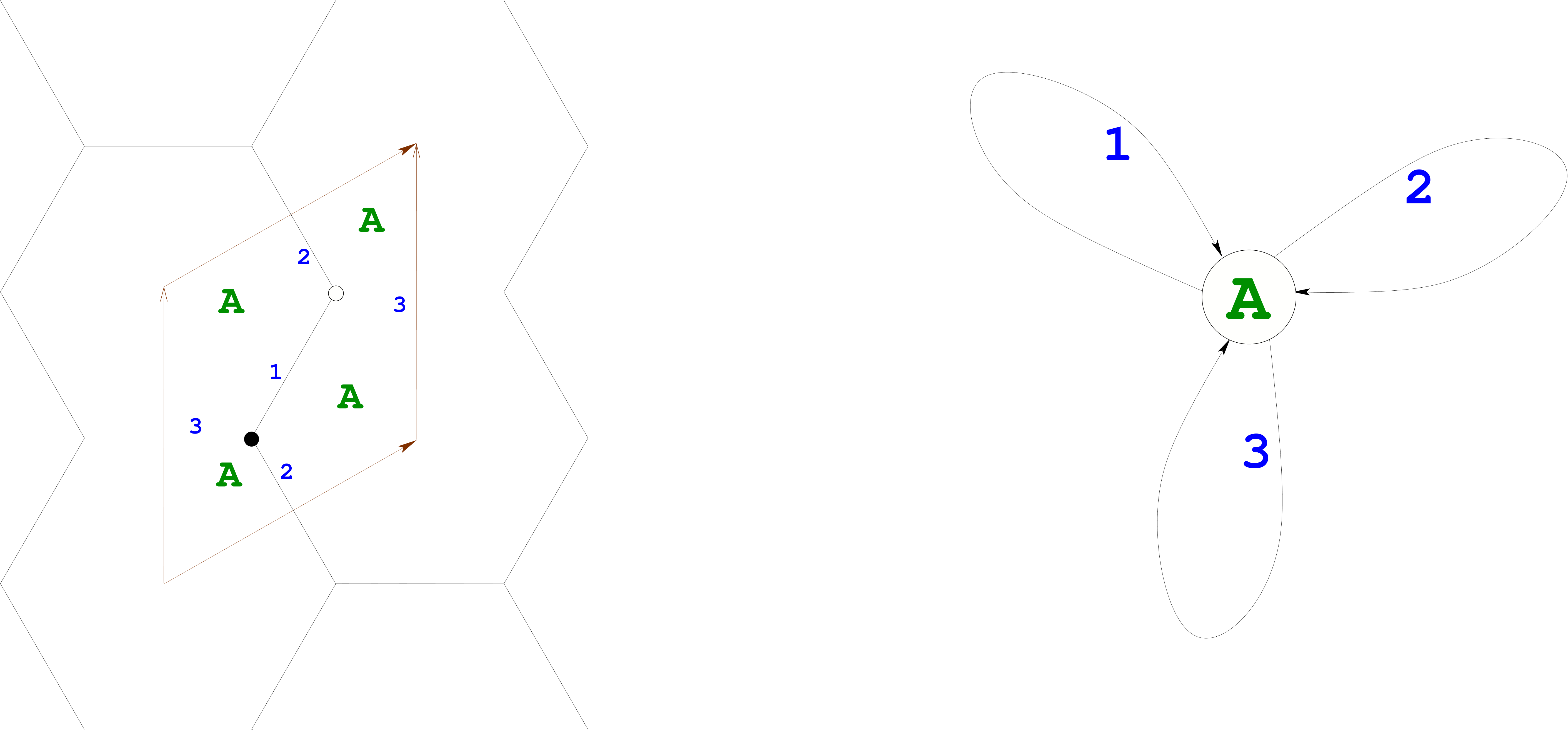}
\end{center}
We have superpotential:
\begin{equation}
W = \tr \left( \Phi_1\Phi_2\Phi_3 - \Phi_1\Phi_3\Phi_2 \right)
\end{equation}
with F-terms
\begin{equation}
[\Phi_i, \Phi_j] = 0 \ .
\end{equation}
We also see from the quiver that the adjoints fields $\Phi_i$ are themselves gauge invariants.
Working  in the case of rank one for the gauge groups and thus over the complex numbers for the fields, the master space is
\begin{equation}
S/I_1 = \mathbb{C} \left[ \Phi_1,\dots\Phi_3\right] / \left< \partial_i W\right>_{i=1,\dots,3} = \mathbb{C} \left[ \Phi_1,\dots\Phi_3\right] \ ,
\end{equation}
and is thus freely generated.

To find the full (mesonic) vacuum moduli space, we consider the ring $R = \mathbb{C} \left[ \Phi_1,\dots\Phi_3,y_1,\dots,y_3 \right]$ and ideal $I_2 = \left< \partial_i W,y_i - \Phi_i \right>_{i=1,\dots,3}$, and we get that $I_2 = \left< y_i -\Phi_i \right>_{i=1,\dots,3}$ as, for $\Phi_i \in \mathbb{C}$, all partials are zero.
We then eliminate all the $\Phi$s and substitute the resulting ideal into ring $R' = \mathbb{C}\left[ y_1,\dots,y_3\right]$ to get ideal $V$ representing the vacuum moduli space. We see that
\begin{equation} 
R'/V \cong \mathbb{C} \left[ y_1,\dots,y_3 \right] \ ,
\end{equation}
which has degree 1, dimension 3 and Hilbert series 
\begin{equation} 
H(t, \cM_{g=1, \ (p,q)=(6,3)}) = H(t, \ \cM_{\IC^3}) = \frac{1}{(1-t)^3} \ ,
\end{equation}
which shows it is the Calabi-Yau 3-fold $\IC^3$, as by construction.

We can also write down a permutation triple such that $\sigma_B \sigma_W \sigma_{\infty} = id$: 
\begin{equation}
\sigma_B = (1\;2\;3) \ , \quad
\sigma_W = (1\;2\;3) \ , \quad
\sigma_{\infty} = (1\;2\;3) \ .
\end{equation}
Hence we have ramification structure $\{3,3,3\}$.
A suitable Belyi pair \cite{Jejjala:2010vb} is: $y^2 = x^3 + 1$, with $\beta(x,y) = \frac{1}{2}(1+y)$.

\paragraph*{\fbox{$\mathbf{\{p,q\} = \{4,4\}}$}}
This is the well-known \emph{conifold} theory:
\begin{center}
\includegraphics[width=15cm]{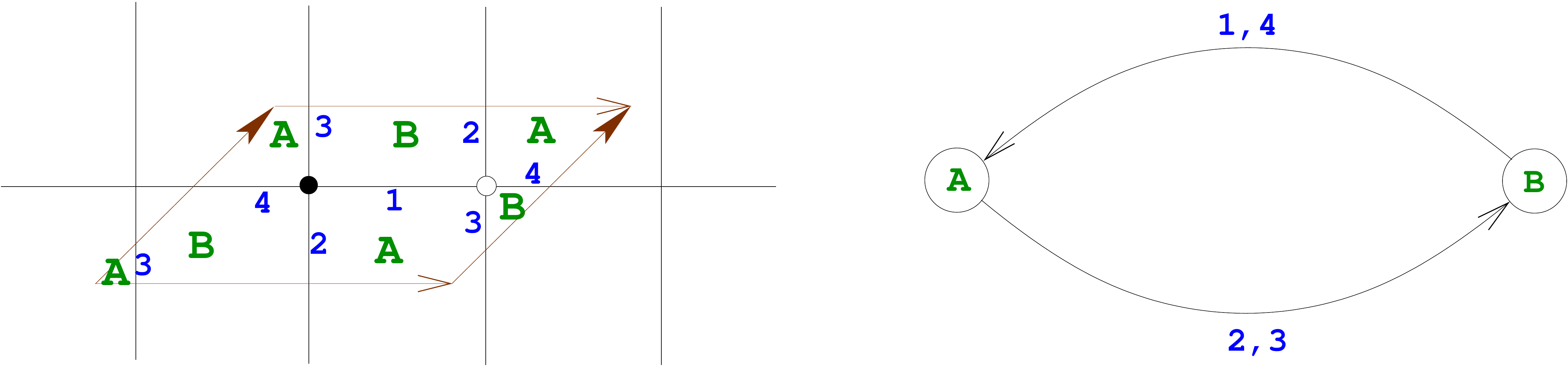}
\end{center}
We have superpotential
\begin{equation}
W = \tr \left( \Phi_1\Phi_2\Phi_4\Phi_3 - \Phi_1\Phi_3\Phi_4\Phi_2  \right) \ .
\end{equation}
As an illustration, the (cyclic) partials are listed here:
\begin{align}
\nonumber
\partial_1 W &= \tr \left( \Phi_2\Phi_4\Phi_3 - \Phi_3\Phi_4\Phi_2 \right) \ , \quad
\nonumber
\partial_2 W = \tr \left( \Phi_4\Phi_3\Phi_1 - \Phi_1\Phi_3\Phi_4 \right) \\
\nonumber
\partial_3 W &= \tr \left( \Phi_1\Phi_2\Phi_4 - \Phi_4\Phi_2\Phi_1 \right) \ , \quad
\partial_4 W = \tr \left( \Phi_3\Phi_1\Phi_2 - \Phi_2\Phi_1\Phi_3 \right) \ ;
\end{align}
Henceforth, these F-terms shall be omitted.
We also have 4 gauge invariants:
\begin{equation}
r_1 = \Phi_1\Phi_2 \ , \quad
r_2 = \Phi_1\Phi_3 \ , \quad
r_3 = \Phi_4\Phi_2 \ , \quad
r_4 = \Phi_4\Phi_3 \ .
\end{equation}
We see that, working over the complex numbers, all partials vanish, so the master space is
\begin{equation}
S/I_1 = \mathbb{C} \left[ \Phi_1,\dots\Phi_4\right] / \left< \partial_i W\right>_{i=1,\dots,4} = \mathbb{C} \left[ \Phi_1,\dots\Phi_4\right] \ .
\end{equation}

To find the full vacuum moduli space, we consider the ring $R = \mathbb{C} \left[ \Phi_1,\dots\Phi_4,y_1,\dots,y_4 \right]$ and ideal $I_2 = \left< \partial_i W,y_j - r_j \right>_{i=1,\dots,4;j=1,\dots,4}$, and we get that $I_2 = \left< y_j -r_j \right>_{j=1,\dots,4}$ as, for $\Phi_i \in \mathbb{C}$, all partials are zero.
We then eliminate all the $\Phi$s and substitute the resulting ideal into ring $R' = \mathbb{C}\left[ y_1,\dots,y_4\right]$ to get ideal $V$ representing the vacuum moduli space. Using \cite{mac2}, we see that V has dimension 3, degree $2$ and after assigning weights to each $y_j$ equal to the degree of the monomial they represent, we get the familiar Hilbert series \cite{Benvenuti:2006qr}
\begin{equation}
H(t, \cM_{g=1, \ (p,q)=(4,4)}) = \frac{1+t^2}{(1-t^2)^3} \ .
\end{equation}

We can also write down a permutation triple such that $\sigma_B \sigma_W \sigma_{\infty} = id$: 
\begin{equation}
\sigma_B = (1243) \ , \quad
\sigma_W = (1243) \ , \quad
\sigma_{\infty} = (14)(23) \ .
\end{equation}
Hence we have ramification structure $\{4, 4, 2^2\}$.
A suitable Belyi pair \cite{Jejjala:2010vb} is:
$ y^2 = x(x-1)(x-\frac{1}{2})$ with $\beta(x,y) = \frac{x^2}{2x-1}$.

In fact, for genus equal to 1, the above are the {\it only} regular tessellations, corresponding to the two symmetric cases of trivalent and quadrivalent tilings. \\
Let us now move onto the next genus.

\subsubsection{Genus 2}
The genus 2 situation is that of the two-handled tilings considered recently in \cite{Cremonesi:2013aba}.
We solve equation \eqref{regtilingeqn} with $g=2$ and obtain the following solution set:
\begin{align}
\{ p,q \} \in& \{ \{3,7\}, \{3,8 \}, \{3,9 \}, \{3,10 \}, \{3,12 \}, \{3,18 \}, \{4,5 \}, \{4,6 \}, \{4,12 \}, \{5,4 \}, \{5,5 \}, \nonumber \\
          & \{5,10 \}, \{6,4 \}, \{6,6 \}, \{7,3 \}, \{8,3 \}, \{8,4 \}, \{9,3 \}, \{10,3 \}, \{10,5 \}, \nonumber \\
& \{12,3 \}, \{12,4 \}, \{18,3 \} \}
\end{align}
We now discard all solutions with $p$ odd, since we need to be able to impose bipartite structure on it and impose that $V$ is an even number, so there can be equal numbers of black and white vertices. We also discard all solutions with $q > 4g = 8$, in accordance with \eqref{qbound}.
Using equation \eqref{eq:RegTilingRelations}, we can compile the summary in Table \ref{t:genus2}.

\begin{table}[t!!!]
\begin{center}
\begin{tabular}{c|c|c|c}

$\{p,q\}$ & $(V,E,F)$ & Exists? & $\mathcal{M}_{mes}$ Calabi-Yau? \\
\hline
$\{4,6\}$ & $(4,12,6)$ & Yes & Yes \\
$\{4,8\}$ & $(16,24,6)$ & Yes & Yes \\
$\{6,6\}$ & $(2,6,2)$ & Yes & Yes \\
$\{8,3\}$ & $(16,24,6)$ & Yes & Yes \\
$\{8,4\}$ & $(4,8,2)$ & Yes & Yes \\
$\{10,5\}$ & $(2,5,1)$ & Yes & Yes \\
\hline
$\{6,4\}$ & $(6,12,4)$ & Yes & No \\
\hline
$\{10,3\}$ & $(10,15,3)$ & No & n/a \\
$\{12,4\}$ & $(3,6,1)$ & ? & - \\
$\{4,12\}$ & $(1,6,3)$ & ? & - \\
\hline
$\{4,5\}$ & $(8,20,10)$ & No & - \\
$\{12,3\}$ & $(8,12,2)$ & No & - \\
$\{18,3\}$ & $(6,9,1)$ & No & - \\
\end{tabular}
\end{center}
\sf{\caption{Regular tilings for genus 2. Here n/a denotes the moduli space is currently beyond computational powers.}
\label{t:genus2}}
\end{table}

There are no balanced, regular bipartite tilings for the following solutions:
\begin{itemize}
\item $\{ p,q \} = \{4,5\}$: this tiling would have $E=20$, so its rotational symmetry group G would be of order 40. Let $n_2$ be the number of Sylow-2-subgroups of G. We note the tiling's faces have $4=2^2$ sides, so by considering rotating the vertices of a single square, we see that $n_2 \geq F = 10$. However, by another Sylow theorem, $n_2 \equiv 1 \mod{2}$. Looking at the prime factorization of $40=2^3\cdot 5$, we see that we must then have $n_2 = 5$, which is impossible. So this tiling does not exist.
\item $\{ p,q \} = \{ 12,3 \}$: we have here that $F=2, q=3$. Noting that $F<q$, we see that at each vertex, at least one face borders itself. However, as $F,q$ are coprime, there must be at least one face that borders a different face. Hence we do not have edge-transitivity.
\item $\{p,q\} = \{10,3\}$: consider the associated graph. Note this is a graph with 10 vertices and where each vertex is of valency 3. A classification of all such graphs exists \cite{wolframRegularGraph} and all but the \emph{Petersen graph} are easily seen to not be edge-transitive. To see the Petersen graph cannot be the required graph, we simply note it is not bipartite as it contains cycles of length 5.
\item $\{p,q \} = \{12,4\}$: using equation \ref{regtilingeqn}, we see that this tiling, if it existed, would have $(V,E,F) = (3,6,1)$. As each vertex is connected to at least one other vertex, we see that by regularity, all vertices are connected to each other. Hence it would be impossible to impose a bipartite structure on it. 
\item $\{p,q \} = \{4,12\}$: using equation \ref{regtilingeqn}, we see that this tiling, if it existed, would have $(V,E,F) = (1,6,3)$. As it only has one vertex, it cannot be a balanced bipartite tiling, so it is not of interest to us.
\item $\{p,q\} = \{18,3\}$: we note that this tiling exists if and only if its dual, with $\{p,q\} = \{3,18\}$, exists. Now note that this dual tiling has $V = 1, E = 9, F = 6$. Consider now the adjacency graph of its triangular faces. No face can border itself (as by edge-transitivity it would then have to do this at least three times, which is impossible) and no face can border the same face twice (otherwise we have again a violation of edge-transitivity). As the graph certainly needs to be regular, we see that it is a regular graph with six vertices of degree three. Only two such graphs exist \cite{wolframRegularGraph} and are pictured below:

\begin{center}
\includegraphics[width=13cm]{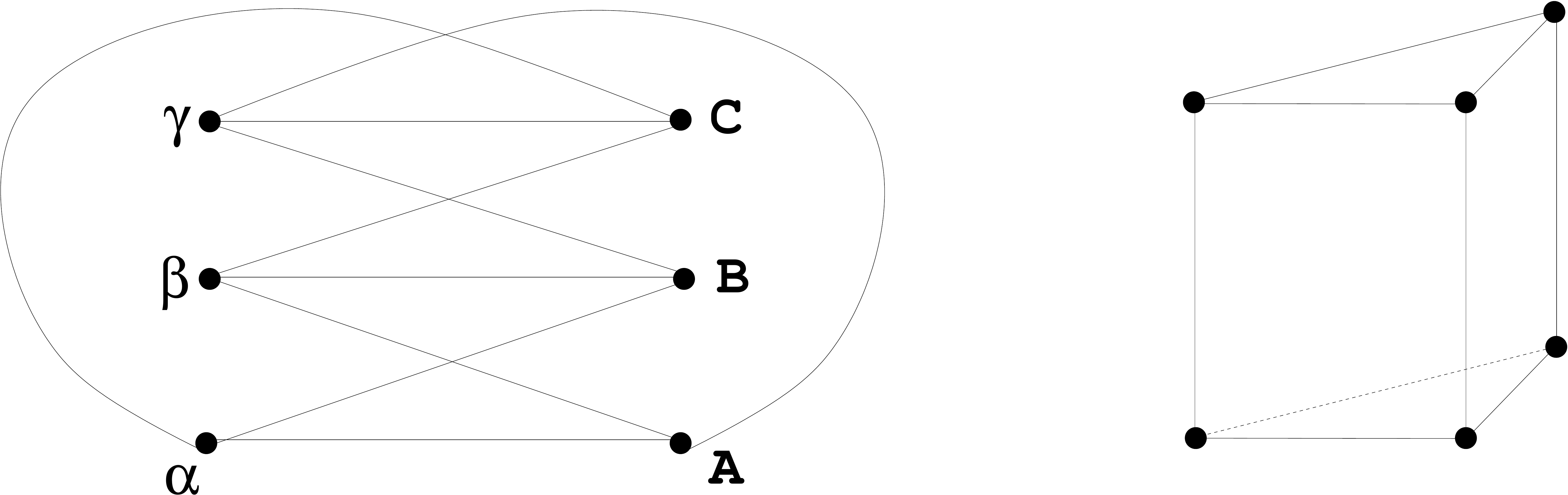}
\end{center}

Note the right-hand graph cannot be our graph, as it violates edge transitivity: three edges seperate two squares, whereas the other six seperate a square and a triangle.

To see the left-hand graph (the \emph{Thomsen graph} or \emph{utility graph}) can also be ruled out, we need to consider the tiling $\{p,q\} = \{3,18\}$. If we go clockwise around its vertex, we must travel through each face three times, traversing each edge exactly once in each direction and never going from face A to face B to face A again (as each face can only border another face once).\\
This path corresponds to an Eulerian circuit in the directed graph pictured below, with 18 edges:

\begin{center}
\includegraphics[width=13cm]{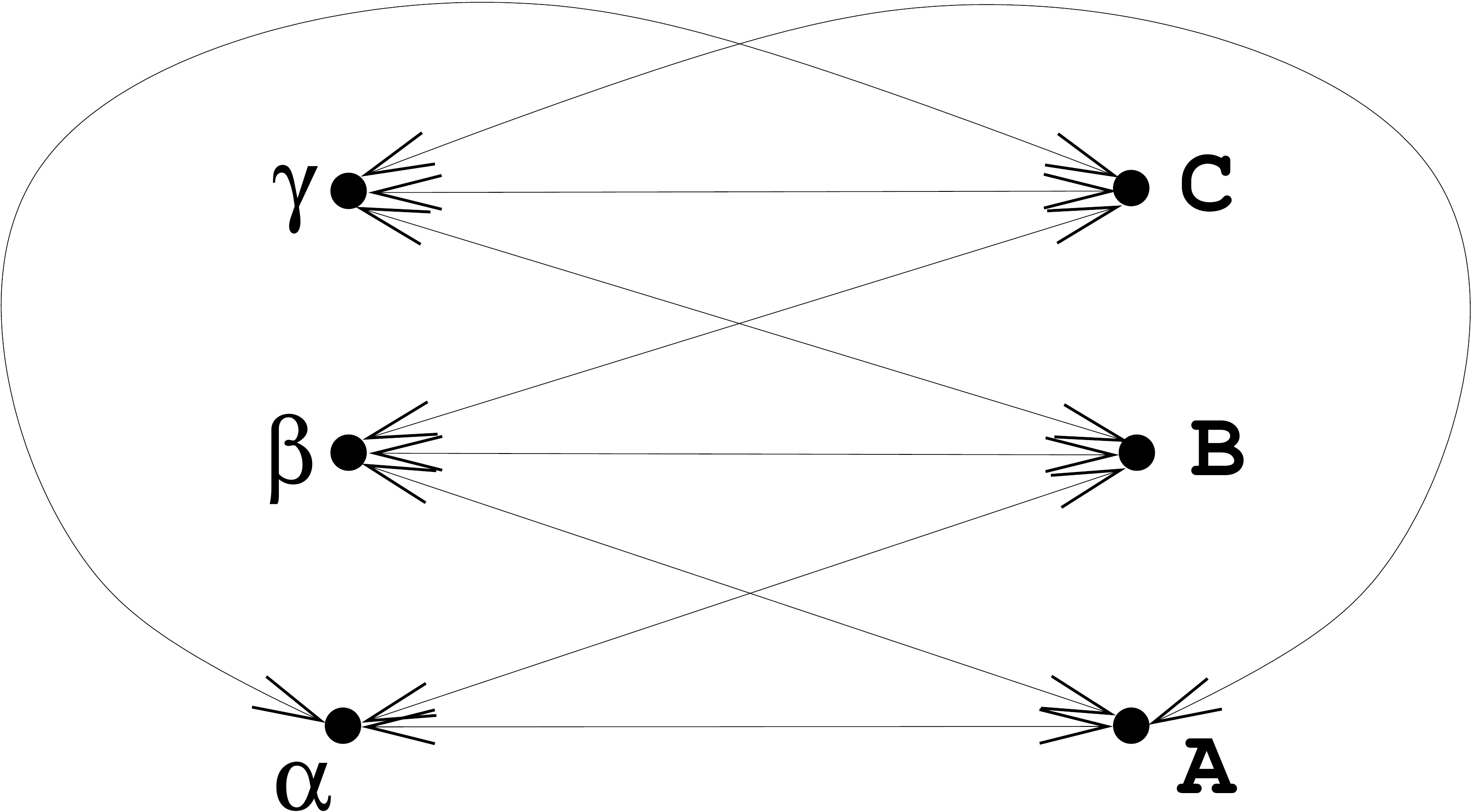}
\end{center}

(An Eulerian circuit is a path that traverses each edge exactly once and has the same start point and end point).

We note that the BEST theorem guarantees the existence of Eulerian paths in this graph \cite{BEST}. However, we will show no such circuit satisfies our constraint of not double-backing on ourselves.

To see this, we pick an arbitrary starting point $A$. We note that our circuit contains $18$ edges and that we need to return to A three times. Hence we can denote the number of edges until return for these three cycles as $n_1, n_2, n_3$, with the constraint $n_1+n_2+n_3 = 18$. We note the $n_i$ are even and $n_i \geq4$ and without loss of generality, we can suppose $n_1 \geq n_2 \geq n_3$.\\
By simple elimination, we see that the only solutions are \\ 
$(n_1,n_2,n_3) \in \{ (6,6,6),(8,6,4),(10,4,4) \}$. We consider these seperately:
\begin{itemize}
\item $(n_1,n_2,n_3) = (6,6,6)$. Starting in $\alpha$, we see that after suitable relabelling of vertices, we start as follows: $(\alpha,A,\beta,B,\gamma)$. To then return to $\alpha$ in six steps, we must visit C and $\alpha$, or visit A and $\alpha$. Note if we did the latter, we have gone from A to itself in four steps, which is impossible as all $n_i = 6$ and the cycles should be the same for each starting vertex. Hence we see our first cycle is of the form $(\alpha,A,\beta,B,\gamma,C,\alpha)$. For the second cycle, we must go to B (C would be double-backing). Continuing similarly with our constraints, we see the second cycle must be $(\alpha,B,\beta,C, \gamma,A,\alpha)$ and the third one $(\alpha,C,\beta,A,\gamma,B,A,\alpha)$ to give total path:
\begin{equation}
(\alpha,A,\beta,B,\gamma,C,\alpha,B,\beta,C, \gamma,A,\alpha,C,\beta,A,\gamma,B,A,\alpha) \ .
\end{equation}
Note now that the cycles of return for $\alpha,\beta,\gamma$ all have length 6, but for A,B,C they are of length $10,4,4$. Hence this path is not the path we are looking for.

\item $(n_1,n_2,n_3) = (8,6,4)$. Using the same procedure as before, we see the first cycle must be, after suitable relabelling of vertices, $(\alpha,A,\beta,B,\gamma,C,\beta,A,\alpha)$. \\
We then wish to perform the 6-cycle. If we first go to B, the cycle must be $(\alpha,B,\beta,C,\gamma,B,\alpha)$. However, if we then look at the final cycle, we see it must start as $(\alpha,C,\alpha)$, which is not allowed.\\
So we must then first go to C. After that we must go to $\gamma$, where we notice we can only go to A. However, we get stuck at A, as we cannot double-back on ourselves.\\
So we see no path with these $n_i$ exists.

\item $(n_1,n_2,n_3) = (10,4,4)$. We proceed as before. We see that, after suitable relabelling, our first cycle has to be $(\alpha,A,\beta,B,\gamma,C,\beta,A,\gamma,B,\alpha)$. Now note that we have returned from B to itself in six steps, which is not allowed (in fact, if we continued, we would create the same path as in the case $(n_1,n_2,n_3) = (6,6,6)$). \\
So we see this path does not suffice.
\end{itemize}

Hence we see that no adjacency graph of the right form exists for $\{p,q\} = \{3,18\}$. Hence it, and its dual $\{p,q\} = \{18,3\}$, cannot exist as regular tilings.
\end{itemize}

\paragraph*{\fbox{$\mathbf{\{p,q\} = \{10,5\}}$}}
This is the solution with $V=2,E=5,F=1$:
\begin{center}
\includegraphics[width=12cm]{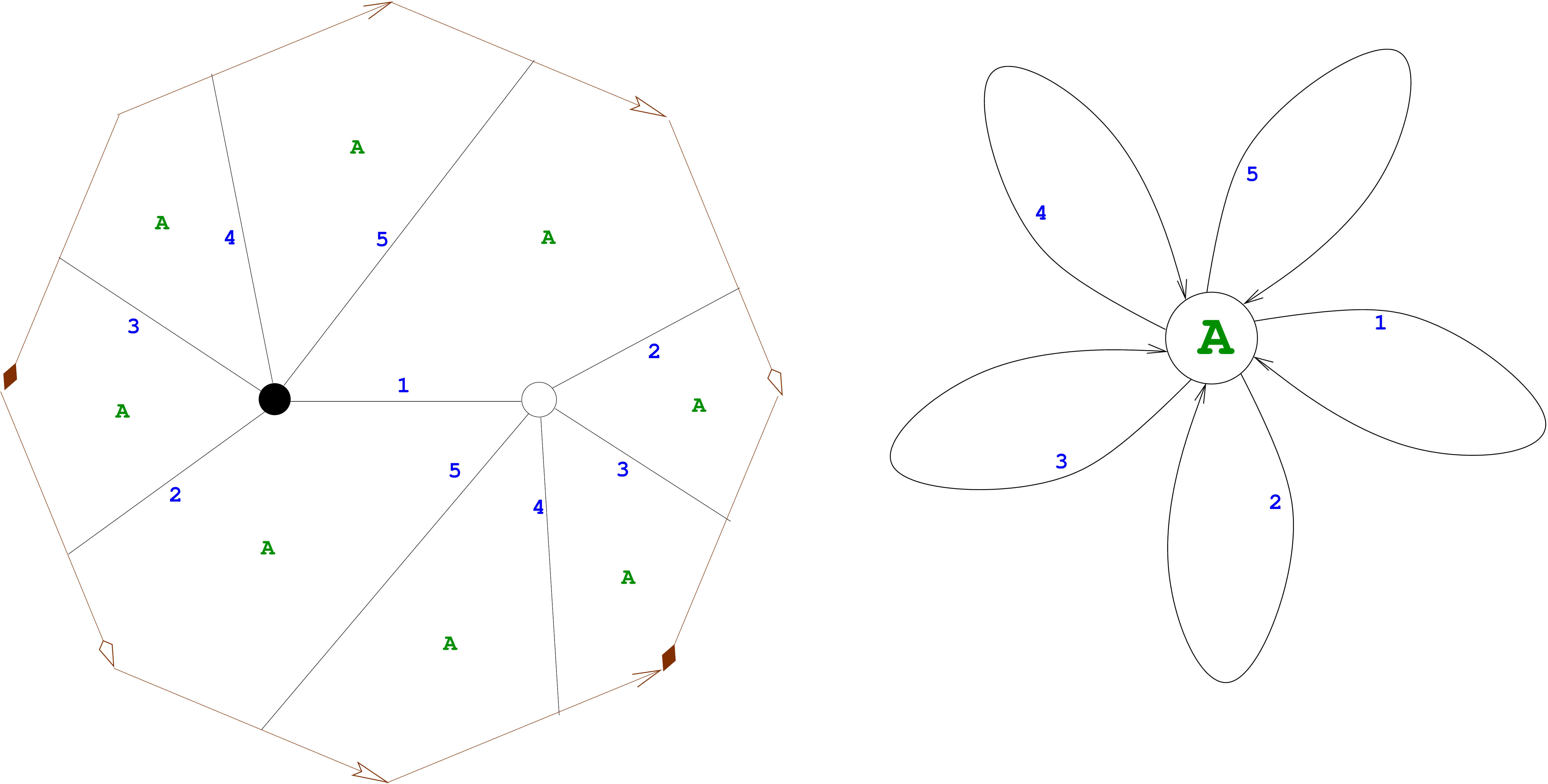}
\end{center}
We have superpotential 
\begin{equation} 
W = \tr \left( \Phi_1\Phi_2\Phi_3\Phi_4\Phi_5 - \Phi_1\Phi_5\Phi_4\Phi_3\Phi_2 \right) \ .
\end{equation}
We also see from the quiver that the $\Phi_i$ are themselves the gauge invariants.
We see that, working over the complex numbers, all partials vanish, so the master space is
\begin{equation}S_1/I_1 = \mathbb{C} \left[ \Phi_1,\dots\Phi_5\right] / \left< \partial_i W\right>_{i=1,\dots,5} = \mathbb{C} \left[ \Phi_1,\dots\Phi_5\right] \ .
\end{equation}
To find the vacuum moduli space, we consider the ring $R = \mathbb{C} \left[ \Phi_1,\dots\Phi_5,y_1,\dots,y_5 \right]$ and ideal $I_2 = \left< \partial_i W,y_i - \Phi_i \right>_{i=1,\dots,5}$, and we get that $I_2 = \left< y_i -\Phi_i \right>_{i=1,\dots,5}$ as, for $\Phi_i \in \mathbb{C}$, all partials are zero.
We then eliminate all the $\Phi$s and substitute the resulting ideal into ring $R' = \mathbb{C}\left[ y_1,\dots,y_5\right]$ to get ideal $V$ representing the vacuum moduli space. We see that
\begin{equation} R'/V \cong \mathbb{C} \left[ y_1,\dots,y_5 \right] \ ,
\end{equation}
which has dimension 5, degree 1 and Hilbert series 
\begin{equation} 
H(t, \cM_{g=2, \ (p,q)=(10,5)}) = \frac{1}{(1-t)^5} \ .
\end{equation}
Of course, the above is an over-kill but is a good check of our algorithms.
The moduli space - as is indicated by the Hilbert series - is nothing but $\IC^5$ and our gauge theory is the pentapetalous generalization of the ``clover'' theory of $\cN=4$ SYM in four dimensions, in agreement with Model 5.2 of \cite{Cremonesi:2013aba}.

We can also write down a permutation triple such that $\sigma_B \sigma_W \sigma_{\infty} = id$: 
\begin{equation}
\sigma_B = (1\;2\;3\;4\;5)\ , \quad
\sigma_W = (1\;2\;3\;4\;5)\ , \quad
\sigma_{\infty} = (1\;4\;2\;5\;3) \ ,
\end{equation}
with ramification structure $\{5,5,5\}$.

\paragraph*{\fbox{$\mathbf{\{p,q\} = \{6,6\}}$}}
Here we have $V=2,E=6,F=2$:
\begin{center}
\includegraphics[width=15cm]{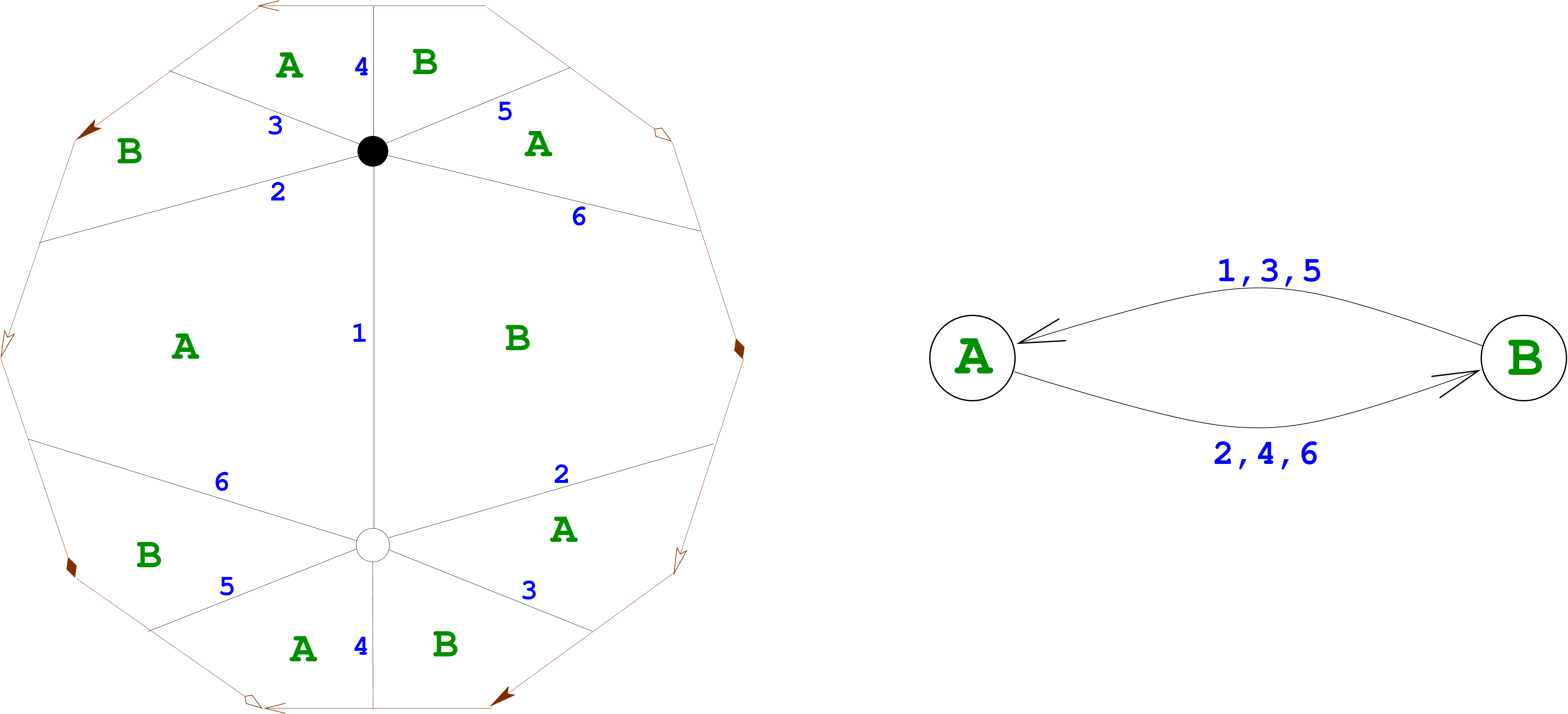}
\end{center}
We have superpotential 
\begin{equation} 
W = \tr \left( \Phi_1\Phi_2\Phi_3\Phi_4\Phi_5\Phi_6 - \Phi_1\Phi_6\Phi_5\Phi_4\Phi_3\Phi_2 \right)
\end{equation}
and 9 gauge invariants:
\begin{align}
\nonumber
&r_1 = \Phi_1\Phi_2 \ , \quad
r_2 = \Phi_1\Phi_4 \ , \quad
r_3 = \Phi_1\Phi_6 \ , \quad
r_4 = \Phi_3\Phi_2 \ , \quad
r_5 = \Phi_3\Phi_4 \ ,
\\
&r_6 = \Phi_3\Phi_6\ , \quad
r_7 = \Phi_5\Phi_2\ , \quad
r_8 =\Phi_5\Phi_4\ , \quad
r_9 = \Phi_5\Phi_6 \ .
\end{align}
We see that, working over the complex numbers, all partials vanish, so the master space is
\begin{equation}
S/I_1 = \mathbb{C} \left[ \Phi_1,\dots\Phi_6\right] / \left< \partial_i W,\right>_{i=1,\dots,6} = \mathbb{C} \left[ \Phi_1,\dots\Phi_6\right] \ .
\end{equation}

To find the vacuum moduli space, we consider the ring $R = \mathbb{C} \left[ \Phi_1,\dots\Phi_6,y_1,\dots,y_9 \right]$ and ideal $I_2 = \left< \partial_i W,y_j - r_j \right>_{i=1,\dots,6;j=1,\dots,9}$, and we get that $I_2 = \left< y_j -r_j \right>_{j=1,\dots,9}$ as, for $\Phi_i \in \mathbb{C}$, all partials are zero.
We then eliminate all the $\Phi$s and substitute the resulting ideal into ring $R' = \mathbb{C}\left[ y_1,\dots,y_5\right]$ to get ideal $V$ representing the vacuum moduli space. Using \cite{mac2}, we see that V has dimension 5, degree 6 and after assigning weights to each $y_j$ equal to the degree of the monomial they represent, we get Hilbert series
\begin{equation}
H(t, \cM_{g=2, \ (p,q)=(6,6)}) = \frac{1+4t^2+t^4}{(1-t^2)^5} \ .
\end{equation}
The palindromic numerator indicates it is a Calabi-Yau 5-fold and is in agreement with the results found in \cite{Cremonesi:2013aba}.
This is, of course, a tri-saggital generalization of the conifold theory.
It is interesting to note that unlike the conifold, which is a quadric hypersurface in $\IC^4$, this is not a complete intersection.

We can also write down a permutation triple such that $\sigma_B \sigma_W \sigma_{\infty} = id$: 
\begin{equation}
\sigma_B = (1\;2\;3\;4\;5\;6)\ , \quad
\sigma_W = (1\;2\;3\;4\;5\;6)\ , \quad
\sigma_{\infty} = (1\;5\;3)(2\;6\;4) \ .
\end{equation}
So we note we have ramification structure $\{6,6,3^2\}$.

\paragraph*{\fbox{$\mathbf{\{p,q\} = \{8,4\}}$}}
Here we have $V=4,E=8,F=2$:

\begin{center}
\includegraphics[width=15cm]{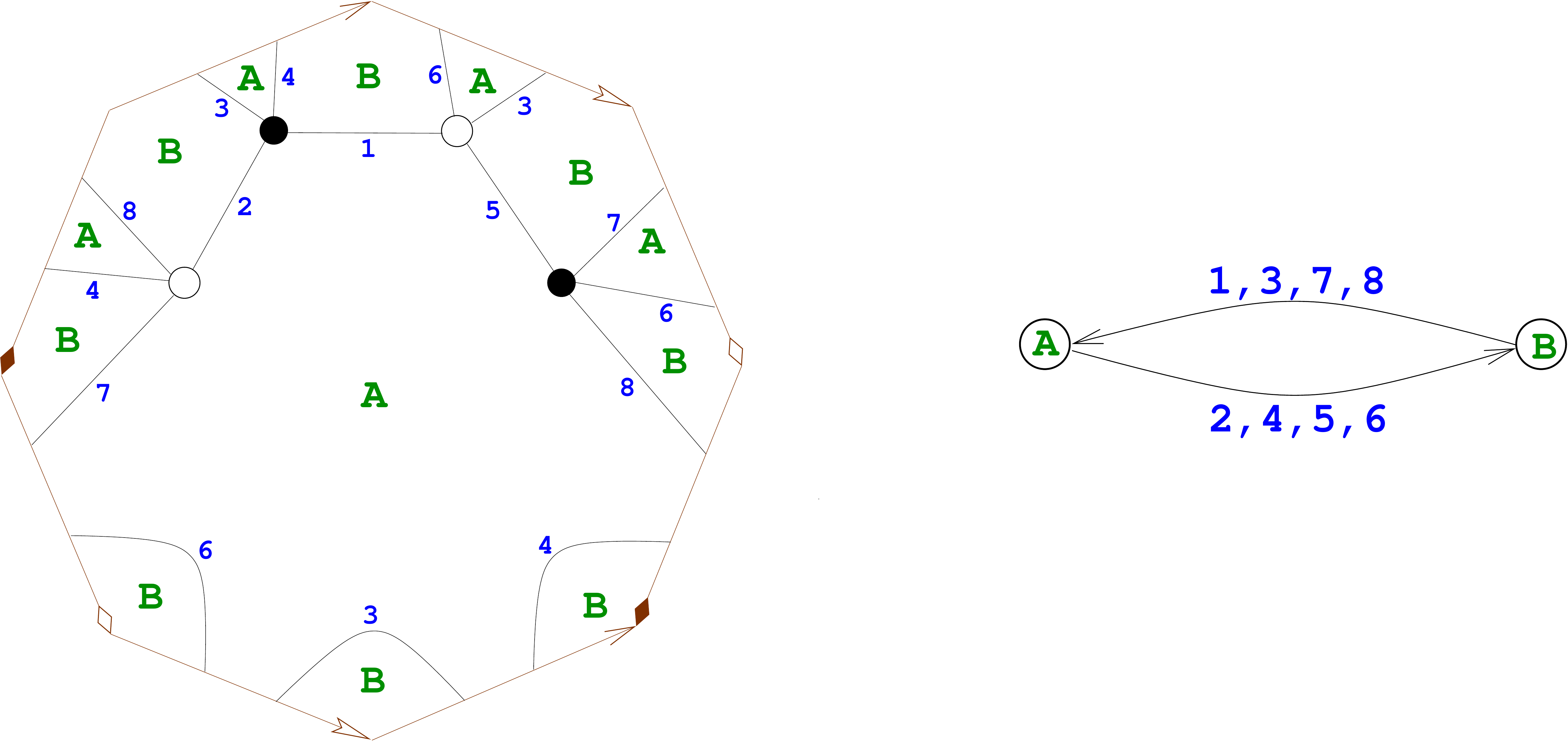}
\end{center}

We have superpotential 
\begin{equation} W = \tr \left( \Phi_1\Phi_2\Phi_3\Phi_4 + \Phi_5\Phi_7\Phi_6\Phi_8 - \Phi_1\Phi_5\Phi_3\Phi_6 - \Phi_2\Phi_8\Phi_4\Phi_7 \right)
\end{equation}
and 16 gauge invariants:
\begin{equation}r _{i,j} = \Phi_i \Phi_j \qquad \qquad \text{with } i\in\{1,3,7,8\}, j\in\{2,4,5,6\} \ . \end{equation}

To find the master space, we define ring $S = \mathbb{C}[\Phi_1,\dots,\Phi_8]$ and ideal $I_1 = \left< \partial_i W \right>_{i=1,\dots,8}$. We then generate master space $R = S /I_1$, and using \cite{mac2} we find that $I_1$ has dimension 6, degree 4 and Hilbert series
\begin{equation} 
H(t, \cF^{\flat}_{g=2, \ (p,q)=(8,4)}) = \frac{ 1 + 2t + 3t^2 - 4t^3 + 2 t^4}{(1-t)^6} \ .
\end{equation}

Using primary decomposition in \cite{mac2}, and we get that the curve given by $I_1$ is the union of those given by ideals of which 2 are trivial with degree 1, dimension 4 and Hilbert series $\frac{1}{(1-t)^4}$. The \emph{coherent component} is of degree 4, dimension 6 and has Hilbert series 
\begin{equation} 
H(t, \ ^{\text{Irr}}\cF^{\flat}_{g=2, \ (p,q)=(8,4)}) =  \frac{1+2t+t^2}{(1-t)^6} \ .
\end{equation}

To find the vacuum moduli space, we consider the ring $R = \mathbb{C} \left[ \Phi_1,\dots\Phi_8,y_1,\dots,y_{16} \right]$ and ideal $I_2 = \left< \partial_i W,y_j - r_j \right>_{i=1,\dots,8;j=1,\dots,16}$. We then eliminate all the $\Phi$s and substitute the resulting ideal into ring $R' = \mathbb{C}\left[ y_1,\dots,y_{16}\right]$ to get ideal $V$ representing the vacuum moduli space. Using \cite{mac2}, we see that V has dimension 5, degree 24 and after assigning weights to each $y_j$ equal to the degree of the monomial they represent, we get Hilbert series
\begin{equation}H(t, \cM_{g=2, \ (p,q)=(8,4)}) = \frac{1+11t^2+11t^4+t^6}{(1-t^2)^5}\ , \end{equation}
its palindromic numerator indicating it is a Calabi-Yau 5-fold.

We can also write down a permutation triple such that $\sigma_B \sigma_W \sigma_{\infty} = id$: 
\begin{equation}
\sigma_B = (1\;2\;3\;4)(5\;7\;6\;8)\ , \quad
\sigma_W = (1\;6\;3\;5)(2\;7\;4\;8)\ , \quad
\sigma_{\infty} = (1\;8\;3\;7)(2\;6\;4\;5) \ .
\end{equation}
So we note we have ramification structure $\{4^2,4^2,4^2\}$.

\paragraph*{\fbox{$\mathbf{\{p,q\} = \{6,4\}}$}}
Here we have $V=6,E=12,F=4$:

\begin{center}
\includegraphics[width=15cm]{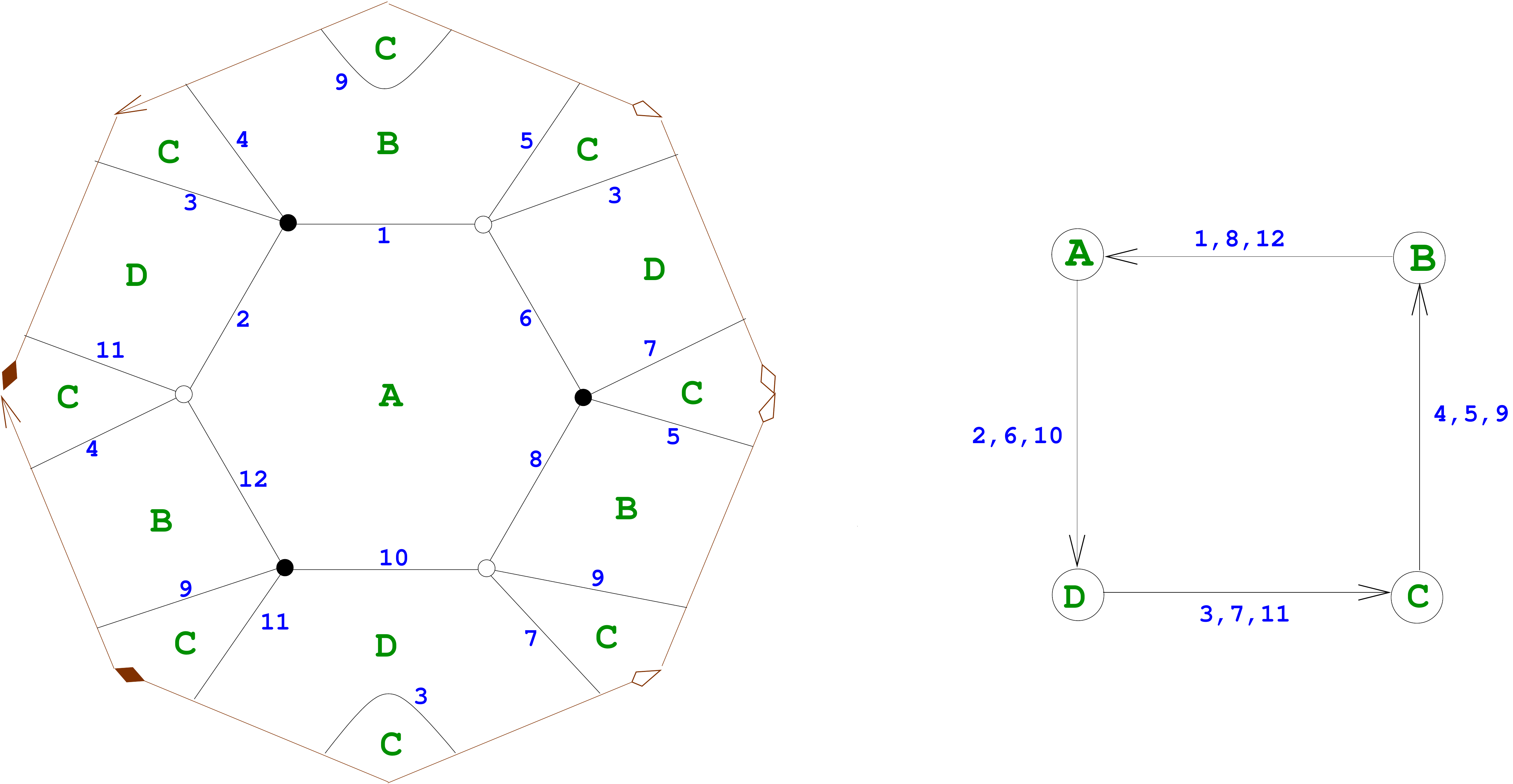}
\end{center}

We have superpotential 
\begin{equation} W = \tr \left( \Phi_1\Phi_2\Phi_3\Phi_4 + \Phi_5\Phi_8\Phi_6\Phi_7 + \Phi_9\Phi_{12}\Phi_{10}\Phi_{11} - \Phi_1\Phi_6\Phi_3\Phi_5 - \Phi_2\Phi_{11}\Phi_4\Phi_{12} -\Phi_7\Phi_9\Phi_8\Phi_{10} \right) 
\end{equation}
and 81 gauge invariants:
\begin{equation}r _{i,j,k,l} = \Phi_i \Phi_j \Phi_k \Phi_l \qquad \qquad \text{with } i\in\{1,8,12\}, j\in \{2,6,10\}, k \in \{ 3,7,11\}, l \in \{4,5,9\}. \end{equation}

To find the master space, we define the ring $S_1 = \mathbb{C}[\Phi_1,\dots,\Phi_{12}]$ and ideal $I_1 = \left< \partial_i W \right>_{i=1,\dots,12}$ to generate the master space $R = S_1 /I_1$. We find using singular that $I_1$ has dimension 8, degree 16 and Hilbert series
\begin{equation}
H(t, \cF^{\flat}_{g=2, \ (p,q)=(6,4)}) = \frac{1 + 4t + 10t^2 + 8t^3 -6t^4 - 20t^5 + 28t^6 - 12t^7 + 3t^8}{(1-t)^8} \ .
\end{equation}

Using primary decomposition in \cite{mac2}, we get that the curve given by $I_1$ is the union of those given by ideals of which:
3 are trivial of degree 1, dimension 4 and Hilbert series $\frac{1}{(1-t)^4}$;
2 are trivial of degree 1, dimension 6 and Hilbert series $\frac{1}{(1-t)^6}$;
12 are of degree 2, dimension 6 and have Hilbert series $\frac{1+t}{(1-t)^6}$.
The \emph{coherent component} is of degree 16, dimension 8 and has Hilbert series 
\begin{equation} H(t, \ ^{\text{Irr}}\cF^{\flat}_{g=2, \ (p,q)=(6,4)}) =  \frac{1+4t+6t^2+4t^3+t^4}{(1-t)^8} \ . \end{equation}

To find the vacuum moduli space, we consider the ring $R = \mathbb{C} \left[ \Phi_1,\dots\Phi_{12},y_1,\dots,y_{81} \right]$ and ideal $I_2 = \left< \partial_i W,y_j - r_j \right>_{i=1,\dots,12;j=1,\dots,81}$. 
We then eliminate all the $\Phi$s and substitute the resulting ideal into ring $R' = \mathbb{C}\left[ y_1,\dots,y_{81}\right]$ to get ideal $V$ representing the vacuum moduli space. Using \cite{mac2}, we see that V has dimension 5, degree 216 and after assigning weights to each $y_j$ equal to the degree of the monomial they represent, we get Hilbert series
\begin{equation}H(t, \cM_{g=2, \ (p,q)=(6,4)}) = \frac{1+47t^4+114t^8+62t^{12}-11t^{16}+3t^{20}}{(1-t^4)^5} \ . \end{equation}

We can also write down a permutation triple such that $\sigma_B \sigma_W \sigma_{\infty} = id$: 
\begin{align}
\nonumber
\sigma_B &= (1\;2\;3\;4)(5\;8\;6\;7)(9\;12\;10\;11)\\
\nonumber
\sigma_W &= (1\;5\;3\;6)(2\;12\;4\;11)(7\;10\;8\;9)\\
\sigma_{\infty} &= (1\;8\;12)(2\;10\;6)(3\;7\;11)(4\;9\;5)
\end{align}
So we note we have ramification structure $\{4^3,4^3,3^4\}$.

\paragraph*{\fbox{$\mathbf{\{p,q\} = \{8,3\}}$}}
Here we have $V=16,E=24,F=6$:

\begin{center}
\includegraphics[width=15cm]{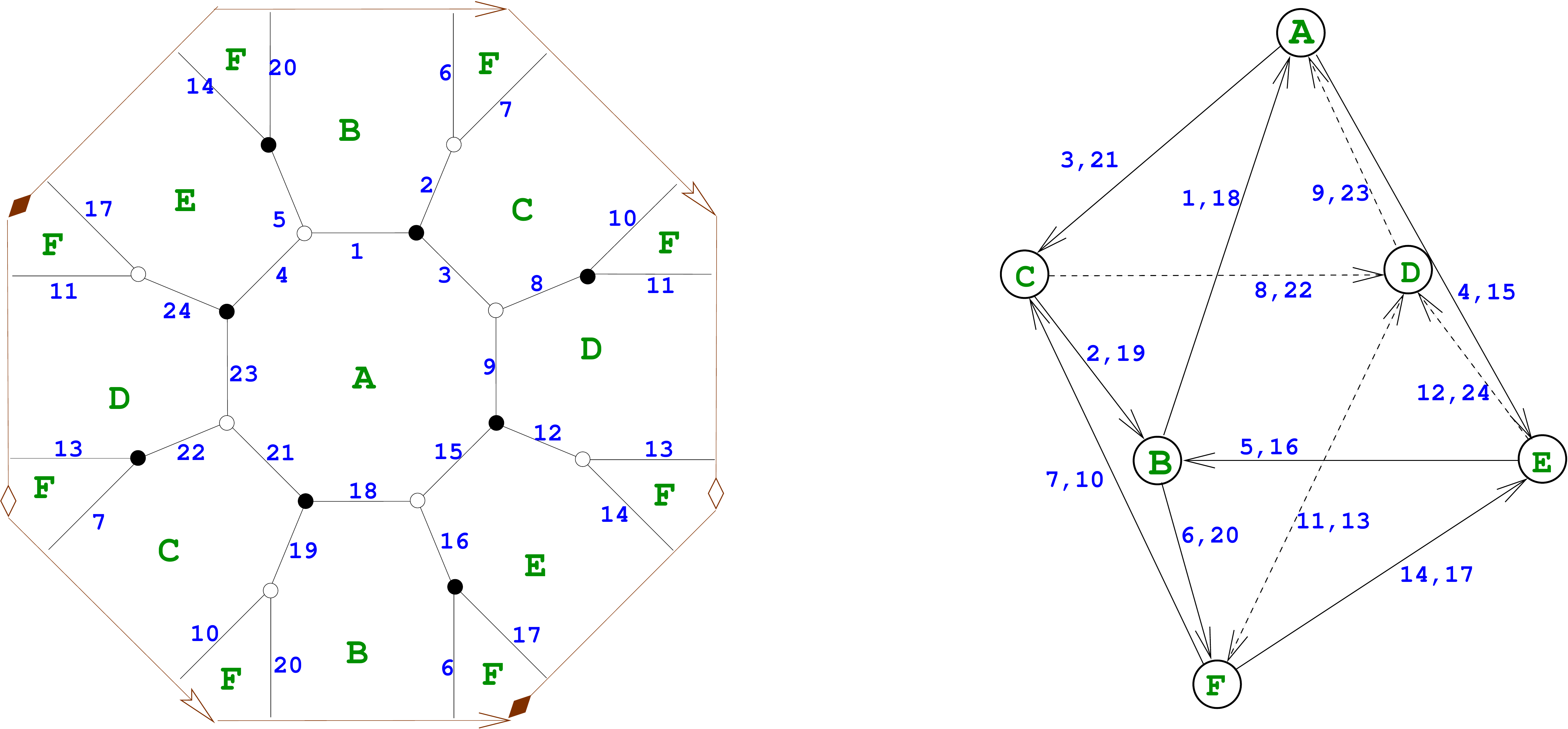}
\end{center}

We have superpotential 
\begin{align}
W = \tr ( & \Phi_{1}\Phi_{2}\Phi_{3} + \Phi_{8}\Phi_{10}\Phi_{11} + \Phi_{9}\Phi_{12}\Phi_{15} + \Phi_{6}\Phi_{16}\Phi_{17} + \Phi_{19}\Phi_{21}\Phi_{18} + \Phi_{7}\Phi_{13}\Phi_{22} \nonumber \\
 & + \Phi_{4}\Phi_{23}\Phi_{24} + \Phi_5\Phi_{14}\Phi_{20} - \Phi_{1}\Phi_{5}\Phi_{4} - \Phi_{2}\Phi_{7}\Phi_{6} - \Phi_{3}\Phi_{9}\Phi_{8} - \Phi_{12}\Phi_{14}\Phi_{13}  \nonumber \\
 & -\Phi_{15}\Phi_{18}\Phi_{16} - \Phi_{10}\Phi_{20}\Phi_{19} - \Phi_{21}\Phi_{23}\Phi_{22} - \Phi_{11}\Phi_{24}\Phi_{17} )
\end{align}
and 64 gauge invariants of the form:
\begin{equation}r _{i,j,k} = \Phi_i \Phi_j \Phi_k \ . \end{equation}

To find the master space, we define ring $S_1 = \mathbb{C}[\Phi_1,\dots,\Phi_{24}]$ and ideal $I_1 = \left< \partial_i W \right>_{i=1,\dots,24}$. We then generate master space $R = S_1 /I_1$, and using singular we find that $I_1$ has dimension 10, degree 594 and Hilbert series
\begin{equation}
H(t, \cF^{\flat}_{g=2, \ (p,q)=(8,3)}) =
\frac{1 + 14 t + 81 t^2 + 233 t^3 + 268 t^4 - 45 t^5 - 63 t^6 + 105 t^7}{(1-t)^{10}} \ .
\end{equation}

To find the vacuum moduli space, we consider the ring $R = \mathbb{C} \left[ \Phi_1,\dots\Phi_{24},y_1,\dots,y_{64} \right]$ and ideal $I_2 = \left< \partial_i W,y_j - r_j \right>_{i=1,\dots,24;j=1,\dots,64}$. 
We then eliminate all the $\Phi$s and substitute the resulting ideal into ring $R' = \mathbb{C}\left[ y_1,\dots,y_{64}\right]$ to get ideal $V$ representing the vacuum moduli space. Using \cite{mac2}, we see that V has dimension 5, degree 96 and after assigning weights to each $y_j$ equal to the degree of the monomial they represent, we get Hilbert series
\begin{equation}H(t, \cM_{g=2, \ (p,q)=(8,3)}) =  \frac{1+20t^3+54t^6+20t^9+t^{12}}{(1-t^3)^5} \ , \end{equation}
its palindromic numerator indicating this is a Calabi-Yau 5-fold.

We can also write down a permutation triple such that $\sigma_B \sigma_W \sigma_{\infty} = id$: 
\begin{align}
\nonumber
\sigma_B &= (1\;2\;3)(4\;23\;24)(5\;14\;20)(6\;16\;17)(7\;13\;22)(8\;10\;11)(9\;12\;15)(18\;19\;21)\\
\nonumber
\sigma_W &= (1\;4\;5)(2\;6\;7)(3\;8\;9)(10\;19\;20)(11\;17\;24)(12\;13\;14)(15\;16\;18)(21\;22\;23)\\
\sigma_{\infty} &= (1\;20\;18\;6)(2\;22\;19\;8)(3\;15\;21\;4)(5\;24\;16\;12)(7\;17\;10\;14)(9\;11\;23\;13) \ .
\end{align}
So we note we have ramification structure $\{3^8,3^8,4^6\}$.

\paragraph*{\fbox{$\mathbf{\{p,q\} = \{4,8\}}$}}
Here we have $V=16,E=24,F=6$:

\begin{center}
\includegraphics[width=15cm]{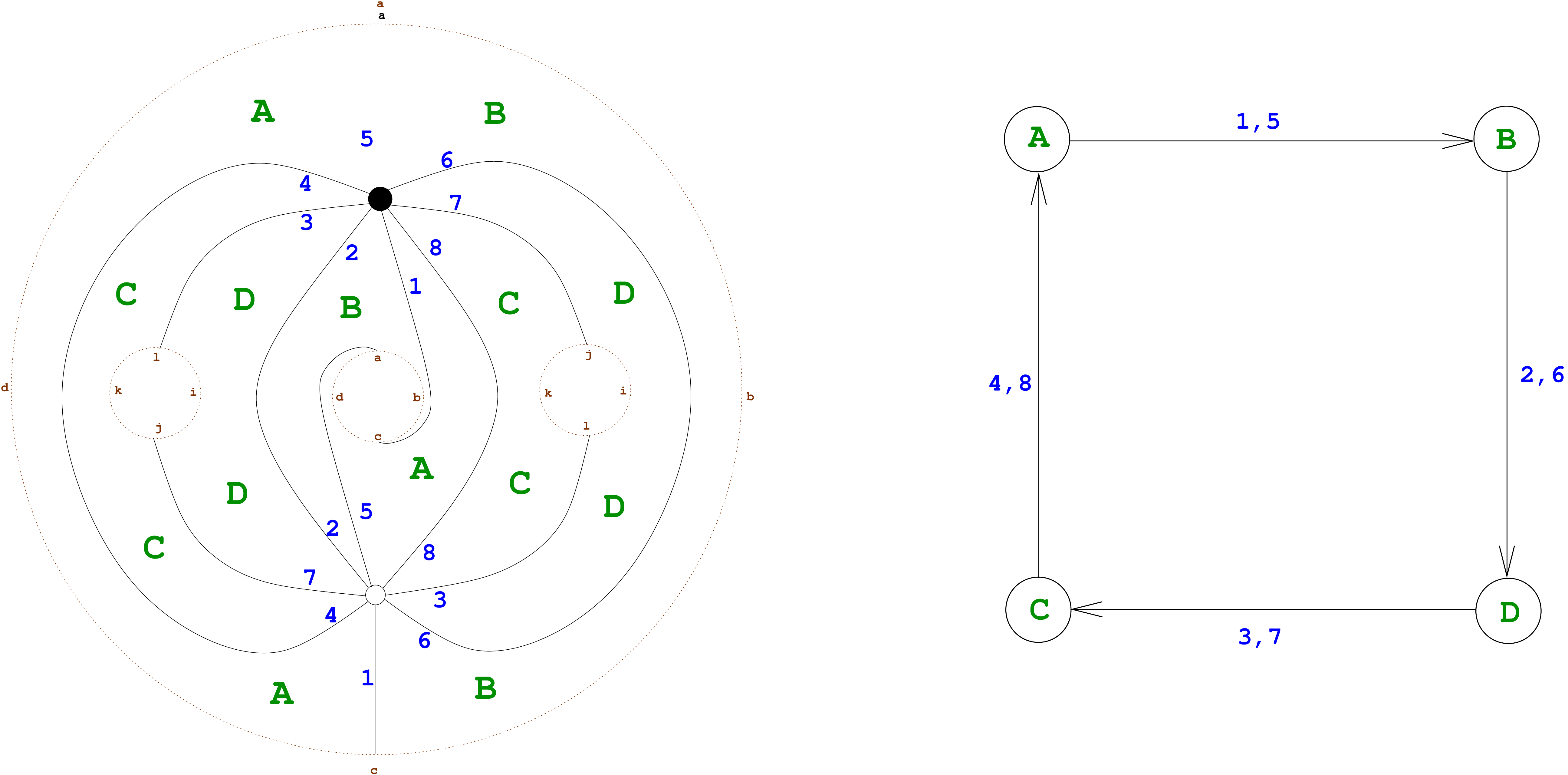}
\end{center}

We have superpotential 
\begin{equation} W = \tr \left( \Phi_1\Phi_2\Phi_3\Phi_4\Phi_5\Phi_6\Phi_7\Phi_8 - \Phi_1\Phi_6\Phi_3\Phi_8\Phi_5\Phi_2\Phi_7\Phi_4 \right)
\end{equation}
and 16 gauge invariants of the form:
\begin{equation}r _{i,j,k,l} = \Phi_i \Phi_j \Phi_k \Phi_l \qquad \qquad \text{with } i\in\{1,5\}, j\in\{2,6\}, k \in \{ 3,7\}, l \in\{4,8 \} \ . \end{equation}

If we consider the ring $S = \mathbb{C} \left[ \Phi_1,\dots\Phi_8,y_1,\dots,y_{16} \right]$ and ideal $I = \left< \partial_i W,y_j - r_j \right>_{i=1,\dots,8;\;j=1,\dots,16}$, we get that $I = \left< y_i -\Phi_i \right>_{i=1,\dots,8}$ as, for $\Phi_i \in \mathbb{C}$, all partials are zero.

To find the vacuum moduli space, we consider the ring $R = \mathbb{C} \left[ \Phi_1,\dots\Phi_8,y_1,\dots,y_{16} \right]$ and ideal $I_2 = \left< \partial_i W,y_j - r_j \right>_{i=1,\dots,8;j=1,\dots,16}$, and we get that $I_2 = \left< y_j -r_j \right>_{j=1,\dots,16}$ as, for $\Phi_i \in \mathbb{C}$, all partials are zero.
We then eliminate all the $\Phi$s and substitute the resulting ideal into ring $R' = \mathbb{C}\left[ y_1,\dots,y_{16}\right]$ to get ideal $V$ representing the vacuum moduli space. Using \cite{mac2}, we see that V has dimension 5, degree 24 and after assigning weights to each $y_j$ equal to the degree of the monomial they represent, we get Hilbert series
\begin{equation}H(t, \cF^{\flat}_{g=2, \ (p,q)=(4,8)}) =\frac{1+11t^4 + 11t^8+t^{12}}{(1-t^4)^5} \ . \end{equation}

Its palindromic numerator indicates this is a Calabi-Yau 5-fold and is in agreement with the results found in \cite{Cremonesi:2013aba}.

We can also write down a permutation triple such that $\sigma_B \sigma_W \sigma_{\infty} = id$: 
\begin{equation}
\sigma_B = (1\;2\;3\;4\;5\;6\;7\;8) \ , \quad
\sigma_W = (1\;4\;7\;2\;5\;8\;3\;6) \ , \quad
\sigma_{\infty} = (1\;5)(2\;6)(3\;7)(4\;8) \ .
\end{equation}
So we note we have ramification structure $\{8,8,2^4\}$.

\paragraph*{\fbox{$\mathbf{\{p,q\} = \{4,6\}}$}}
Here we have $V=4,E=12,F=6$:

\begin{center}
\includegraphics[width=15cm]{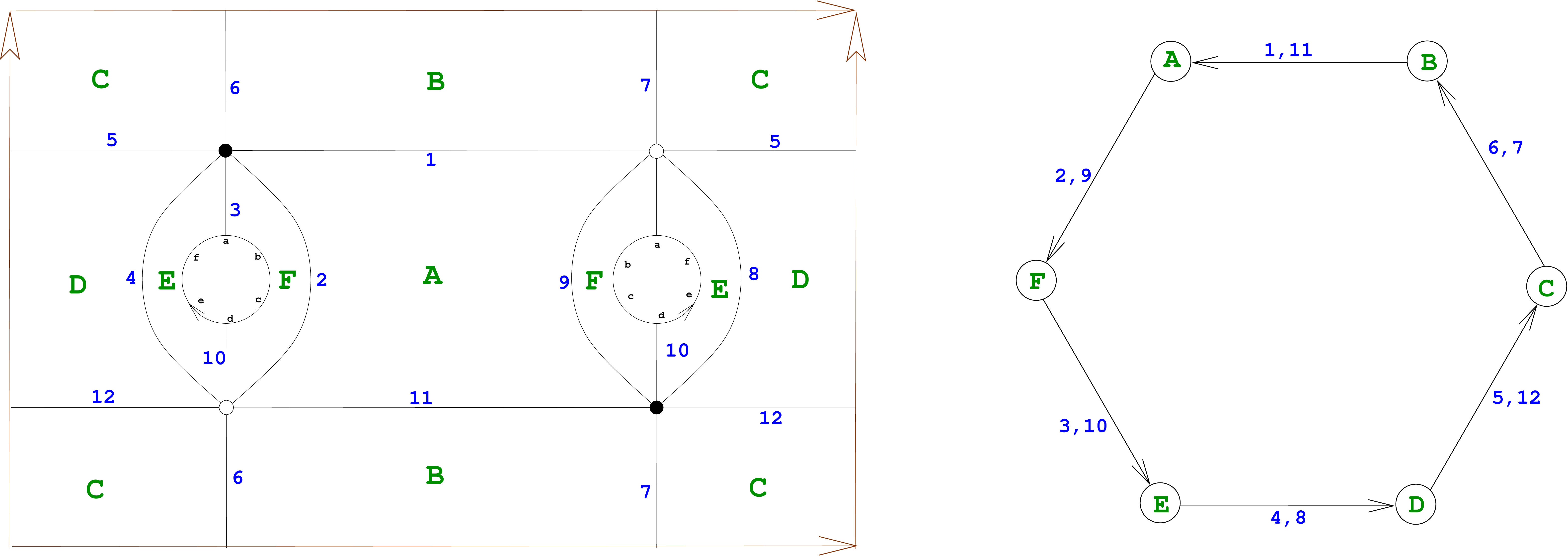}
\end{center}

We have superpotential 
\begin{equation} W = \tr \left( \Phi_1\Phi_2\Phi_3\Phi_4\Phi_5\Phi_6 + \Phi_{7}\Phi_{11}\Phi_9\Phi_{10}\Phi_{8}\Phi_{12} - \Phi_{1}\Phi_{9}\Phi_{3}\Phi_{8}\Phi_{5}\Phi_{7} - \Phi_{2}\Phi_{10}\Phi_{4}\Phi_{12}\Phi_{6}\Phi_{11} \right)
\end{equation}
and 64 gauge invariants of the form:
\begin{align}
&r_{i,j,k,l,m,n} = \Phi_i \Phi_j \Phi_k \Phi_l\Phi_m\Phi_n \nonumber \\
& \text{with } i\in\{1,11\}, j\in\{2,9\}, k \in \{3,10 \}, l \in\{4,8\}, m \in\{ 5,12\} , n \in  \{6,7 \} \ .
\end{align}
To find the master space, we define the ring $S = \mathbb{C}[\Phi_1,\dots,\Phi_{12}]$ and ideal $I_1 = \left< \partial_i W \right>_{i=1,\dots,12}$ to generate the master space $R = S /I_1$. We find using \cite{mac2} that $I_2$ has dimension 10, degree 9 and Hilbert series
\begin{equation}H(t, \cF^{\flat}_{g=2, \ (p,q)=(4,6)}) = \frac{1+2t+3t^2+4t^3+5t^4-6t^5-8t^6+8t^7}{(1-t)^{10}} \ . \end{equation}

Using primary decomposition in \cite{mac2}, we get that the curve given by $I_1$ is the union of those given by ideals of which
18 are trivial of degree 1, dimension 8 and Hilbert series $\frac{1}{(1-t)^8}$.
The \emph{coherent component} is of degree 9, dimension 10 and has Hilbert series 
\begin{equation} 
H(t, \ ^{\text{Irr}}\cF^{\flat}_{g=2, \ (p,q)=(4,6)}) =  \frac{1 - 2t^3+t^6}{(1-t)^{12}} = \frac{(1+t+t^2)^2}{(1-t)^{10}} \ . 
\end{equation}

To find the vacuum moduli space, we consider the ring $R = \mathbb{C} \left[ \Phi_1,\dots\Phi_{12},y_1,\dots,y_{64} \right]$ and ideal $I_2 = \left< \partial_i W,y_j - r_j \right>_{i=1,\dots,12;j=1,\dots,64}$.
We then eliminate all the $\Phi$s and substitute the resulting ideal into ring $R' = \mathbb{C}\left[ y_1,\dots,y_{64}\right]$ to get ideal $V$ representing the vacuum moduli space. Using \cite{mac2}, we see that V has dimension 5, degree 216 and after assigning weights to each $y_j$ equal to the degree of the monomial they represent, we get Hilbert series
\begin{equation} 
H(t, \cM_{g=2, \ (p,q)=(4,6)}) = \frac{1+44t^6 + 126t^{12}+44t^{18}+t^{24}}{(1-t^6)^5} \ ,
\end{equation}
its palindromic numerator indicating it is a Calabi-Yau 5-fold.

We can also write down a permutation triple such that $\sigma_B \sigma_W \sigma_{\infty} = id$: 
\begin{align}
\nonumber
\sigma_B &= (1\;2\;3\;4\;5\;6)(7\;11\;9\;10\;8\;12)\\
\nonumber
\sigma_W &= (1\;7\;5\;8\;10\;9)(2\;11\;6\;12\;4\;10) \\
\sigma_{\infty} &= (1\;11)(2\;9)(3\;10)(4\;8)(5\;12)(6\;7) \ .
\end{align}
So we note we have ramification structure $\{6^2,6^2,2^6\}$.

\subsubsection{Genus 3}

We use Matlab to find solutions to equation \eqref{regtilingeqn} with $g=3$, discarding all solutions with $p$ odd, as we need to be able to impose bipartite structure on it. We impose that $V$ is an even number, so there can be equal numbers of black and white vertices. We also discard all solutions with $q> 4g = 12$, in accordance with \eqref{qbound}. \\
As before, we compile the summary solutions in Table \ref{t:genus3}. 
\begin{table}[t!!]
\begin{center}
\begin{longtable}{cc|c|c|c}

$\{p,q\}$ & & $(V,E,F)$ & Exists? & $\cM_{mes}$ Calabi-Yau? \\
\hline
\{4,12\} & & (2,12,6)  &  Yes & Yes \\ 
\{8,8\} & & (2,8,2)  &  Yes & Yes \\ 
\{14,7\} & & (2,7,1)  &  Yes & Yes \\ 
\hline
\{4,6\} & & (8,24,12) &  Yes & $E$ too large  \\ 
\{6,4\} & & (12,24,8)  &  Yes & $E$ too large  \\ 
\{8,3\} & & (32,48,12)  &  Yes & $E$ too large  \\ 
\{12,3\} & & (16,24,4)  &  Yes & $E$ too large  \\ 
\{4,8\} & Model A & (4,16,8)  &  Yes & No \\
        & Model B & (4,16,8)  &  Yes & $|R|$ too large \\
\{14,3\} & & (14,21,3) & Yes &$ |R|$ too large  \\
\{8,4\} & Model A& (8,16,4)  &  Yes & $|R|$ too large \\ 
        & Model B& (8,16,4)  &  Yes & $|R|$ too large \\ 
\hline
\{12,4\} & & (6,12,2)  &  Yes & No \\ 
\hline
\{6,6\} & & (4,12,4)  &  Yes & Not bipartite \\ 
\hline
\{4,5\} & & (16,40,20) & No & n/a  \\ 
\{6,5\} & & (6,15,5) & No & n/a  \\ 
\{6,9\} & & (2,9,3) & No & n/a  \\
\{10,3\} & & (20,30,6) & No & n/a  \\ 
\{10,5\} & & (4,10,2)  &  No & n/a \\ 
\{18,3\} & & (12,18,2) & No & n/a  \\ 
\{30,3\} & & (10,15,1) & No & n/a  \\ 
\end{longtable}
\end{center}
\sf{\caption{Regular tilings for genus 3. By ``E too large" or ``$|R|$ too large" we mean that the number of edges, or the number of gauge invariants, respectively, was too big to complete calculations on the moduli space.
\label{t:genus3}}}
\end{table}

To see no regular tiling exists with:
\begin{itemize}
\item $\{ p,q \} = \{4,5\}$: same reasoning as for $\{ 4,5\} $ in the genus 2 case, now with $E=40$.
\item $\{ p,q \} = \{ 6,5\}$: this exists if and only if its dual $\{ 5,6\}$ exists. To see this does not exist, let $G$ be its rotational symmetry group. Then note that $|G|$ divides $2E = 30$. Noting its faces have $5$ sides, we see there are $n_5$ Sylow 5-subgroups of $G$, where $n_5 \geq F = 6$. However, a Sylow theorem states that $n_5 \equiv  1 \mod{5}$ and that $n_5$ divides $|G| / 5 = 6$. This together means $n_5 = 1$, which is a contradiction.
\item $\{ p,q \} = \{ 6,9\}$: same as above. Consider its dual $\{9,6\}$ and look at $n_3$, the number of Sylow 3-subgroups of its rotational groups. As the dual has 2 faces, $n_3 \geq 2$ but a Sylow theorem implies $n_3 = 1$.
\item $\{ p,q \} = \{ 10,3\}$: this exists if and only if its dual $\{ 3,10\}$ exists. To see this does not exist, let $G$ be its rotational symmetry group. Then note that $|G|$ divides $2E = 60$. Noting its faces have $3$ sides, we see there are $n_3$ Sylow 3-subgroups of $G$, where $n_3 \geq F = 20$. However, a Sylow theorem states that $n_3 \equiv  1 \mod{3}$ and that $n_3$ divides $|G| / 3 = 20$. It is clearly impossible to satisfy all three of these conditions simultaneously.
\item $\{ p,q \} = \{10,5\}$: similar to $\{12,3\}$ with genus 2. We have $F=2, q=5$, so $F<q$ means that at each vertex, some face must border itself. However, $F,q$ are coprime, so at least one face must border a different face, so we don't have edge-transitivity.
\item $\{ p,q \} = \{18,3\}$: we note $F=2, q=3$, so $F<q$ and $F,q$ are coprime, meaning there is no edge-transitivity.
\item $\{p,q\} = \{ 30,3 \}$: consider the associated graph. This has 10 vertices, all of valency 3. We saw in the case of genus $g=2$, $\{p,q\} = \{10,3\}$ that no regular bipartite graph of this form exists. Hence no such bipartite tiling exists.
\end{itemize}

\paragraph*{\fbox{$\mathbf{\{p,q\} = \{14,7\}}$}}
This is the regular tiling with $V=2, E=7, F = 1$:
\begin{center}
\includegraphics[width=13cm]{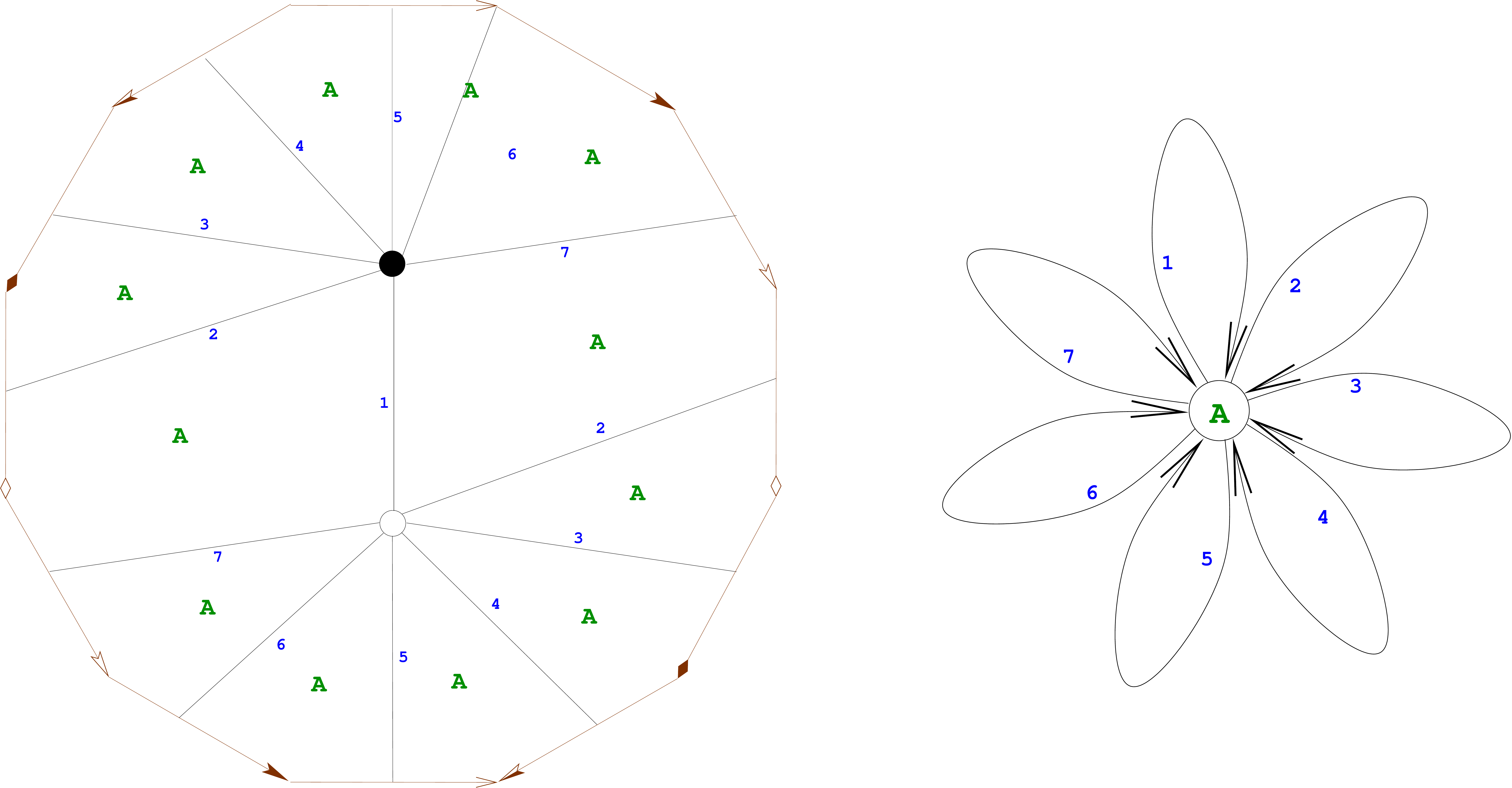}
\end{center}
We have superpotential 
\begin{equation} 
W = \tr \left( \Phi_1\Phi_2\Phi_3\Phi_4\Phi_5\Phi_6\Phi_7 - \Phi_1\Phi_7\Phi_6\Phi_5\Phi_4\Phi_3\Phi_2 \right) \ .
\end{equation}
We note the gauge invariants are simply the $\Phi_i$.
We see that, working over the complex numbers, all partials vanish, so the master space is
\begin{equation}
S/I_1 = \mathbb{C} \left[ \Phi_1,\dots\Phi_7\right] / \left< \partial_i W\right>_{i=1,\dots,7} = \mathbb{C} \left[ \Phi_1,\dots\Phi_7\right] \ .
\end{equation}

To find the vacuum moduli space, we consider the ring $R = \mathbb{C} \left[ \Phi_1,\dots\Phi_7,y_1,\dots,y_7 \right]$ and ideal $I_2 = \left< \partial_i W,y_i - \Phi_i \right>_{i=1,\dots,7}$, and we get that $I_2 = \left< y_i -\Phi_i \right>_{i=1,\dots,7}$ as, for $\Phi_i \in \mathbb{C}$, all partials are zero.
We then eliminate all the $\Phi$s and substitute the resulting ideal into ring $R' = \mathbb{C}\left[ y_1,\dots,y_7\right]$ to get ideal $V$ representing the vacuum moduli space. We see that
$R'/V \cong \mathbb{C} \left[ y_1,\dots,y_7 \right]$, which has degree 1, dimension 7 and Hilbert series 
\begin{equation} 
H(t, \cM_{g=3, \ (p,q)=(14,7)}) = \frac{1}{(1-t)^7} \ .
\end{equation}

We can also write down a permutation triple such that $\sigma_B \sigma_W \sigma_{\infty} = id$: 
\begin{equation}
\sigma_B = (1\;2\;3\;4\;5\;6\;7)\ , \quad
\sigma_W = (1\;2\;3\;4\;5\;6\;7) \ , \quad
\sigma_{\infty} = (1\;6\;4\;2\;7\;5\;3) \ .
\end{equation}
So we note we have ramification structure $\{7, 7, 7 \}$.

\paragraph*{\fbox{$\mathbf{ \{ p,q \} = \{ 4,12 \}}$}}

This is the tiling with $V=2, E=12, F=6$:

\begin{center}
\includegraphics[width=15cm]{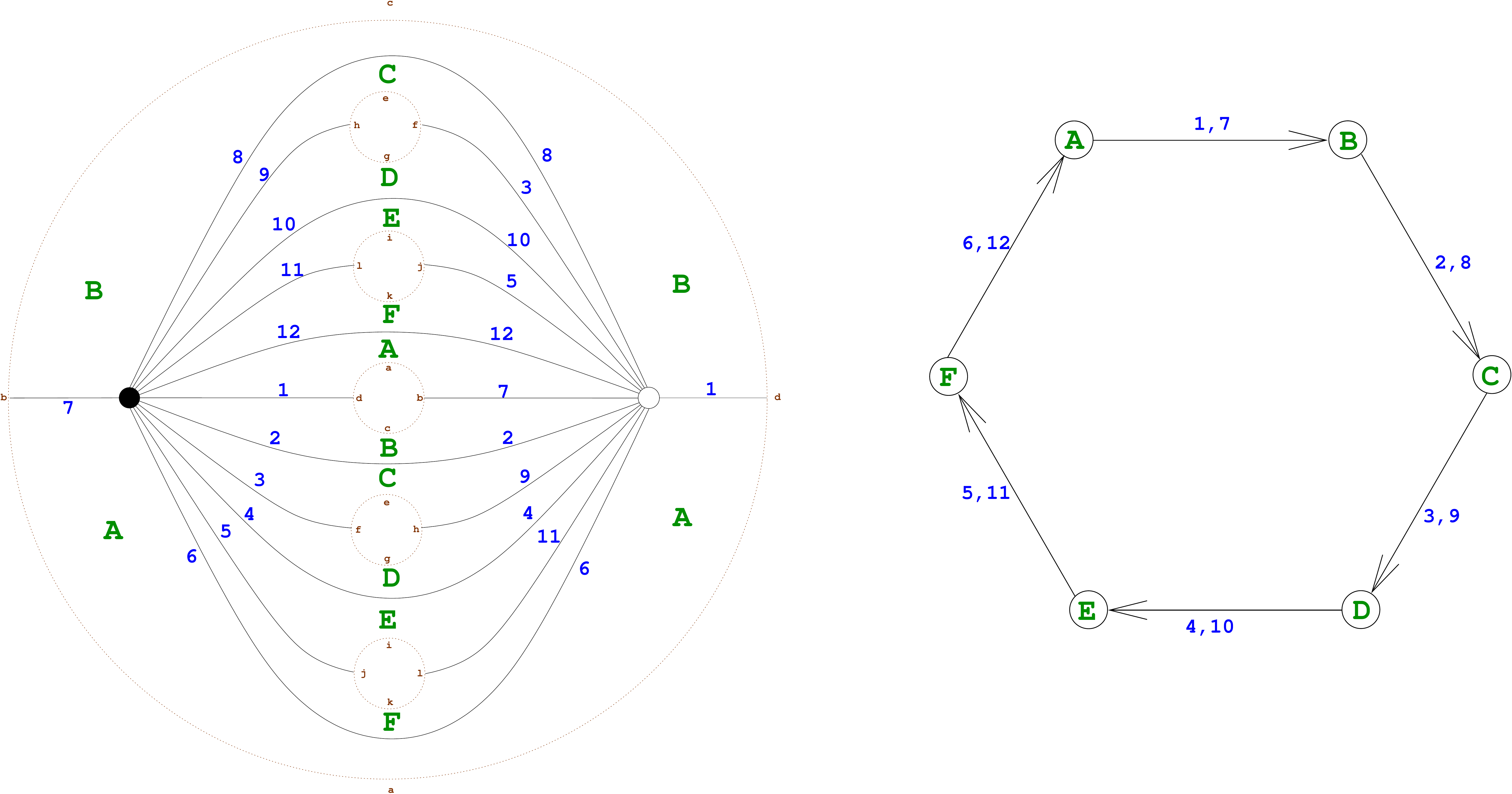}
\end{center}

We have superpotential 
\begin{equation}
W = \tr \left( \Phi_1\Phi_2\Phi_3\Phi_4\Phi_5\Phi_6\Phi_7\Phi_8\Phi_9\Phi_{10}\Phi_{11}\Phi_{12} - \Phi_1\Phi_8\Phi_3\Phi_{10}\Phi_5\Phi_{12}\Phi_7\Phi_2\Phi_9\Phi_4\Phi_{11}\Phi_6 \right)
\end{equation}
and 64 gauge invariants of the form
\begin{align}
&r_{i,j,k,l,m,n} = \Phi_i \Phi_j \Phi_k \Phi_l\Phi_m\Phi_n \nonumber \\ 
&\text{with } i\in\{1,7\}, j\in\{2,8\}, k \in \{3,9  \}, l \in\{4,10\}, m \in\{ 5,11\} , n \in \{6,12\} \ .
\end{align}
We see that, working over the complex numbers, all partials vanish, so the master space is
\begin{equation}
S/I_1 = \mathbb{C} \left[ \Phi_1,\dots\Phi_{12}\right] / \left< \partial_i W\right>_{i=1,\dots,12} = \mathbb{C} \left[ \Phi_1,\dots\Phi_{12}\right] \ .
\end{equation}
To find the vacuum moduli space, we consider the ring $R = \mathbb{C} \left[ \Phi_1,\dots\Phi_{12},y_1,\dots,y_{64} \right]$ and ideal $I_2 = \left< \partial_i W,y_j - r_j \right>_{i=1,\dots,12;j=1,\dots,64}$.
We then eliminate all the $\Phi$s and substitute the resulting ideal into ring $R' = \mathbb{C}\left[ y_1,\dots,y_{64}\right]$ to get ideal $V$ representing the vacuum moduli space. Using \cite{mac2}, we see that V has dimension 7, degree 720 and after assigning weights to each $y_j$ equal to the degree of the monomial they represent, we get Hilbert series
\begin{equation} 
H(t, \cM_{g=3, \ (p,q)=(4,12)}) = \frac{1+57t^6 + 302t^{12}+302t^{18}+57t^{24}+t^{30}}{(1-t^6)^7} \ ,
\end{equation}
its palindromic numerator indicating it is a Calabi-Yau 7-fold.

We can also write down a permutation triple such that $\sigma_B \sigma_W \sigma_{\infty} = id$: 
\begin{align}
\nonumber
\sigma_B &= (1\;2\;3\;4\;5\;6\;7\;8\;9\;10\;11\;12)\\
\nonumber
\sigma_W &= (1\;6\;11\;4\;9\;2\;7\;12\;5\;10\;3\;8) \\
\sigma_{\infty} &= (1\;7)(2\;8)(3\;9)(4\;10)(5\;11)(6\;12) \ .
\end{align}
So we note we have ramification structure $\{12,12,2^6\}$.

\paragraph*{\fbox{$\mathbf{\{p,q\} = \{12,4\}}$}}
This is the regular tiling with $V=6, E=12, F = 2$:

\begin{center}
\includegraphics[width=15cm]{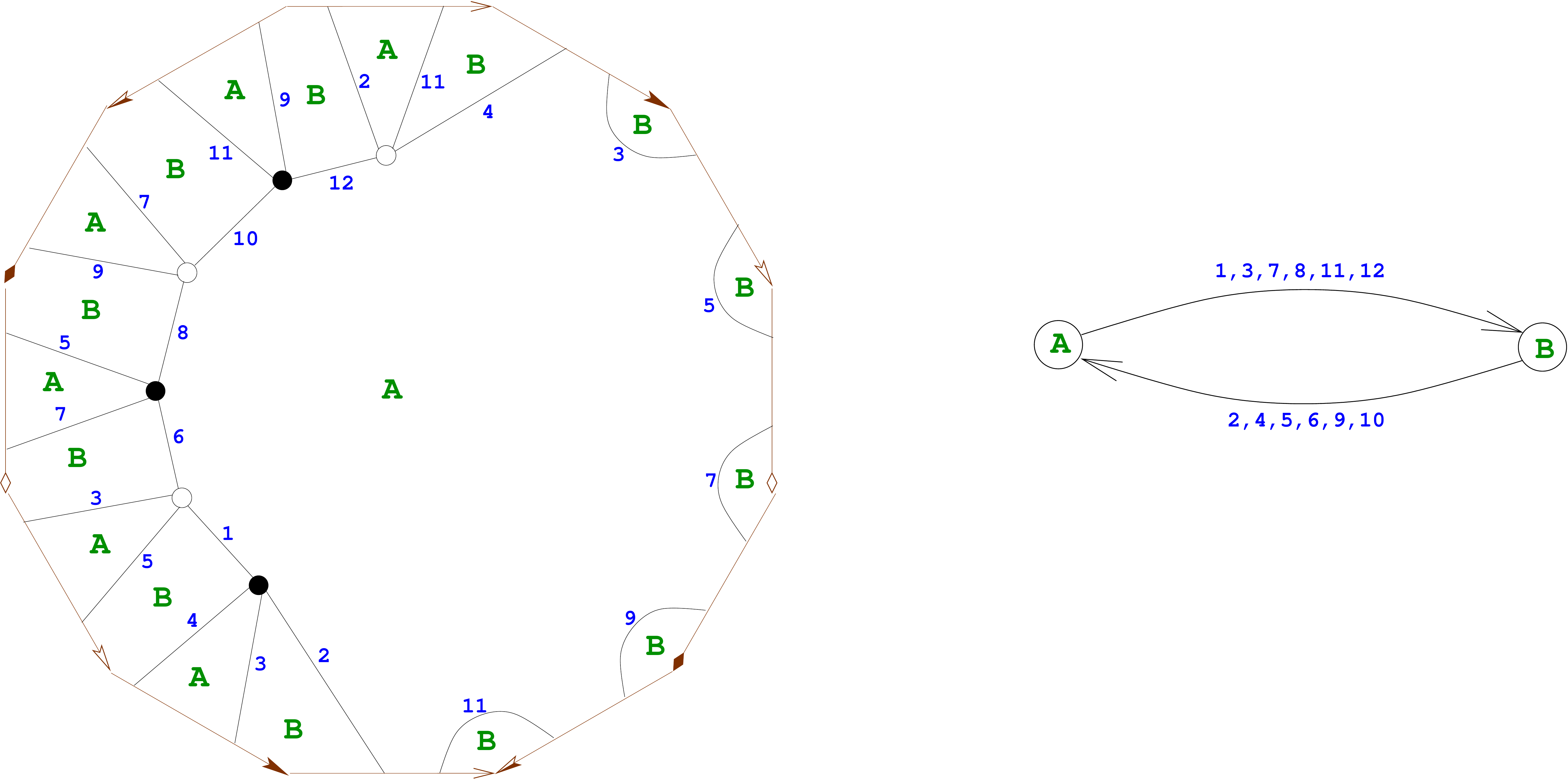}
\end{center}

We have superpotential 
\begin{equation} 
W =  \tr \left( \Phi_1\Phi_2\Phi_3\Phi_4 + \Phi_5\Phi_8\Phi_6\Phi_7 + \Phi_9\Phi_{12}\Phi_{10}\Phi_{11} - \Phi_1\Phi_6\Phi_3\Phi_5 - \Phi_2\Phi_{12}\Phi_4\Phi_{11}  - \Phi_7\Phi_9\Phi_8\Phi_{10} \right)
\end{equation}
and 36 gauge invariants:
\begin{equation}
r_{i,j} = \Phi_i\Phi_j \qquad \qquad \text{with } i\in \{ 1,3,7,8,11,12 \}, j\in \{ 2,4,5,6,9,10 \} \ . 
\end{equation}
To find the master space, we define the ring $S = \mathbb{C}[\Phi_1,\dots,\Phi_{12}]$ and ideal $I_1 = \left< \partial_i W \right>_{i=1,\dots,12}$ to generate the master space $R = S /I_1$. We find using \cite{mac2} that $I_1$ has dimension 8, degree 16 and Hilbert series
\begin{equation} 
H(t, \cF^{\flat}_{g=3, \ (p,q)=(12,4)}) = \frac{1+4t+10t^2+8t^3-6t^4-20t^5+28t^6-12t^7+3t^8}{(1-t)^8} \ .
\end{equation}

Using primary decomposition in \cite{mac2}, we get that the curve given by $I_1$ is the union of those given by ideals of which:
3 are trivial of dimension 4, degree 1 and have Hilbert series $\frac{1}{(1-t)^4}$;
2 are trivial of dimension 6, degree 1 and have Hilbert series $\frac{1}{(1-t)^6}$;
12 are non-trivial of dimension 6, degree 2 and have Hilbert series $\frac{1+t}{(1-t)^6}$.
The \emph{coherent component} is of dimension 8, degree 16 and has Hilbert series 
\begin{equation} 
H(t, \ ^{\text{Irr}}\cF^{\flat}_{g=3, \ (p,q)=(12,4)}) =  \frac{1+4t+6t^2+4t^3+t^4}{(1-t)^8} \ . 
\end{equation}

To find the vacuum moduli space, we consider the ring $R = \mathbb{C} \left[ \Phi_1,\dots\Phi_{12},y_1,\dots,y_{36} \right]$ and ideal $I_2 = \left< \partial_i W,y_j - \Phi_j \right>_{i=1,\dots,12;\; j=1,\dots,36}$.
We then eliminate all the $\Phi$s and substitute the resulting ideal into ring $R' = \mathbb{C}\left[ y_1,\dots,y_{36}\right]$ to get ideal $V$ representing the vacuum moduli space. Using \cite{mac2}, we see that V has dimension 7, degree 320 and after assigning weights to each $y_j$ equal to the degree of the monomial they represent, we get Hilbert series
\begin{equation} 
H(t, \cM_{g=3, \ (p,q)=(12,4)}) = \frac{1+29t^2+145t^4+109t^6+23t^8+19t^{10}-9t^{12}+3t^{14}}{(1-t^2)^7} \ .
\end{equation}
We can also write down a permutation triple such that $\sigma_B \sigma_W \sigma_{\infty} = id$: 
\begin{align}
\nonumber
\sigma_B &= (1\;2\;3\;4)(5\;8\;6\;7)(9\;12\;10\;11)\\
\nonumber
\sigma_W &= (1\;5\;3\;6)(2\;11\;4\;12)(7\;10\;8\;9) \\
\sigma_{\infty} &= (1\;11\;7\;3\;12\;8)(2\;6\;10\;4\;5\;9) \ .
\end{align}
So we note we have ramification structure $\{4^3, 4^3, 6^2\}$.

\paragraph*{\fbox{$\mathbf{\{p,q\} = \{8,8\}}$}}
This is the tiling with $V=2, E=2, F=8$.
Now we encounter two different theories, which though sharing the same moduli space, have different bipartite graphs which are non-isomorphic.
We will call the models A and B to distinguish them.
\subparagraph{Model A} We begin with the first theory, which has the following tessellation and quiver:

\begin{center}
\includegraphics[width=12cm]{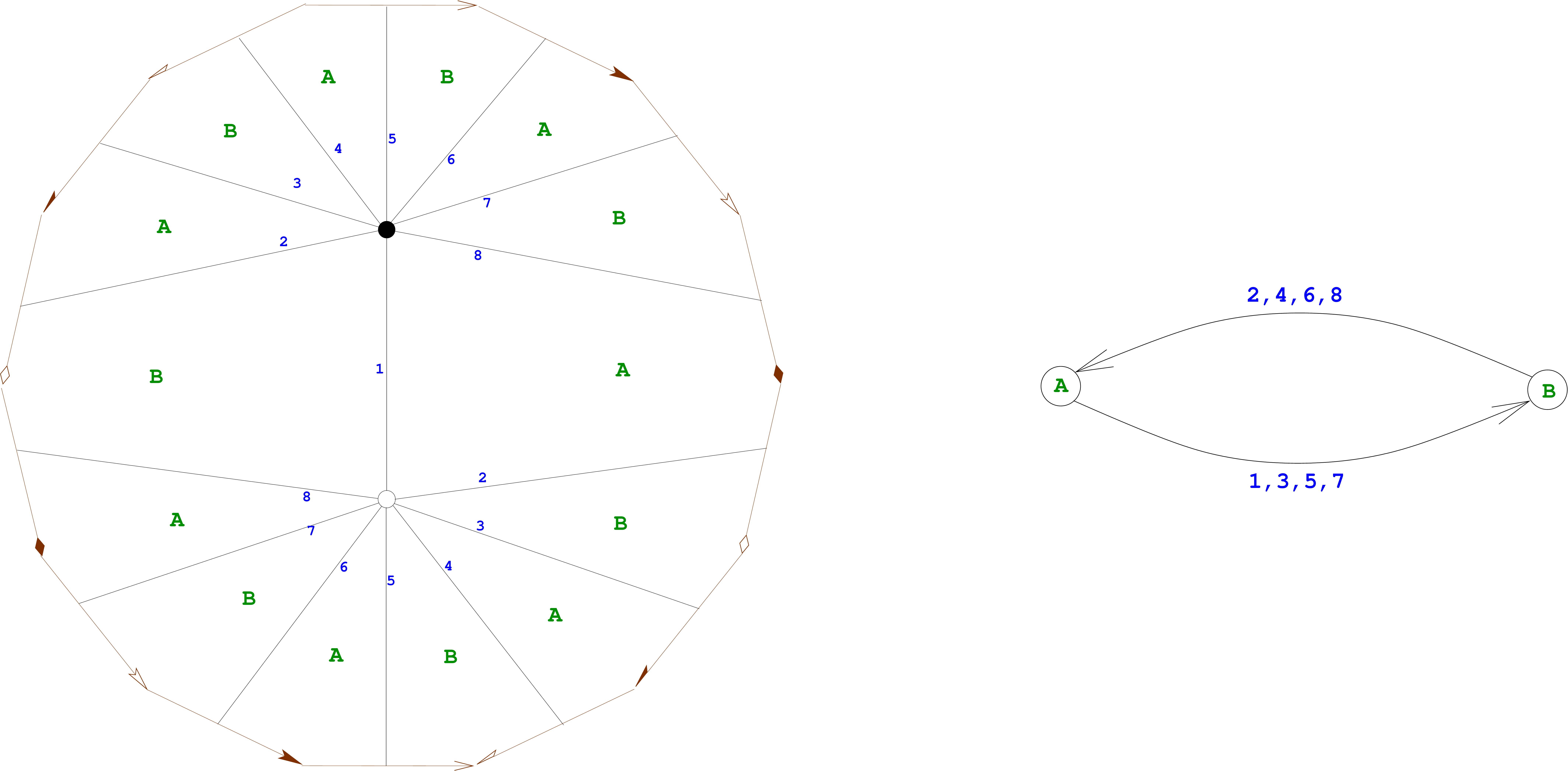}
\end{center}
We have superpotential:
\begin{equation}
W=\tr \left( \Phi_1\Phi_2\Phi_3\Phi_4\Phi_5\Phi_6\Phi_7\Phi_8 - \Phi_1\Phi_8\Phi_7\Phi_6\Phi_5\Phi_4\Phi_3\Phi_2 \right)
\end{equation}
and 16 gauge invariants:
\begin{equation}
r_{i,j} = \Phi_i \Phi_j \qquad\qquad \text{with } i \in \{1,3,5,7\}, j\in \{2,4,6,8\} \ .
\end{equation}
We see that, working over the complex numbers, all partials vanish, so the master space is
\begin{equation}
S/I_1 = \mathbb{C} \left[ \Phi_1,\dots\Phi_8\right] / \left< \partial_i W\right>_{i=1,\dots,8} = \mathbb{C} \left[ \Phi_1,\dots\Phi_8\right] \ . 
\end{equation}

To find the vacuum moduli space, we consider the ring $R = \mathbb{C} \left[ \Phi_1,\dots\Phi_8,y_1,\dots,y_{16} \right]$ and ideal $I_2 = \left< \partial_i W,y_j - \Phi_j \right>_{i=1,\dots,8;\;j=1,\dots,16}$, and we get that $I_2 = \left< y_j -\Phi_j \right>_{j=1,\dots,16}$ as, for $\Phi_i \in \mathbb{C}$, all partials are zero.
We then eliminate all the $\Phi$s and substitute the resulting ideal into ring $R' = \mathbb{C}\left[ y_1,\dots,y_8\right]$ to get ideal $V$ representing the vacuum moduli space. We see that V has dimension 7, degree 20 and has Hilbert series
\begin{equation}
H(t, \cM_{g=3, \ (p,q)=(8,8)}) = \frac{1+9t^2+9t^4+t^6}{(1-t^2)^7} \ ,
\end{equation}
its palindromic numerator indicating it is a Calabi-Yau 7-fold.
We can also write down a permutation triple such that $\sigma_B \sigma_W \sigma_{\infty} = id$: 
\begin{equation}
\sigma_B = (1\;2\;3\;4\;5\;6\;7\;8)\ , \quad
\sigma_W = (1\;2\;3\;4\;5\;6\;7\;8) \ , \quad
\sigma_{\infty} = (1\;7\;5\;3)(2\;8\;6\;4) \ .
\end{equation}
So we note we have ramification structure $\{8,8,4^2\}$.


\subparagraph{Model B} Next, we have the alternative theory, with tessellation and quiver

\begin{center}
\includegraphics[width=13cm]{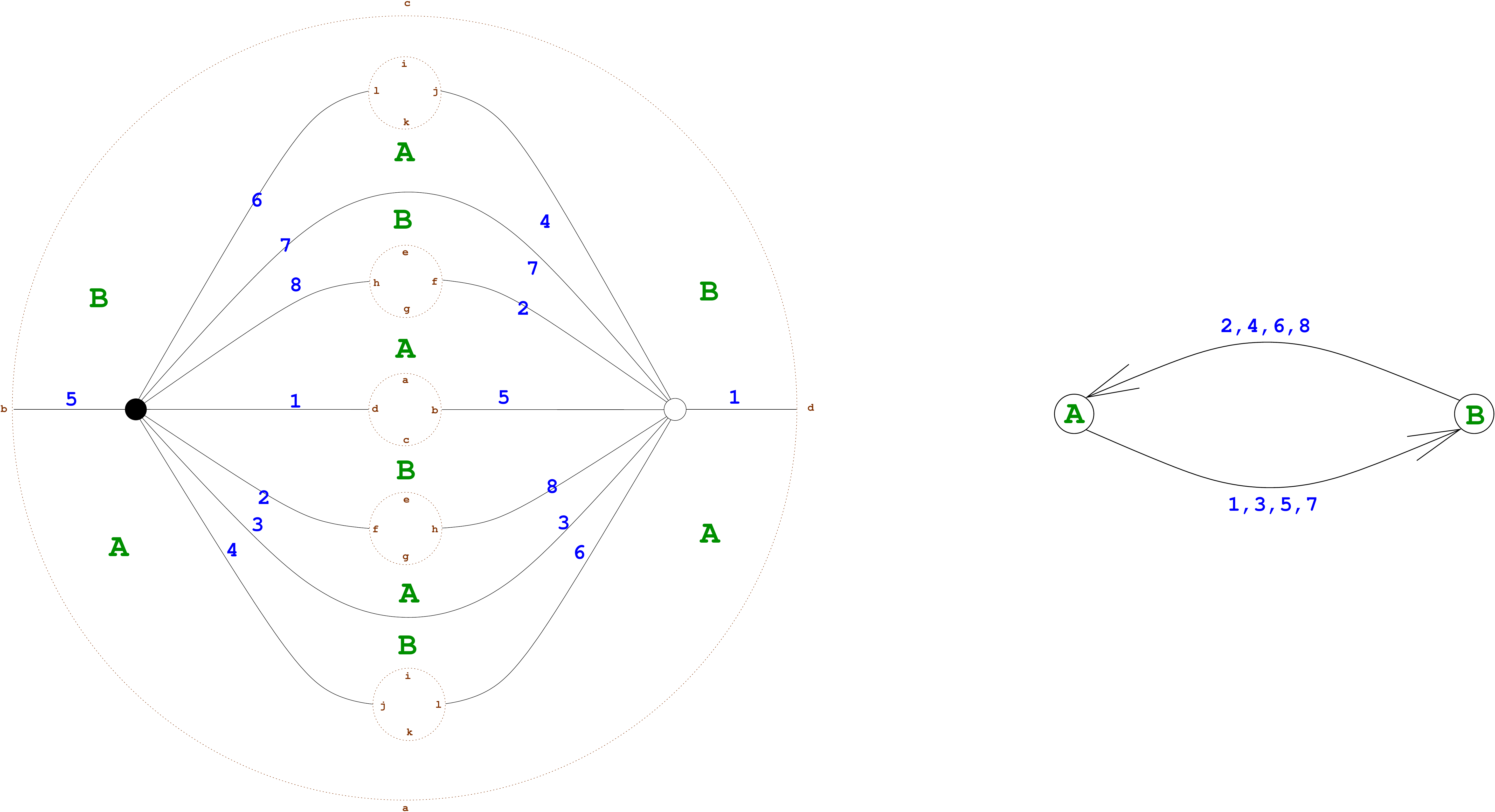}
\end{center}
\begin{equation}
W  = \tr \left( \Phi_1\Phi_2\Phi_3\Phi_4\Phi_5\Phi_6\Phi_7\Phi_8 - \Phi_1\Phi_8\Phi_3\Phi_6\Phi_5\Phi_4\Phi_7\Phi_2 \right) \ .
\end{equation}
We note that, over the complex numbers, all partials vanish. Also, all gauge invariants are the same as those in model A (up to permutation of the indices) and hence the master and vacuum moduli space are the same.

We can also write down a permutation triple such that $\sigma_B \sigma_W \sigma_{\infty} = id$: 
\begin{align}
\nonumber
\sigma_B &= (1\;2\;3\;4\;5\;6\;7\;8)\\
\nonumber
\sigma_W &= (1\;6\;3\;8\;5\;2\;7\;4) \\
\sigma_{\infty} &= (1\;3\;5\;7)(2\;4\;6\;8) \ .
\end{align}
So we note we have ramification structure $\{8,8,4^2\}$.
We see that though the ramification structure is the same, the actual triple, as elements of the permutation group, cannot be changed to that of Model A via redefinition.

\paragraph*{\fbox{$\mathbf{ \{p,q\} = \{4,8\}}$}}
This is the tiling with $V=4, E=16, F=8$.
Again, we have two theories, which we will call Models A and B.

\subparagraph{Model A}
Here the quiver and tessellation are:
\begin{center}
\includegraphics[width=15cm]{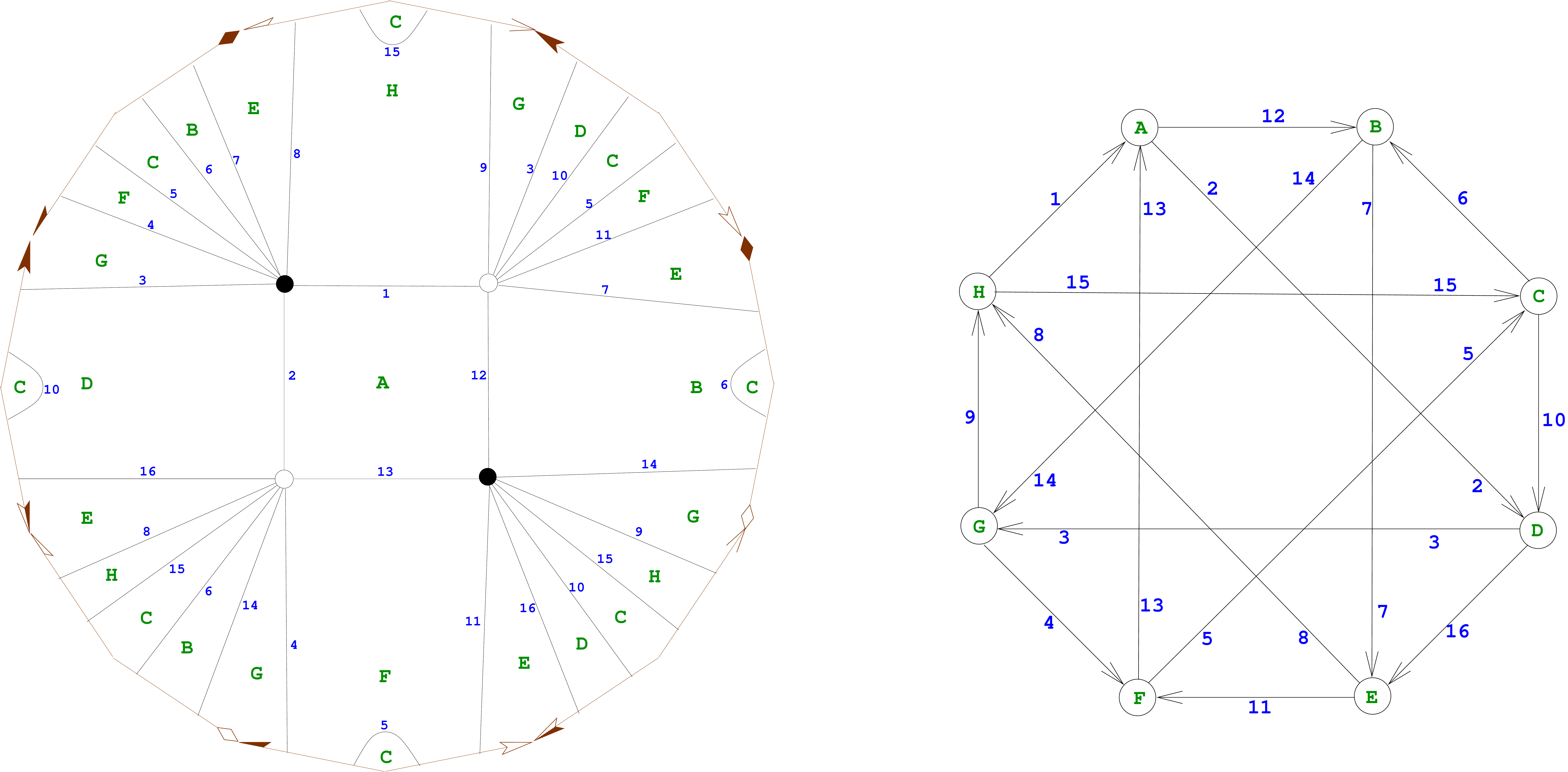}
\end{center}

We have superpotential
\begin{align}
W = \tr (&\Phi_1\Phi_2\Phi_3\Phi_4\Phi_5\Phi_6\Phi_7\Phi_8 + \Phi_9\Phi_{15}\Phi_{10}\Phi_{16}\Phi_{11}\Phi_{13}\Phi_{12}\Phi_{14} \nonumber \\
& - \Phi_1\Phi_{12}\Phi_7\Phi_{11}\Phi_5\Phi_{10}\Phi_3\Phi_9 - \Phi_2\Phi_{16}\Phi_8\Phi_{15}\Phi_6\Phi_{14}\Phi_4\Phi_{13})
\end{align}
and 16 gauge invariants:
\begin{align}
\nonumber
&
r_1 = \Phi_1\Phi_{12}\Phi_7\Phi_8\ , \quad
r_2 = \Phi_1\Phi_2\Phi_{16}\Phi_8\ , \quad
r_3 = \Phi_1\Phi_2\Phi_3\Phi_9\ , \quad  
r_4 = \Phi_1\Phi_{12}\Phi_{14}\Phi_9\ , \quad
\\
\nonumber
&
r_5 = \Phi_2\Phi_3\Phi_4\Phi_{13}\ , \quad
r_6 = \Phi_2\Phi_{16}\Phi_{11}\Phi_{13}\ , \quad
r_7 = \Phi_3\Phi_4\Phi_5\Phi_{10}\ , \quad 
r_8 = \Phi_3\Phi_9\Phi_{15}\Phi_{10}\ , \quad
\\
\nonumber
&
r_9 = \Phi_4\Phi_5\Phi_6\Phi_{14}\ , \quad
r_{10} = \Phi_4\Phi_{13}\Phi_{12}\Phi_{14}\ , \quad
r_{11} = \Phi_5\Phi_6\Phi_7\Phi_{11}\ , \quad
r_{12} = \Phi_5\Phi_{10}\Phi_{16}\Phi_{11}\ , \quad
\\
&
r_{13} = \Phi_6\Phi_7\Phi_8\Phi_{15}\ , \quad
r_{14} = \Phi_6\Phi_{14}\Phi_9\Phi_{15}\ , \quad
r_{15} = \Phi_7\Phi_{11}\Phi_{13}\Phi_{12}\ , \quad
r_{16} = \Phi_8\Phi_{15}\Phi_{10}\Phi_{16} \ .
\end{align}

To find the master space, we define the ring $S = \mathbb{C}[\Phi_1,\dots,\Phi_{16}]$ and ideal $I_1 = \left< \partial_i W \right>_{i=1,\dots,16}$ to generate the master space $R = S/I_1$. We find using \cite{mac2} that $I_1$ has dimension 14, degree 16 and Hilbert series
\begin{equation}
H(t, \cF^{\flat}_{g=3, \ (p,q)=(4,8)_A}) = \frac{1 + 2t+3t^2 + 4t^3 + 5t^4 + 6t^5 + 7t^6 - 8t^7 - 10t^8 - 12t^9 +18t^{10}}{(1-t)^{14}} \ .
\end{equation}

To find the vacuum moduli space, we consider the ring $R = \mathbb{C} \left[ \Phi_1,\dots\Phi_{16},y_1,\dots,y_{16} \right]$ and ideal $I_2 = \left< \partial_i W,y_j - r_j \right>_{i=1,\dots,16;j=1,\dots,16}$.
We then eliminate all the $\Phi$s and substitute the resulting ideal into ring $R' = \mathbb{C}\left[ y_1,\dots,y_{16}\right]$ to get ideal $V$ representing the vacuum moduli space. Using \cite{mac2}, we see that V has dimension 7, degree 88 and after assigning weights to each $y_j$ equal to the degree of the monomial they represent, we get Hilbert series
\begin{equation}
H(t, \cM_{g=3, \ (p,q)=(4,8)_A}) = \frac{1 +9t^4 + 37t^8 + 29t^{12} + 32t^{20} - 35t^{24} + 21t^{28} - 7t^{32} + t^{36}}{(1-t^4)^7} \ .
\end{equation}

We can also write down a permutation triple such that $\sigma_B \sigma_W \sigma_{\infty} = id$: 
\begin{align}
\nonumber
\sigma_B &= (1\;2\;3\;4\;5\;6\;7\;8)(9\;15\;10\;16\;11\;13\;12\;14)\\
\nonumber
\sigma_W &= (1\;9\;3\;10\;5\;11\;7\;12)(2\;13\;4\;14\;6\;15\;8\;16) \\
\sigma_{\infty} &= (1\;15)(2\;12)(3\;16)(4\;9)(5\;13)(6\;10)(7\;14)(8\;11) \ .
\end{align}
So we note we have ramification structure $\{8^2,8^2,2^8\}$.

\subparagraph{Model B} The tesselation and quiver are:

\begin{center}
\includegraphics[width=15cm]{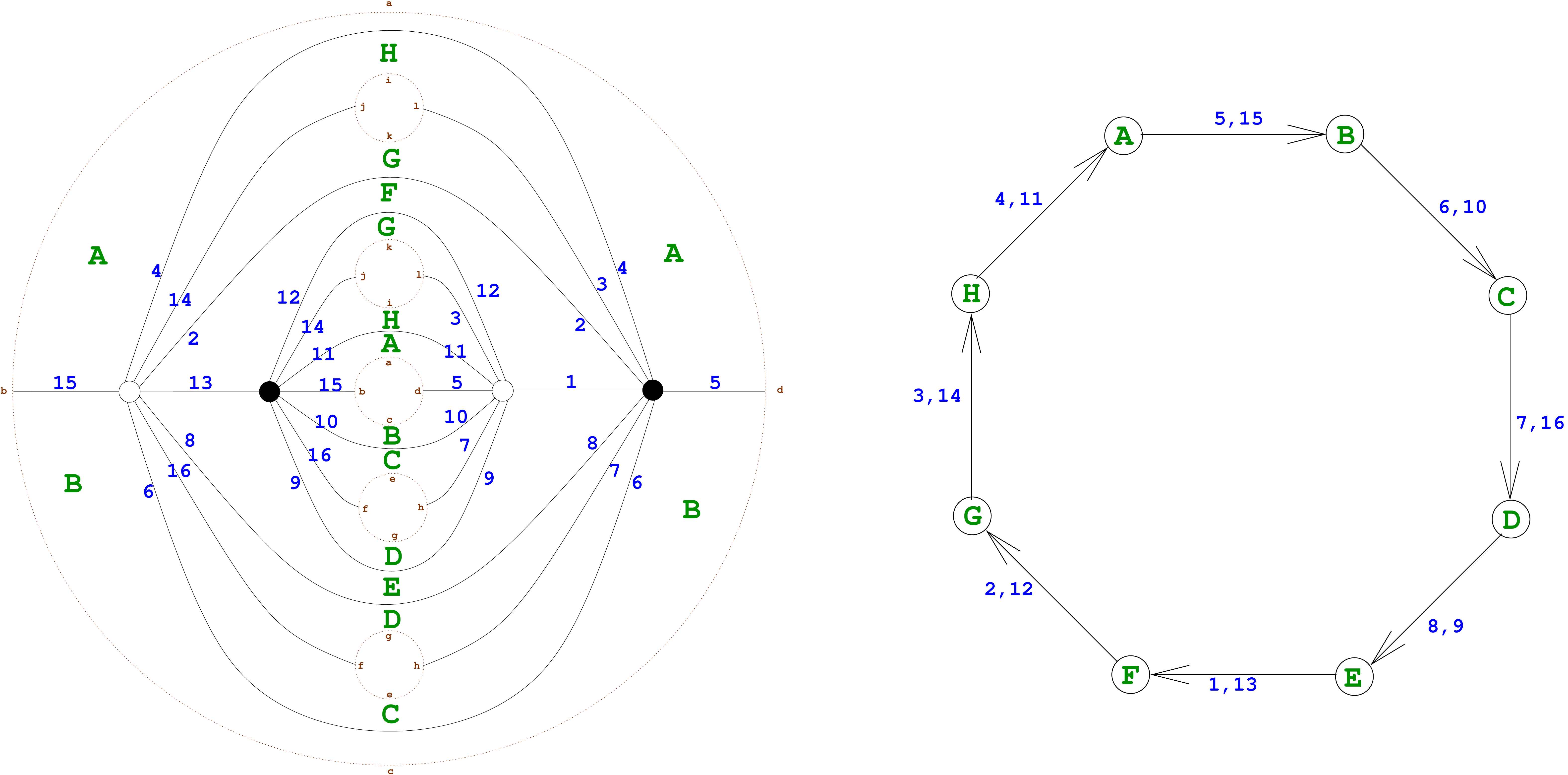}
\end{center}

We have superpotential
\begin{align} 
W = \tr ( & \Phi_1\Phi_2\Phi_3\Phi_4\Phi_5\Phi_6\Phi_7\Phi_8 + \Phi_9\Phi_{13}\Phi_{12}\Phi_{14}\Phi_{11}\Phi_{15}\Phi_{10}\Phi_{16} \nonumber \\
&- \Phi_1\Phi_{12}\Phi_3\Phi_{11}\Phi_5\Phi_{10}\Phi_7\Phi_9 - \Phi_2\Phi_{14}\Phi_4\Phi_{15}\Phi_6\Phi_{16}\Phi_8\Phi_{13} )
\end{align}
and 256 gauge invariants:
\begin{align}
&r_{i_1,\dots i_8} = \prod_{j=1}^8 \Phi_{i_j} \nonumber \\
&\text{with } i_1\in\{1,13\}, i_2\in\{2,12\}, i_3\in\{3,14\}, i_4\in\{4,11\}, \nonumber \\
&i_5\in\{5,15\}, i_6\in\{6,10\}, i_7\in\{7,16\}, i_8\in\{8,9\} \ .
\end{align}
To find the master space, we define the ring $S = \mathbb{C}[\Phi_1,\dots,\Phi_{16}]$ and ideal $I_1 = \left< \partial_i W \right>_{i=1,\dots,16}$ to generate the master space $R = S/I_1$. We find using singular that $I_1$ has dimension 14, degree 16 and Hilbert series
\begin{equation}
H(t, \cF^{\flat}_{g=3, \ (p,q)=(4,8)_B}) = \frac{1 +2t+3t^2 + 4t^3 + 5t^4 + 6t^5 + 7t^6 - 8t^7 - 10t^8 - 12t^9 +18t^{10}}{(1-t)^{14}} \ .
\end{equation}
So we see we have the same master space as in model A. This should be expected, as we see that for $\Phi_i \in \mathbb{C}$, the superpotentials are the same in models A and B.

We can also write down a permutation triple such that $\sigma_B \sigma_W \sigma_{\infty} = id$: 
\begin{align}
\nonumber
\sigma_B &= (1\;2\;3\;4\;5\;6\;7\;8)(9\;13\;12\;14\;11\;15\;10\;16)\\
\nonumber
\sigma_W &= (1\;9\;7\;10\;5\;11\;3\;12)(2\;13\;8\;16\;6\;15\;4\;14) \\
\sigma_{\infty} &= (1\;13)(2\;12)(3\;14)(4\;11)(5\;15)(6\;10)(7\;16)(8\;9) \ .
\end{align}
So we note we have ramification structure $\{8^2,8^2,2^8\}$.

\paragraph*{\fbox{$\mathbf{ \{p,q\} = \{8,4\}}$}}
This is the tiling with $V=8, E=16, F=4$. There are again two models, A and B:

\subparagraph{Model A} The tesselation and quiver are:

\begin{center}
\includegraphics[width=15cm]{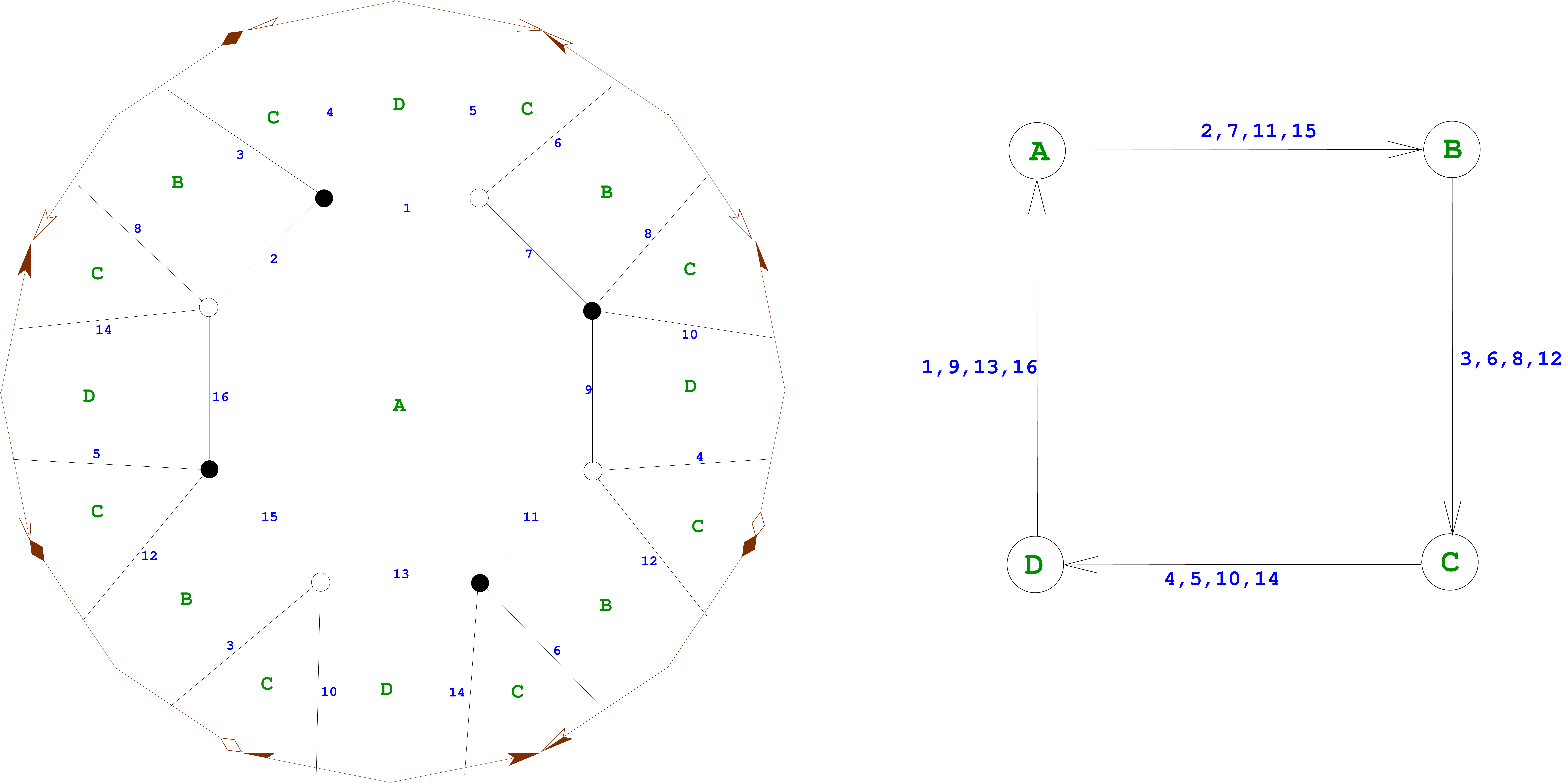}
\end{center}

We have superpotential
\begin{align}
W = \tr ( & \Phi_1\Phi_2\Phi_3\Phi_4 + \Phi_5\Phi_{16}\Phi_{15}\Phi_{12} + \Phi_6\Phi_{14}\Phi_{13}\Phi_{11} + \Phi_7\Phi_8\Phi_{10}\Phi_9 \nonumber \\
& - \Phi_1\Phi_7\Phi_6\Phi_5 - \Phi_2\Phi_8\Phi_{14}\Phi_{16} - \Phi_3\Phi_{10}\Phi_{13}\Phi_{15} - \Phi_4\Phi_9\Phi_{11}\Phi_{12})
\end{align}
and 256 gauge invariants:
\begin{align}
&r_{i,j,k,l} = \Phi_i\Phi_j\Phi_k\Phi_l \nonumber \\
& \text{with } i \in \{ 1,9,13,16 \}, j\in\{2,7,11,15\}, k \in \{ 3,6,8,12 \}, l \in \{ 4,5,10,14 \} \ .
\end{align}

To find the master space, we define the ring $S = \mathbb{C}[\Phi_1,\dots,\Phi_{16}]$ and ideal $I_1 = \left< \partial_i W \right>_{i=1,\dots,16}$ to generate the master space $R = S/I_1$. We find using \cite{mac2}	 that $I_1$ has dimension 10, degree 96 and Hilbert series
\begin{equation}
H(t, \cF^{\flat}_{g=3, \ (p,q)=(8,4)_A}) = \frac{1 + 6t + 21t^2 + 40t^3 + 39t^4 -30t^5 + 19t^6}{(1-t)^{10}} \ .
\end{equation}

We can also write down a permutation triple such that $\sigma_B \sigma_W \sigma_{\infty} = id$: 
\begin{align}
\nonumber
\sigma_B &= (1\;2\;3\;4)(5\;16\;15\;12)(6\;14\;13\;11)(7\;8\;10\;9)\\
\nonumber
\sigma_W &= (1\;5\;6\;7)(2\;16\;14\;8)(3\;15\;13\;10)(4\;12\;11\;9)\\
\sigma_{\infty} &= (1\;9\;13\;16)(2\;7\;11\;15)(3\;8\;6\;12)(4\;10\;14\;5) \ .
\end{align}
So we note we have ramification structure $\{4^4,4^4,4^4\}$.

\subparagraph{Model B} The tesselation and quiver are:

\begin{center}
\includegraphics[width=15cm]{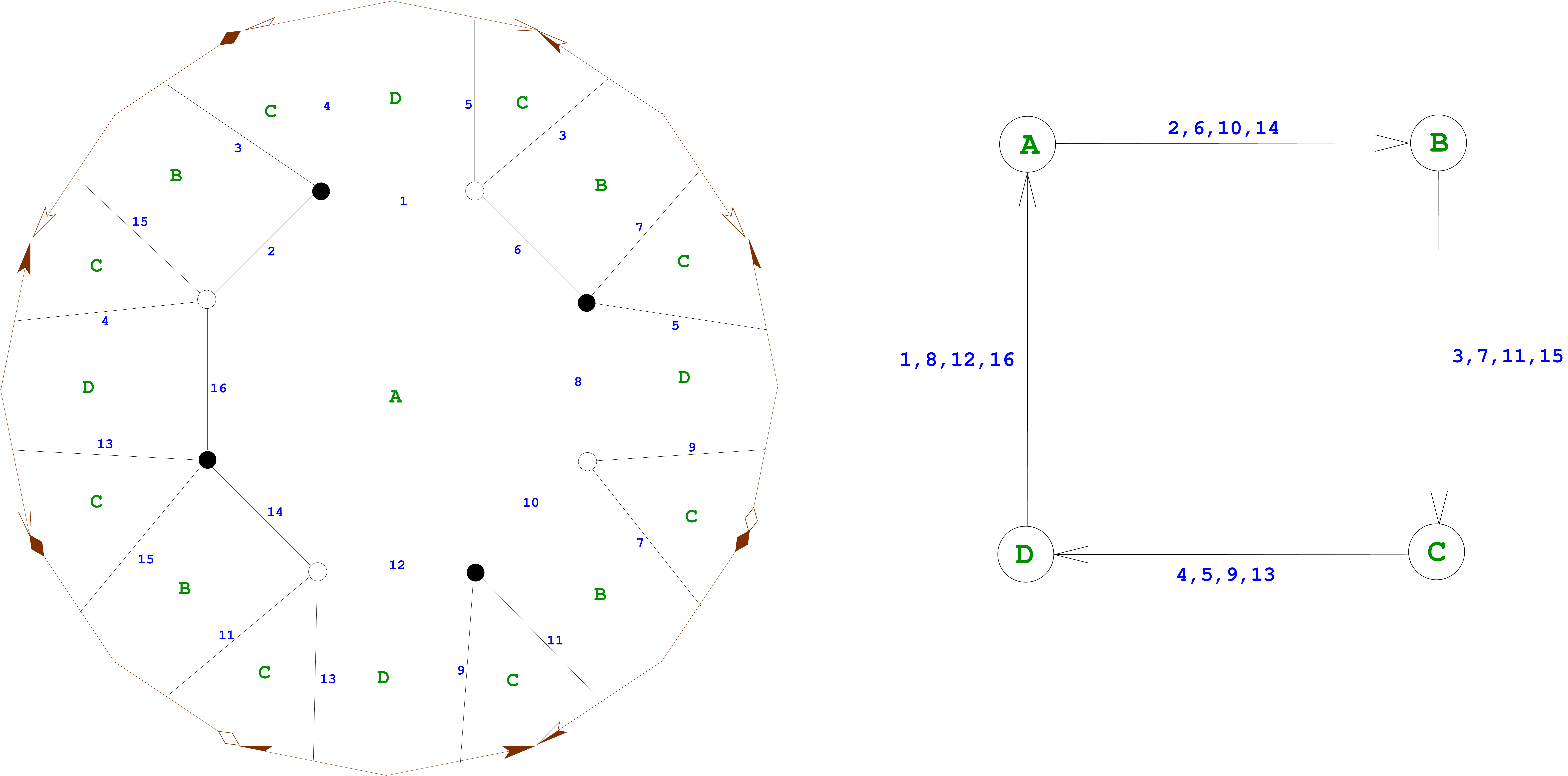}
\end{center}

We have superpotential
\begin{align}
W = \tr ( & \Phi_1\Phi_2\Phi_3\Phi_4 + \Phi_5\Phi_8\Phi_6\Phi_7 + \Phi_9\Phi_{12}\Phi_{10}\Phi_{11} + \Phi_{13}\Phi_{16}\Phi_{14}\Phi_{15} \nonumber \\
&- \Phi_1\Phi_6\Phi_3\Phi_5 - \Phi_2\Phi_{15}\Phi_4\Phi_{16}-\Phi_7\Phi_9\Phi_8\Phi_{10} - \Phi_{11}\Phi_{13}\Phi_{12}\Phi_{14})
\end{align}
and 256 gauge invariants:
\begin{align}
&r_{i,j,k,l} = \Phi_i\Phi_j\Phi_k\Phi_l \nonumber \\
&\text{with } i \in \{ 1,8,12,16 \}, j\in\{2,6,10,14\}, k \in \{ 3,7,11,15 \}, l \in \{ 4,5,9,13 \} \ .
\end{align}

To find the master space, we define the ring $S = \mathbb{C}[\Phi_1,\dots,\Phi_{16}]$ and ideal $I_1 = \left< \partial_i W \right>_{i=1,\dots,16}$ to generate the master space $R = S/I_1$. We find using \cite{mac2} that $I_1$ has dimension 10, degree 64 and Hilbert series
\begin{align}
& H(t, \cF^{\flat}_{g=3, \ (p,q)=(8,4)_B}) = \nonumber \\
& \frac{1 + 6t +21t^2 + 40t^3 + 39t^4 - 30t^5 - 99t^6 + 44t^7 + 106t^8 - 96t^9 + 32t^{10}}{(1-t)^{10}} \ .
\end{align}

We can also write down a permutation triple such that $\sigma_B \sigma_W \sigma_{\infty} = id$: 
\begin{align}
\nonumber
\sigma_B &= (1\;2\;3\;4)(5\;8\;6\;7)(9\;12\;10\;11)(13\;16\;14\;15)\\
\nonumber
\sigma_W &= (1\;5\;3\;6)(2\;16\;4\;15)(7\;10\;8\;9)(11\;14\;12\;13)\\
\sigma_{\infty} &= (1\;16\;12\;8)(2\;6\;10\;14)(3\;15\;11\;7)(4\;5\;9\;13) \ .
\end{align}
So we note we have ramification structure $\{4^4,4^4,4^4\}$.

\paragraph*{\fbox{$\mathbf{ \{p,q\} = \{14,3\}}$}}

This is the tiling with $V=14, E=21, F=3$:\\

\begin{center}
\includegraphics[width=15cm]{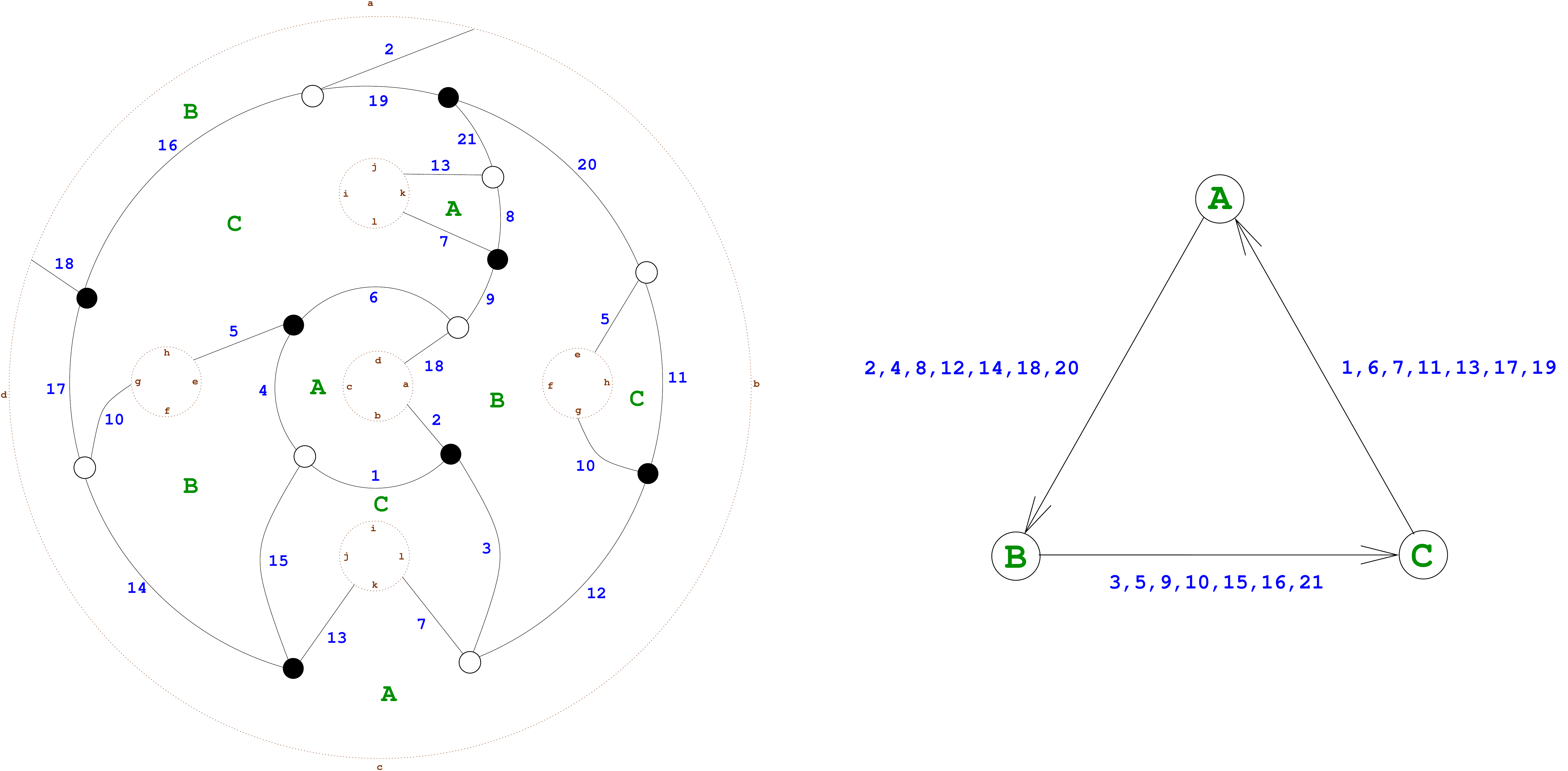}
\end{center}

We have superpotential
\begin{align}
W = \tr ( &\Phi_1\Phi_2\Phi_3 + \Phi_4\Phi_5\Phi_6 + \Phi_7\Phi_8\Phi_9 + \Phi_{10}\Phi_{11}\Phi_{12} + \Phi_{13}\Phi_{14}\Phi_{15} \nonumber \\
&+  \Phi_{16}\Phi_{17}\Phi_{18} + \Phi_{19}\Phi_{20}\Phi_{21} - \Phi_1\Phi_4\Phi_{15} - \Phi_2\Phi_{16}\Phi_{19} - \Phi_3\Phi_7\Phi_{12} \nonumber \\
&- \Phi_5\Phi_{11}\Phi_{20} - \Phi_6\Phi_{18}\Phi_9 - \Phi_8\Phi_{21}\Phi_{13} - \Phi_{10}\Phi_{17}\Phi_{14} )
\end{align}
and 343 gauge invariants:
\begin{align}
& r _{i,j,k} = \Phi_i \Phi_j \Phi_k  \nonumber \\
&\text{with } i\in\{1,6,7,11,13,17,19\}, j\in\{ 2,4,8,12,14,18,20\}, k \in \{3,5,9,10,15,16,21\} \ .
\end{align}

To find the master space, we define the ring $S = \mathbb{C}[\Phi_1,\dots,\Phi_{21}]$ and ideal $I_1 = \left< \partial_i W \right>_{i=1,\dots,21}$ to generate the master space $R = S/I_1$. We find using \cite{mac2} that $I_1$ has dimension 9, degree 232 and Hilbert series 
\begin{equation}
H(t, \cF^{\flat}_{g=3, \ (p,q)=(14,3)}) = (1-t)^{-9}(1 + 12t + 57t^2 + 120t^3 + 57t^4 -72t^5 + 57t^6) \ .
\end{equation}

We can also write down a permutation triple such that $\sigma_B \sigma_W \sigma_{\infty} = id$: 
\begin{align}
\nonumber
\sigma_B &= (1\;2\;3)(4\;5\;6)(7\;8\;9)(10\;11\;12)(13\;14\;15)(16\;17\;18)(19\;20\;21)\\
\nonumber
\sigma_W &= (1\;15\;4)(2\;19\;16)(3\;12\;7)(5\;20\;11)(6\;9\;18)(8\;13\;21)(10\;14\;17)\\
\sigma_{\infty} &= (1\;6\;17\;13\;7\;11\;19)(2\;18\;8\;20\;4\;14\;12)(3\;9\;5\;10\;16\;21\;15) \ .
\end{align}
So we note we have ramification structure $\{3^7,3^7,7^3\}$.

\paragraph*{\fbox{$\mathbf{ \{p,q\} = \{6,4\}}$}}

This is the tiling with $V=12, E=24, F=8$:\\

\begin{center}
\includegraphics[width=15cm]{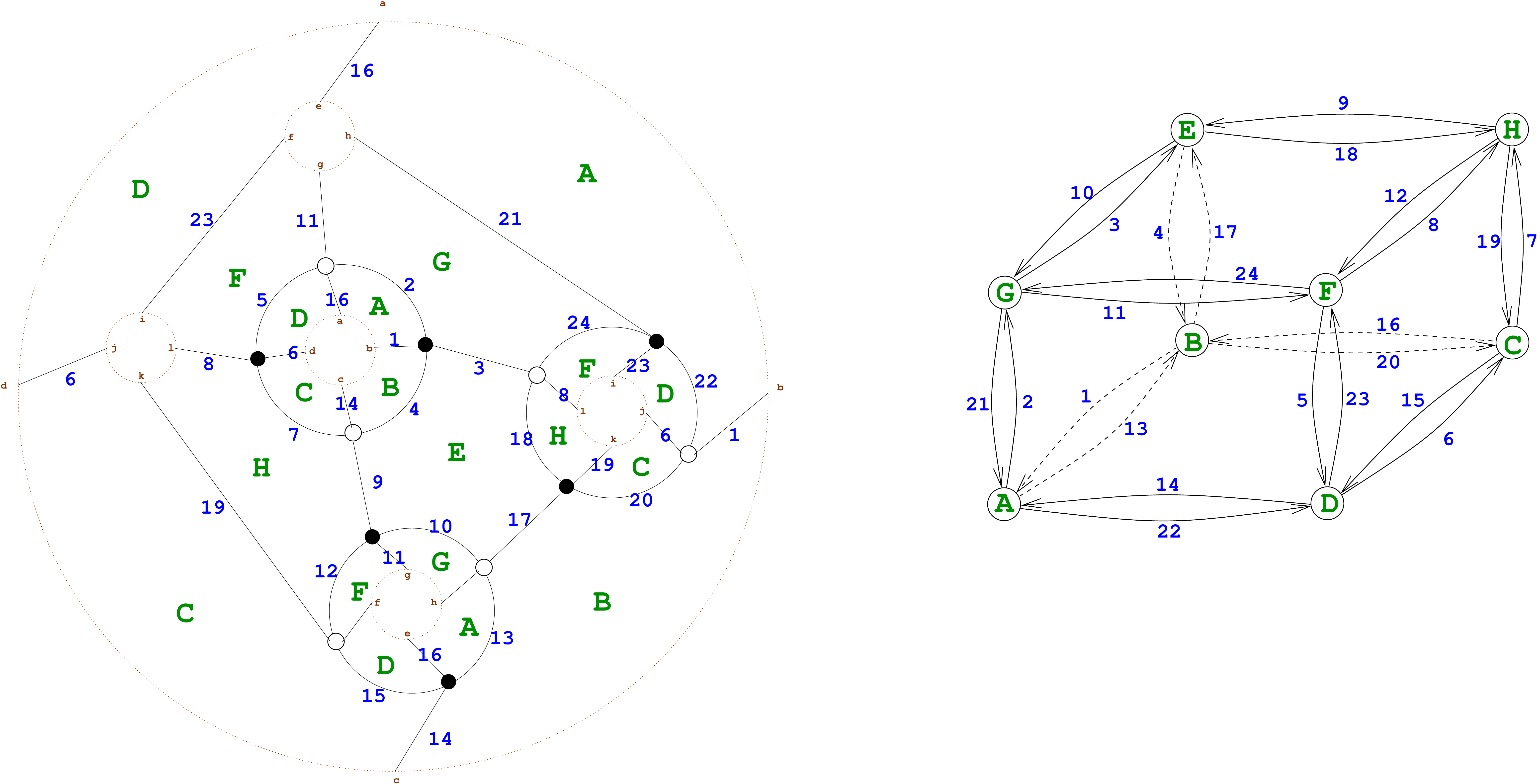}
\end{center}

We have superpotential
\begin{align}
W = \tr ( &\Phi_1\Phi_2\Phi_3\Phi_4 + \Phi_5\Phi_6\Phi_7\Phi_8 + \Phi_9\Phi_{10}\Phi_{11}\Phi_{12} + \Phi_{13}\Phi_{14}\Phi_{15}\Phi_{16} \nonumber \\
&+ \Phi_{17}\Phi_{18}\Phi_{19}\Phi_{20} + \Phi_{21}\Phi_{22}\Phi_{23}\Phi_{24} - \Phi_1\Phi_{22}\Phi_6\Phi_{20} - \Phi_2\Phi_{11}\Phi_5\Phi_{16} \nonumber \\
&- \Phi_3\Phi_{18}\Phi_8\Phi_{24} - \Phi_4\Phi_{14}\Phi_7\Phi_9 - \Phi_{12}\Phi_{19}\Phi_{15}\Phi_{23} - \Phi_{10}\Phi_{21}\Phi_{13}\Phi_{17})
\end{align}
and 24 gauge invariants:
\begin{align}
\nonumber
&
r_1 = \Phi_1\Phi_{13}\ , \quad  
r_2 = \Phi_2\Phi_{21}\ , \quad  
r_3 = \Phi_3\Phi_{10}\ , \quad  
r_4 = \Phi_4\Phi_{17}\ , \quad  
\\ \nonumber &
r_5 = \Phi_5\Phi_{23}\ , \quad  
r_6 = \Phi_6\Phi_{15}\ , \quad  
r_7 = \Phi_7\Phi_{19}\ , \quad  
r_8 = \Phi_8\Phi_{12}\ , \quad  
\\ \nonumber &
r_9 = \Phi_9\Phi_{18}\ , \quad  
r_{10} = \Phi_{11}\Phi_{24}\ , \quad  
r_{11} = \Phi_{14}\Phi_{22}\ , \quad  
r_{12} = \Phi_{16}\Phi_{20}\ , \quad  
\\ \nonumber &
r_{13} = \Phi_1\Phi_2\Phi_3\Phi_4 \ , \quad 
r_{14} = \Phi_{10}\Phi_{21}\Phi_{13}\Phi_{17} \ , \quad 
r_{15} = \Phi_1\Phi_{22}\Phi_6\Phi_{16} \ , \quad 
r_{16} = \Phi_{13}\Phi_{20}\Phi_{15}\Phi_{14} \ , \quad 
\\ \nonumber &
r_{17} = \Phi_2\Phi_{11}\Phi_5\Phi_{14} \ , \quad 
r_{18} = \Phi_{21}\Phi_{22}\Phi_{23}\Phi_{24} \ , \quad 
r_{19} = \Phi_3\Phi_{18}\Phi_{12}\Phi_{24} \ , \quad 
r_{20} = \Phi_8\Phi_9\Phi_{10}\Phi_{11} \ , \quad 
\\ &
r_{21} = \Phi_4\Phi_{20}\Phi_7\Phi_9 \ , \quad 
r_{22} = \Phi_{16}\Phi_{17}\Phi_{18}\Phi_{19} \ , \quad 
r_{23} = \Phi_5\Phi_6\Phi_7\Phi_{12} \ , \quad 
r_{24} = \Phi_8\Phi_{19}\Phi_{15}\Phi_{23} \ .
\end{align}

To find the master space, we define the ring $S = \mathbb{C}[\Phi_1,\dots,\Phi_{24}]$ and ideal $I_1 = \left< \partial_i W \right>_{i=1,\dots,24}$ to generate the master space $R = S/I_1$. We find using \cite{mac2} that $I_1$ has dimension 14, degree 2048 and Hilbert series 
\begin{align}
&H(t, \cF^{\flat}_{g=3, \ (p,q)=(6,4)}) = \nonumber \\
&(1-t)^{-14}(1 + 10t + 55t^2 + 196t^3 + 488t^4 + 812t^5 + 716t^6 -284t^7 \nonumber \\
&- 484t^8 + 212t^9 + 500t^{10} -276t^{11} + 117t^{12} - 18t^{13} + 3t^{14}) \ .
\end{align}

We can also write down a permutation triple such that $\sigma_B \sigma_W \sigma_{\infty} = id$: 
\begin{align}
\sigma_B &= (1\;2\;3\;4)(5\;6\;7\;8)(9\;10\;11\;12)(13\;14\;15\;16)(17\;18\;19\;20)(21\;22\;23\;24)\nonumber\\
\sigma_W &= (1\;20\;6\;22)(2\;16\;5\;11)(3\;24\;8\;18)(4\;9\;7\;14)(10\;17\;13\;21)(12\;23\;15\;19)\nonumber\\
\sigma_{\infty} &= (1\;21\;16)(2\;10\;24)(3\;17\;9)(4\;13\;20)(5\;15\;22)(6\;19\;14)(7\;12\;18)(8\;23\;11) \ .
\end{align}
So we note we have ramification structure $\{4^6,4^6,3^8\}$.

\paragraph*{\fbox{$\mathbf{ \{p,q\} = \{4,6\}}$}}

This is the tiling with $V=8, E=24, F=12$:\\

\begin{center}
\includegraphics[width=13cm]{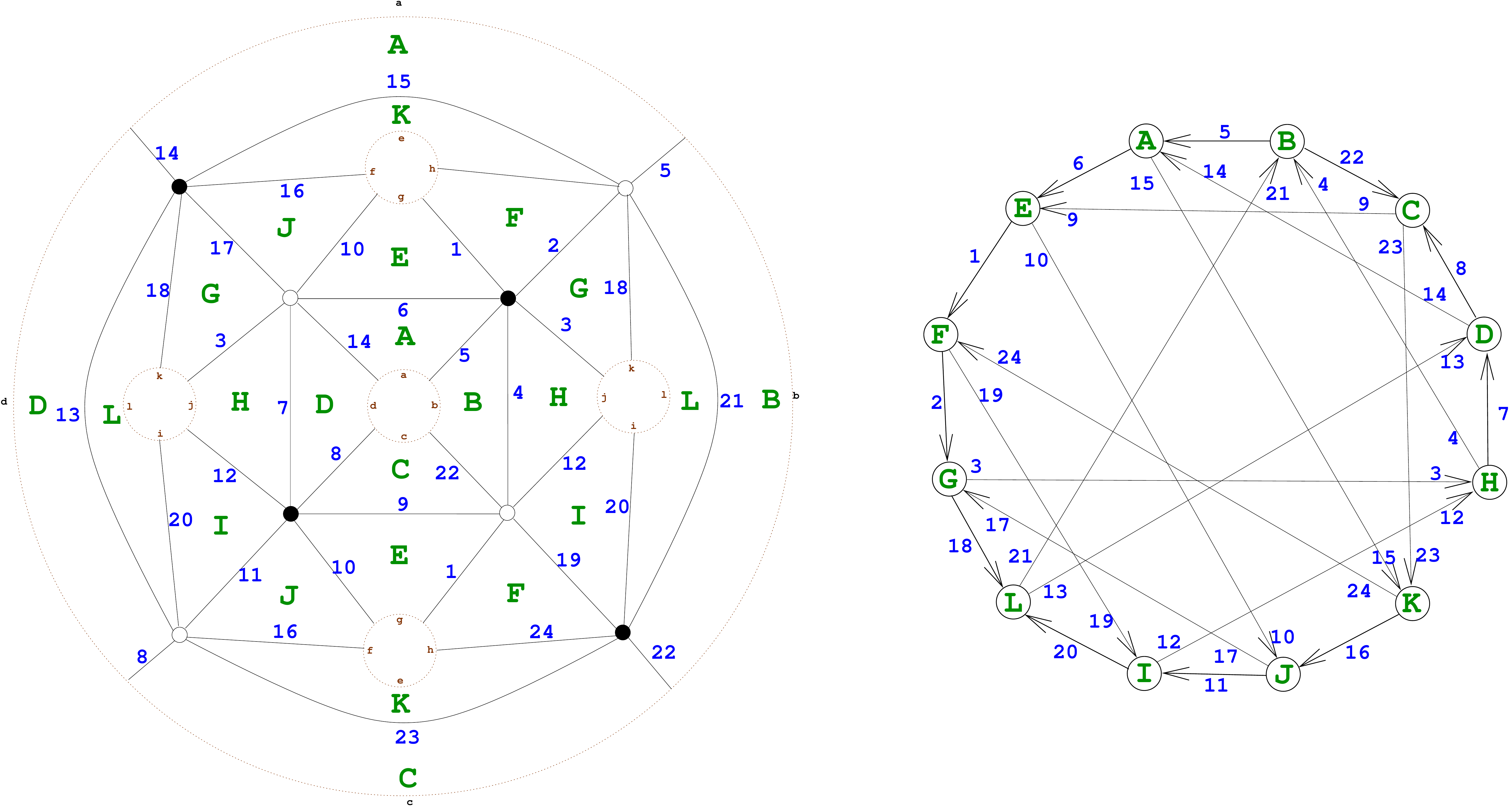}
\end{center}

We have superpotential
\begin{align}
W = \tr (&\Phi_1\Phi_2\Phi_3\Phi_4\Phi_5\Phi_6 + \Phi_7\Phi_8\Phi_9\Phi_{10}\Phi_{11}\Phi_{12} + \Phi_{13}\Phi_{14}\Phi_{15}\Phi_{16}\Phi_{17}\Phi_{18} \nonumber \\
& + \Phi_{19}\Phi_{20}\Phi_{21}\Phi_{22}\Phi_{23}\Phi_{24} - \Phi_1\Phi_{19}\Phi_{12}\Phi_4\Phi_{22}\Phi_9 - \Phi_2\Phi_{18}\Phi_{21}\Phi_5\Phi_{15}\Phi_{24} \nonumber \\ 
&- \Phi_3\Phi_7\Phi_{14}\Phi_6\Phi_{10}\Phi_{17} - \Phi_8\Phi_{23}\Phi_{16}\Phi_{11}\Phi_{20}\Phi_{13}) \ .
\end{align}
We have 64 gauge invariants, all of degree $6$. Due to the high number, we will only show those containing $\Phi_5$, to illustrate the loops in the quiver:
\begin{align}
r_1 &= \Phi_5\Phi_6\Phi_1\Phi_2\Phi_3\Phi_4, 			& r_9 &= \Phi_5\Phi_6\Phi_1\Phi_2\Phi_{18}\Phi_{21}  \nonumber \\
r_2 &= \Phi_5\Phi_{15}\Phi_{24}\Phi_2\Phi_3\Phi_4, 		& r_{10} &= \Phi_5\Phi_{15}\Phi_{24}\Phi_2\Phi_{18}\Phi_{21} \nonumber \\
r_3 &= \Phi_5\Phi_6\Phi_{10}\Phi_{17}\Phi_3\Phi_4, 		& r_{11} &= \Phi_5\Phi_6\Phi_{10}\Phi_{17}\Phi_{18}\Phi_{21}  \nonumber \\
r_4 &= \Phi_5\Phi_{15}\Phi_{16}\Phi_{17}\Phi_3\Phi_4, 	& r_{12} &= \Phi_5\Phi_{15}\Phi_{16}\Phi_{17}\Phi_{18}\Phi_{21} \nonumber \\
r_5 &= \Phi_5\Phi_6\Phi_{10}\Phi_{11}\Phi_{12}\Phi_4, 	& r_{13} &= \Phi_5\Phi_6\Phi_{10}\Phi_{11}\Phi_{20}\Phi_{21}  \nonumber \\
r_6 &= \Phi_5\Phi_{15}\Phi_{12}\Phi_{11}\Phi_{16}\Phi_4, & r_{14} &= \Phi_5\Phi_{15}\Phi_{16}\Phi_{11}\Phi_{20}\Phi_{21}  \nonumber \\
r_7 &= \Phi_5\Phi_6\Phi_1\Phi_{19}\Phi_{12}\Phi_4 , 	& r_{15} &= \Phi_5\Phi_6\Phi_1\Phi_{19}\Phi_{20}\Phi_{21} \nonumber \\
r_8 &= \Phi_5\Phi_{15}\Phi_{24}\Phi_{19}\Phi_{12}\Phi_4, & r_{16} &= \Phi_5\Phi_{15}\Phi_{24}\Phi_{19}\Phi_{20}\Phi_{21} \ .
\end{align}
To find the master space, we define the ring $S = \mathbb{C}[\Phi_1,\dots,\Phi_{24}]$ and ideal $I_1 = \left< \partial_i W \right>_{i=1,\dots,24}$ to generate the master space $R = S/I_1$. We find using \cite{mac2} that $I_1$ has dimension 18, degree 896 and Hilbert series
\begin{align}
&H(t, \cF^{\flat}_{g=3, \ (p,q)=(4,6)}) = \nonumber \\
&(1-t)^{-18}(1 + 6t +21t^2 + 56t^3 + 126t^4 + 228t^5 + 335t^6 + 390t^7 + 300t^8 - 70t^9 - 543t^{10}\nonumber \\
& -660t^{11} - 187t^{12} + 282t^{12} + 1329t^{14} - 1340 t^{15} + 894t^{16} - 384t^{17} + 139t^{18} - 30t^{19} + 3t^{20}) \ .
\end{align}
Due to the high number - and high degrees - of the partials and gauge invariants, we were unable to complete computations on primary decomposition of the master space and on the moduli space.

We can write down a permutation triple such that $\sigma_B \sigma_W \sigma_{\infty} = id$: 
\begin{align}
\sigma_B &= (1\;2\;3\;4\;5\;6)(7\;8\;9\;10\;11\;12)(13\;14\;15\;16\;17\;18)(19\;20\;21\;22\;23\;24)\nonumber\\
\sigma_W &= (1\;9\;22\;4\;12\;19)(2\;24\;15\;5\;21\;18)(3\;17\;10\;6\;14\;7)(8\;13\;20\;11\;16\;23) \nonumber\\
\sigma_{\infty} &= (1\;24)(2\;17)(3\;12)(4\;21)(5\;14)(6\;9)(7\;13)(8\;22)(10\;16)(11\;19)(15\;23)(18\;20) \ .
\end{align}
So we note we have ramification structure $\{6^4,6^4,2^{12}\}$.

\paragraph*{\fbox{$\mathbf{ \{p,q\} = \{8,3\}}$}}

This is the tiling with $V=32, E=48, F=12$:\\

\begin{center}
\includegraphics[width=12cm]{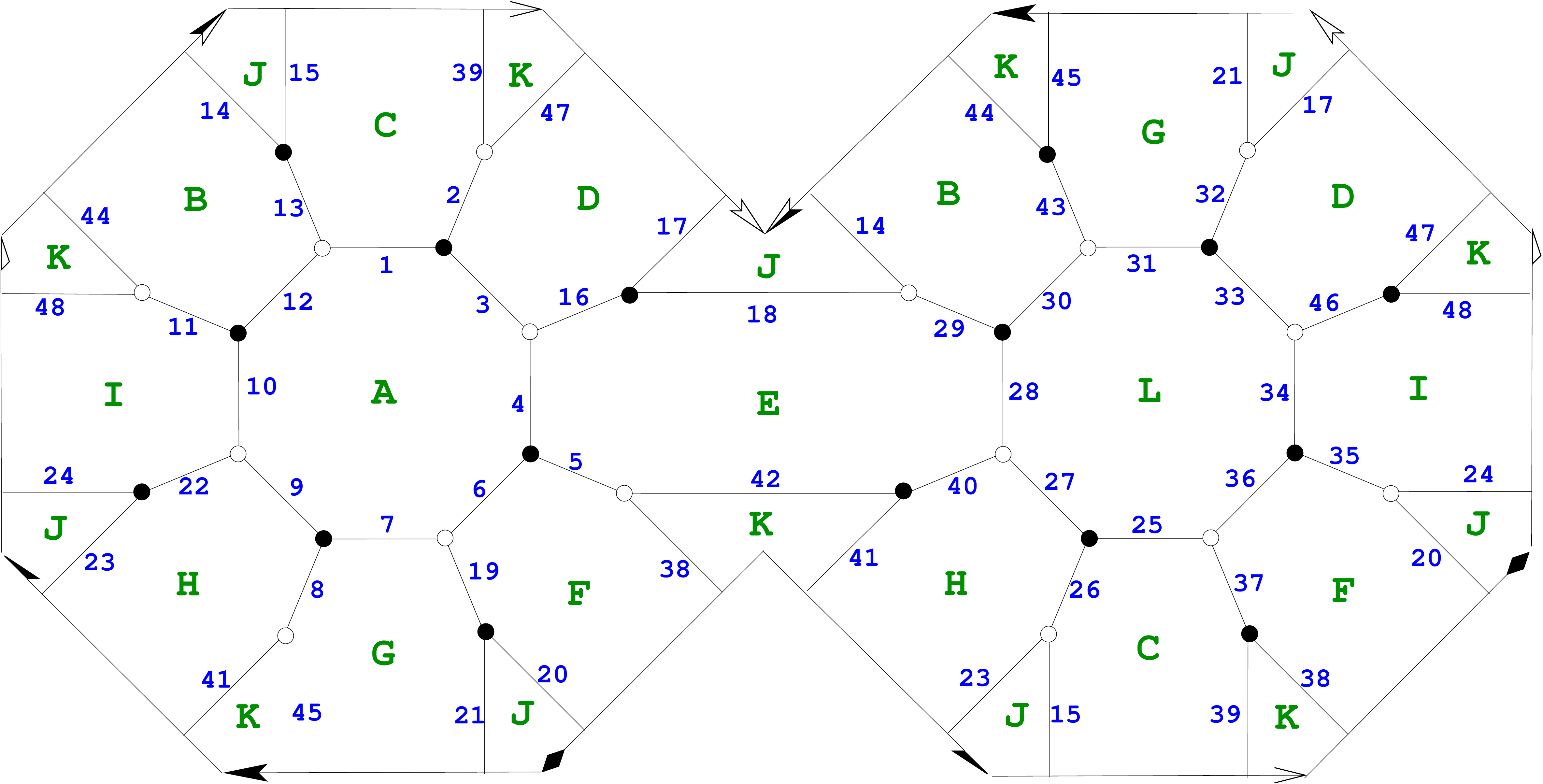}
\end{center}

We have superpotential
\begin{align}
W = \tr (&\Phi_1\Phi_2\Phi_3 + \Phi_4\Phi_5\Phi_6 + \Phi_7\Phi_8\Phi_9 + \Phi_{10}\Phi_{11}\Phi_{12} + \Phi_{13}\Phi_{14}\Phi_{15} + \Phi_{16}\Phi_{17}\Phi_{18} \nonumber \\
& + \Phi_{19}\Phi_{20}\Phi_{21} + \Phi_{22}\Phi_{24}\Phi_{23} + \Phi_{25}\Phi_{26}\Phi_{27} + \Phi_{28}\Phi_{29}\Phi_{30} + \Phi_{31}\Phi_{32}\Phi_{33} + \Phi_{34}\Phi_{35}\Phi_{36} \nonumber \\
& + \Phi_{37}\Phi_{38}\Phi_{39} + \Phi_{40}\Phi_{41}\Phi_{42} + \Phi_{43}\Phi_{44}\Phi_{45} + \Phi_{46}\Phi_{47}\Phi_{48} - \Phi_1\Phi_{13}\Phi_{12} - \Phi_2\Phi_{47}\Phi_{39} \nonumber \\
& - \Phi_3\Phi_4\Phi_{16} - \Phi_5\Phi_{38}\Phi_{42} - \Phi_6\Phi_7\Phi_{19} - \Phi_8\Phi_{41}\Phi_{45} - \Phi_9\Phi_{10}\Phi_{22} - \Phi_{11}\Phi_{44}\Phi_{48} \nonumber \\
& - \Phi_{14}\Phi_{18}\Phi_{29} - \Phi_{30}\Phi_{31}\Phi_{43} - \Phi_{17}\Phi_{21}\Phi_{32} - \Phi_{33}\Phi_{34}\Phi_{46} - \Phi_{20}\Phi_{24}\Phi_{35} - \Phi_{25}\Phi_{36}\Phi_{37} \nonumber \\
&- \Phi_{15}\Phi_{26}\Phi_{23} - \Phi_{27}\Phi_{28}\Phi_{40})
\end{align}
and quiver:
\begin{center}
\includegraphics[width=15cm]{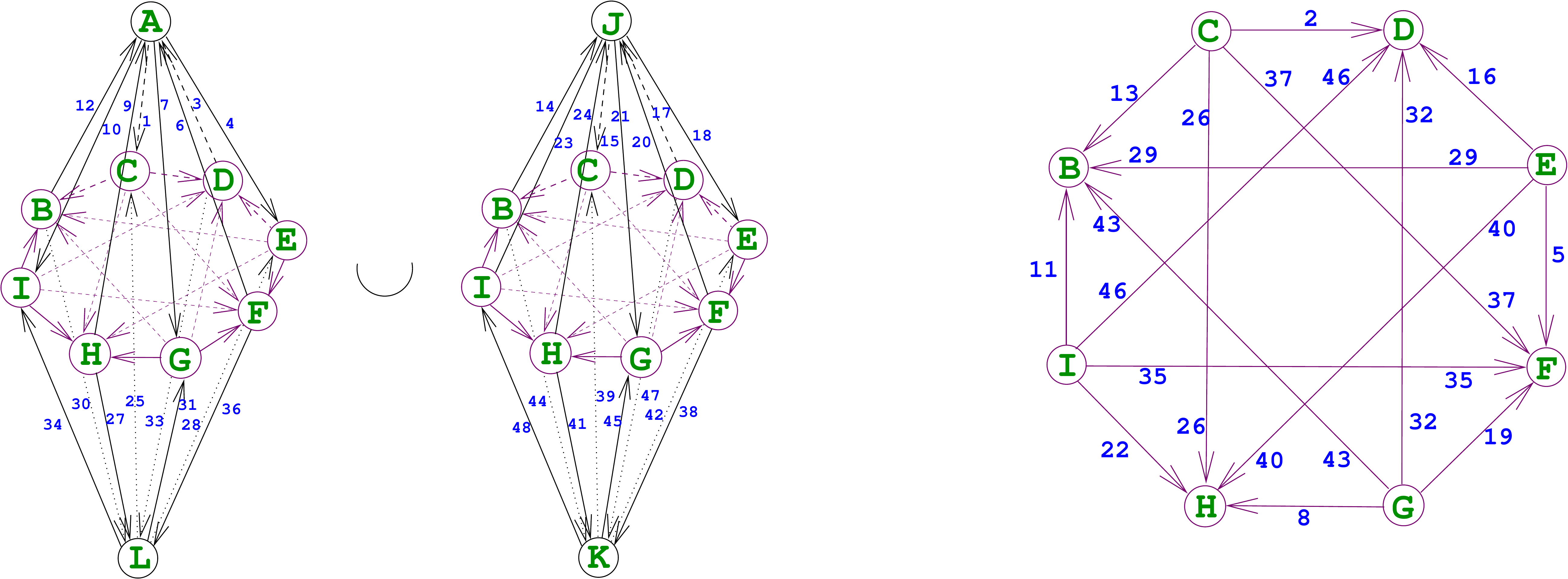}
\end{center}

We have drawn the tesselation and quiver in this way to emphasise the similarities of this tiling with the $g = 2$, $\{p,q\} = \{8,3\}$ case.

We have 8960 gauge invariants of degrees $d \in \{3,6,9,12\}$. \\
To find how many there are, we fix a degree $d$ and count the number $N_d$ of gauge invariants $r_j$ of degree $d$ that start at quiver node A. We then multiply $N_d$ by 12 (the number of quiver nodes) and divide by $d$ to obtain the number of gauge invariants of degree $d$. Note we divide by $d$ as two loops $F_1F_2\dots F_d$, $F'_1F'_2\dots F'_d$ in the quiver give the same gauge invariant if and only if they are related by a cyclic permutation, such as ABCDE and CDEAB.

\begin{itemize}
\item $d = 3$: Example: ACB(A) \\
Starting at A, we have 4 choices to leave A (e.g. to C). We then have 4 choices from the next node (e.g. B) to then have 1 choice to return to A. So we have $N_3 = 16$.\\
This gives a contribution of $16\cdot \frac{12}{3} = 64$ gauge invariants.
\item $d=6$: Example AEDLCB(A) \\
Starting at A, we have 4 choices to leave A (e.g. to E). We then have 4 choices from the next node (e.g. to D). We can then choose from 3 choices (avoiding A), to e.g. L. From there we have 3 choices (avoiding E), to e.g. C. Then we have 3 choices (avoiding D), to e.g. B, after which we have one way to return to A. So we have $N_6 = 4^2\cdot 3^3 = 432$.\\
This gives a contribution of $432\cdot \frac{12}{6} = 864$ gauge invariants.
\item $d=9$: Example AGFJEDLCB(A) \\
We continue the process set out in $d=6$, avoiding any nodes we have already passed, to get $N_9 = 4^2\cdot 3^3 \cdot 2^3 = 3456$ \\
This gives a contribution of $3456\cdot \frac{12}{8} = 5184$ gauge invariants.
\item $d=12$: Example AIHKGFJEDLCB(A) \\
We get that $N_{12} = N_9 = 3456$. \\
This gives a contribution of $3456\cdot \frac{12}{12} = 3456$ gauge invariants.
\end{itemize}

We can also write down a permutation triple such that $\sigma_B \sigma_W \sigma_{\infty} = id$: 
\begin{align}
\sigma_B =& (1\;2\;3)(4\;5\;6)(7\;8\;9)(10\;11\;12)(13\;14\;15)(16\;17\;18)(19\;20\;21)(22\;23\;24)\nonumber \\
&(25\;26\;27)(28\;29\;30)(31\;32\;33)(34\;35\;36)(37\;38\;39)(40\;41\;42)(43\;44\;45)(46\;47\;48)\nonumber \\
\sigma_W =& (1\;12\;13)(2\;39\;47)(3\;16\;4)(5\;42\;38)(6\;19\;7)(8\;45\;41)(9\;22\;10)(11\;48\;44)\nonumber \\
&(14\;29\;18)(15\;23\;26)(17\;32\;21)(20\;35\;24)(25\;37\;36)(27\;40\;28)(30\;43\;31)(33\;46\;34)\nonumber \\
\sigma_{\infty} =& (1\;15\;25\;39)(2\;46\;32\;16)(3\;6\;9\;12)(4\;18\;28\;42)(5\;37\;35\;19)(7\;21\;31\;45)\nonumber \\
&(8\;40\;26\;22)(10\;24\;34\;48)(11\;43\;29\;13)(14\;17\;20\;23)(27\;30\;33\;36)(38\;41\;44\;47) \ .
\end{align}
So we note we have ramification structure $\{3^{16},3^{16},4^{12}\}$.

This concludes our treatment of the regular cases up to genus 3. We now move onto the more intricate semi-regular cases.


\newpage 

\section{Semi-Regular Tessellations} \label{sec:semiregulartilings}\setall

\subsection{Results} \label{sec:SemiRegTilingResults}

As in section \ref{sec:RegTilingResults}, the table below shows a summary of the results, with more detailed results in the relevant subsections.
Note that with the exception of $g=0$ and $g=1$, this is not a complete classification.

\begin{longtable}{cc|c|rl}
$g$			& 	$\mathbf{x}$			& $(dim$,$deg)_{\mathcal{F}^{\flat}}$		&						& Hilbert series  \\
				&											&																				&						& \\
				&											&	$(dim$,$deg)_{\cM_{mes}}$							&						& \\
\cline{1-5}
$0$ 		& 	$ (6,6,4)$  		 	& $(13$,$38200)$ 	& $\mathcal{F}^{\flat}:$ 					& $(1-t)^{-13}(1+23t+240t^2+1484t^3+5923t^4 $\\
				&											&									&																	& $+15381t^5+24218t^6 +17689t^7 - 4145t^8 $ \\
				&											&									&																	& $- 14970t^9- 6947t^{10}-604t^{11} - 53t^{12} - 40t^{13})$  \\
				&											&									& $^{\text{Irr}}\mathcal{F}^{\flat}:$ 		& $n/a$ \\
				& 							 			& $n/a$						&	$ \cM_{mes}:$ 									& $n/a$  \\
\cline{1-5}
$1$ 		&	  $(8,8,4)$  				& $(6$,$14)$			&$\mathcal{F}^{\flat}:$ 					& $(1-t)^{-6}(1 + 6t +9t^2 -5t^3 +3t^4)$  \\
				&											&									&$^{\text{Irr}}\mathcal{F}^{\flat}:$ 		& $(1-t)^{-6}(1+6t+6t^2+t^3)$ \\
				&											& $(3$,$7)$				&$ \cM_{mes}:$										& $(1-t)^{-3}(1+5t^3+t^6)$  \\
				\cline{2-5}
				&	  $(12,6,4)$  			& $(8$,$92)$			&$\mathcal{F}^{\flat}:$ 					& $(1-t)^{-8}(1+10t + 37t^2+47t^3$ \\
				&											&									&																	& $ -15t^4+7t^5+5t^6)$  \\
				&											&									&$^{\text{Irr}}\mathcal{F}^{\flat}:$ 		& $n/a$ \\
				&											& $(3$,$6)$				&$ \cM_{mes}:$										& $(1-t)^{-3}(1+4t^3+t^6)$  \\
\cline{1-5}
$2$			&   $(12,12,12,4)$  	& $(6$,$4)$				&$\mathcal{F}^{\flat}:$ 					& $(1-t)^{-6}(1+2t+3t^2-4t^3+2t^4)$  \\
				&											&									&$^{\text{Irr}}\mathcal{F}^{\flat}:$ 		& $(1-t)^{-6}(1+2t+t^2)$\\
				&											&	$(5$,$10)$			&$ \cM_{mes}:$ 										& $(1-t^2)^{-3}(1-t)^{-2}(1+2t+4t^2+2t^3+t^4) $ \\
				\cline{2-5}
				&   $(8,4,8,8,4)$   	& $(8$,$6)$				&$\mathcal{F}^{\flat}:$ 					& $(1-t)^{-8}(1 + 2t + 3t^4 + 4t^4 - 5t^4 - 3t^5 + 4t^6)$  \\
				&											&									&$^{\text{Irr}}\mathcal{F}^{\flat}:$ 		& $(1-t)^{-8}(1+2t+2t^2+t^3)$\\
				&											&	$(5$,$197)$			&$ \cM_{mes}:$ 										& $(1-t^3)^4(1-t^2)^4(1 + 4t^3 - 13t^5 + 2t^6 + 8t^7$ \\
				&											&									&																	& $ - 8t^8 - 2t^9 + 13t^{10} - 4t^{12} - t^{15})  $ \\
\cline{1-5}
$3$			&   $(18,18,18,6)$   	& $(8$,$16)$			&$\mathcal{F}^{\flat}:$ 					& $(1-t)^{-8}(1 + 4t + 10t^2 + 8t^3 - 6t^4 $ \\
				&											&									&																	& $- 20t^5 + 28t^6 - 12t^7 + 3t^8)$  \\
				&											&									&$^{\text{Irr}}\mathcal{F}^{\flat}:$ 		& $(1-t)^{-8}(1+4t+6t^2+4t^3+t^4)$\\
				&											&	$(7$,$118)$			&$ \cM_{mes}:$ 										& $(1-t^2)^{-5}(1-t)^{-2}(1 + 4t + 14t^2 + 30t^3 + 41t^4 $ \\
				&											&									&																	& $+ 18t^5 - 10t^6 - 2t^7 + 25t^8 + 6t^9 -12t^{10} + 3t^{12}) $ \\
				\cline{2-5}
				&  $(12,4,12,4,12,4)$	& $(10$,$9)$			&$\mathcal{F}^{\flat}:$ 					& $(1-t)^{-10}(1 + 2t + 3t^2 + 4t^3 + 5t^4 $ \\
				&											&									&																	& $ -6t^5 - 8t^6 + 8t^7)$  \\
				&											&									&$^{\text{Irr}}\mathcal{F}^{\flat}:$ 		& $(1-t)^{-10}(1+ 2t + 3t^2 + 2t^3 + t^4)$\\
				&											&	$(7$,$24)$			&$ \cM_{mes}:$ 										& $(1-t^2)^{-7}(1 + 5t^2 + 12t^4 + 5t^6 + t^8)$
\end{longtable}

\subsubsection{Genus 0}

\paragraph*{\fbox{$\mathbf{x = (6,6,4)}$}}

This is a tiling with $V=24,E=36,F=14$.

\begin{center}
\includegraphics[width=15cm]{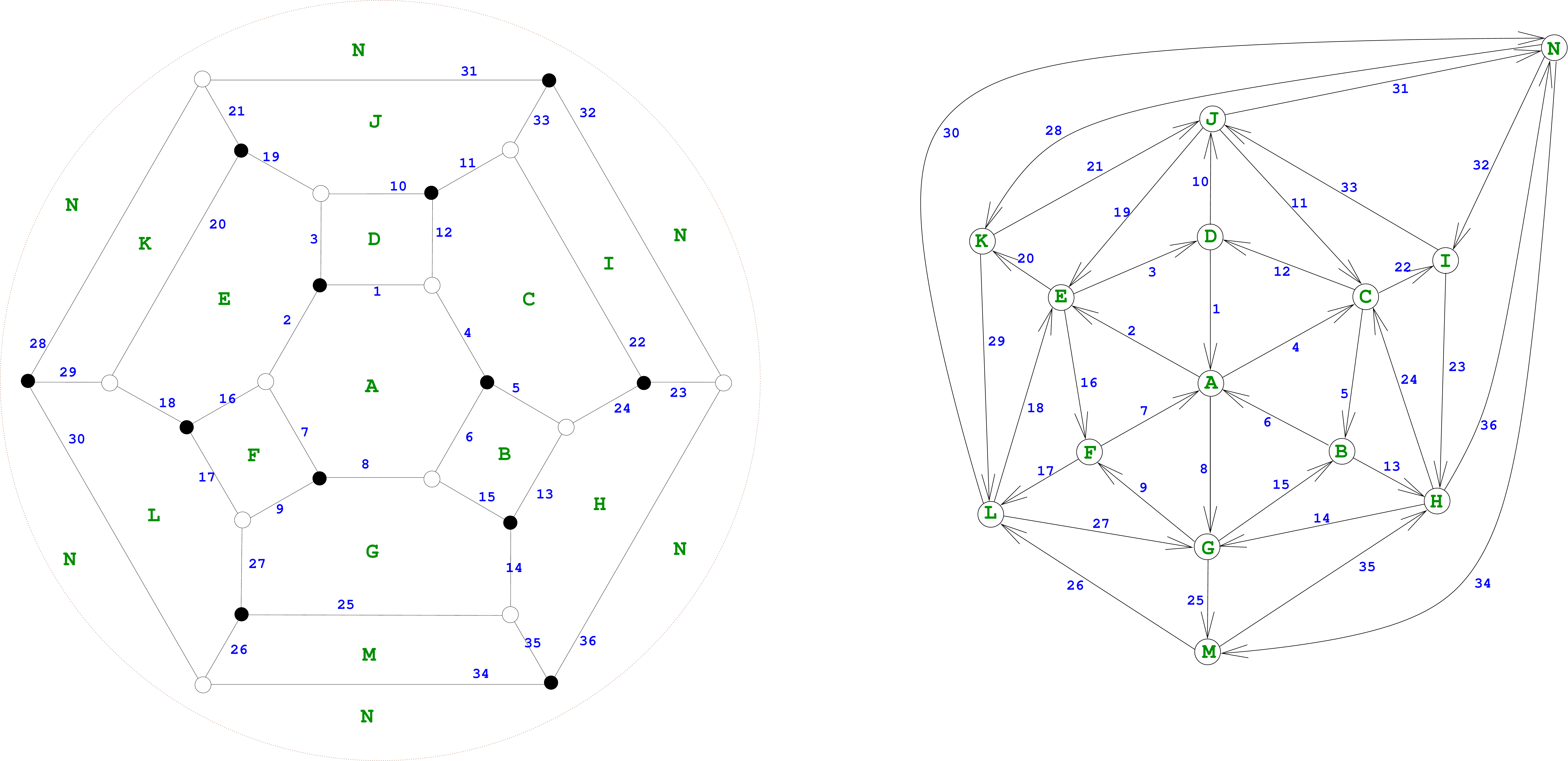}
\end{center}

We have superpotential
\begin{align}
W = \tr (&\Phi_1\Phi_2\Phi_3 + \Phi_4\Phi_5\Phi_6 + \Phi_7\Phi_8\Phi_9 + \Phi_{10}\Phi_{11}\Phi_{12} + \Phi_{13}\Phi_{14}\Phi_{15} + \Phi_{16}\Phi_{17}\Phi_{18} \nonumber \\
&+ \Phi_{19}\Phi_{20}\Phi_{21} + \Phi_{22}\Phi_{23}\Phi_{24} + \Phi_{25}\Phi_{26}\Phi_{27} + \Phi_{28}\Phi_{29}\Phi_{30} + \Phi_{31}\Phi_{32}\Phi_{33} + \Phi_{34}\Phi_{35}\Phi_{36} \nonumber \\
& - \Phi_1\Phi_4\Phi_{12} - \Phi_2\Phi_{16}\Phi_7 - \Phi_3\Phi_{10}\Phi_{19} - \Phi_5\Phi_{13}\Phi_{24} - \Phi_6\Phi_8\Phi_{15} - \Phi_9\Phi_{17}\Phi_{27} \nonumber \\
& - \Phi_{11}\Phi_{22}\Phi_{33} - \Phi_{14}\Phi_{25}\Phi_{35} - \Phi_{18}\Phi_{20}\Phi_{29} - \Phi_{21}\Phi_{31}\Phi_{28} - \Phi_{23}\Phi_{36}\Phi_{32} - \Phi_{26}\Phi_{30}\Phi_{34}) \ .
\end{align}
To find the gauge invariants, we simply note that the monomial terms in the superpotential are exactly the 24 gauge invariants.

To find the master space, we define the ring $S = \mathbb{C}[\Phi_1,\dots,\Phi_{36}]$ and ideal $I_1 = \left< \partial_i W \right>_{i=1,\dots,36}$ to generate the master space $R = S/I_1$. Using \cite{mac2}, we find that $I_1$ has dimension 13, degree 38200 and Hilbert series
\begin{align}
H(t, \cF^{\flat}_{g=0, \ x=(6,6,4,)}) =
(1-t)^{-13}&(1+23t+240t^2+1484t^3+5923t^4+15381t^5+24218t^6 \nonumber \\
&+17689t^7 - 4145t^8 - 14970t^9 - 6947t^{10}-604t^{11} - 53t^{12} - 40t^{13}) \ .
\end{align}

We can also write down a permutation triple such that $\sigma_B \sigma_W \sigma_{\infty} = id$: 
\begin{align}
\sigma_B =& (1\;2\;3)(4\;5\;6)(7\;8\;9)(10\;11\;12)(13\;14\;15)(16\;17\;18)(19\;20\;21)(22\;23\;24)\nonumber \\
&(25\;26\;27)(28\;29\;30)(31\;32\;33)(34\;35\;36)\nonumber \\
\sigma_W =& (1\;12\;4)(2\;7\;16)(3\;19\;10)(5\;24\;13)(6\;15\;8)(9\;27\;17)(11\;33\;22)(14\;35\;25) \nonumber \\
& (18\;29\;20)(21\;28\;31)(23\;32\;36)(26\;34\;30) \nonumber \\
\sigma_{\infty} =& (1\;6\;7)(2\;18\;19)(3\;12)(4\;11\;24)(5\;15)(8\;14\;27)(9\;16)(10\;21\;33)(13\;23\;35) \nonumber\\
& (17\;26\;29)(20\;28)(22\;32)(25\;34)(30\;36\;31) \ .
\end{align}
Hence we have ramification structure $\{3^{12}, 3^{12}, 2^7 \ 3^7\}$.

\subsubsection{Genus 1}

\paragraph*{\fbox{$\mathbf{ x=(8,8,4)}$}}

This is the tiling with $V=8,E=12,F=4$.

\begin{center}
\includegraphics[width=15cm]{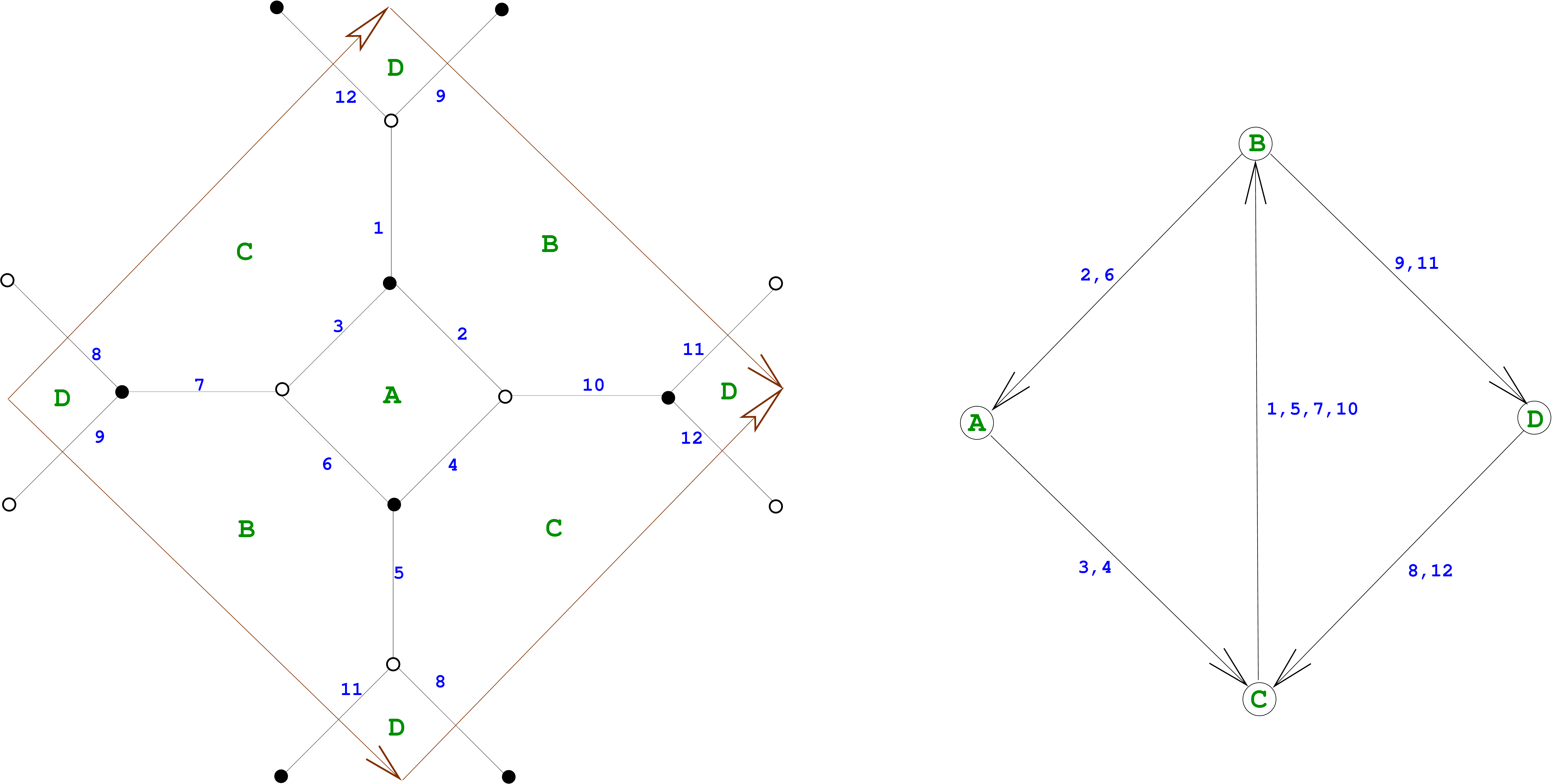}
\end{center}
We can recognise this dimer as the previously studied Hirzebruch Phase II, sometimes denoted $\mathbf{F_0}$ \cite{Franco:2005sm}.

We have superpotential $W$ and 32 gauge invariants $r$ as:
\begin{align}
\nonumber
W = \tr (& \Phi_1\Phi_2\Phi_3 + \Phi_4\Phi_5\Phi_6 + \Phi_7\Phi_8\Phi_9 + \Phi_{10}\Phi_{11}\Phi_{12} \nonumber \\
&- \Phi_1\Phi_9\Phi_{12} - \Phi_2\Phi_4\Phi_{10} - \Phi_3\Phi_7\Phi_6 - \Phi_5\Phi_{11}\Phi_8);
\\ \nonumber
&
r_{i,j,k} = \Phi_i\Phi_j\Phi_k \text{ with } \begin{cases} i \in \{ 1,5,7,10 \}, &j \in \{3,4\},\; k\in \{2,6\} \\
																													i \in \{ 1,5,7,10 \}, &j \in \{8,12\},\; k\in \{9,11\} \ .
																													\end{cases}
\end{align}

To find the master space, we define the ring $S = \mathbb{C}[\Phi_1,\dots,\Phi_{12}]$ and ideal $I_1 = \left< \partial_i W \right>_{i=1,\dots,12}$ to generate the master space $R = S/I_1$. Using \cite{mac2}, we find that $I_1$ has dimension 6, degree 14 and Hilbert series
\begin{equation}
H(t, \cF^{\flat}_{g=1, \ x=(8,8,4)}) =
\frac{1 + 6t +9t^2 -5t^3 +3t^4}{(1-t)^6} \ .
\end{equation}

Using primary decomposition in \cite{mac2}, we get that the curve given by $I_1$ is the union of those given by ideals of which
2 are trivial of dimension 4, degree 1 and have Hilbert series $\frac{1}{(1-t)^4}$.
The \emph{coherent component} is of dimension 6, degree 14 and has Hilbert series 
\begin{equation} 
H(t, \ ^{\text{Irr}}\cF^{\flat}_{g=1, \ x=(8,8,4)}) = \frac{1+6t+6t^2+t^3}{(1-t)^6} \ .
\end{equation}

To find the vacuum moduli space, we consider the ring $R = \mathbb{C} \left[ \Phi_1,\dots\Phi_{12},y_1,\dots,y_{32} \right]$ and ideal $I_2 = \left< \partial_i W,y_j - r_j \right>_{i=1,\dots,12;j=1,\dots,32}$.
We then eliminate all the $\Phi$s and substitute the resulting ideal into ring $R' = \mathbb{C}\left[ y_1,\dots,y_{32}\right]$ to get ideal $V$ representing the vacuum moduli space. Using \cite{mac2}, we see that V has dimension 3, degree $7$ and after assigning weights to each $y_j$ equal to the degree of the monomial they represent, we get Hilbert series
\begin{equation}
H(t, \cM_{g=1, \ x=(8,8,4)}) = \frac{1+5t^3+t^6}{(1-t^3)^3} \ ,
\end{equation}
its palindromic numerator indicating it is a Calabi-Yau 3-fold. However, we can check that neither $cM$ nor $^{\text{Irr}}\cF^{\flat}$ is a complete intersection.

We can also write down a permutation triple such that $\sigma_B \sigma_W \sigma_{\infty} = id$: 
\begin{align}
\nonumber
\sigma_B &= (1\;2\;3)(4\;5\;6)(7\;9\;8)(10\;11\;12)\\
\nonumber
\sigma_W &= (1\;12\;9)(2\;10\;4)(3\;6\;7)(5\;8\;11)\\
\sigma_{\infty} &= (1\;7\;5\;10)(3\;8\;4\;12)(2\;6)(9\;11) \ .
\end{align}
So we note we have ramification structure $\{3^4,3^4,4^2\ 2^2\}$.

\paragraph*{\fbox{$\mathbf{ x = (12,6,4)}$}}

This is the tiling with $V=12,E=18,F=6$.

\begin{center}
\includegraphics[width=12cm]{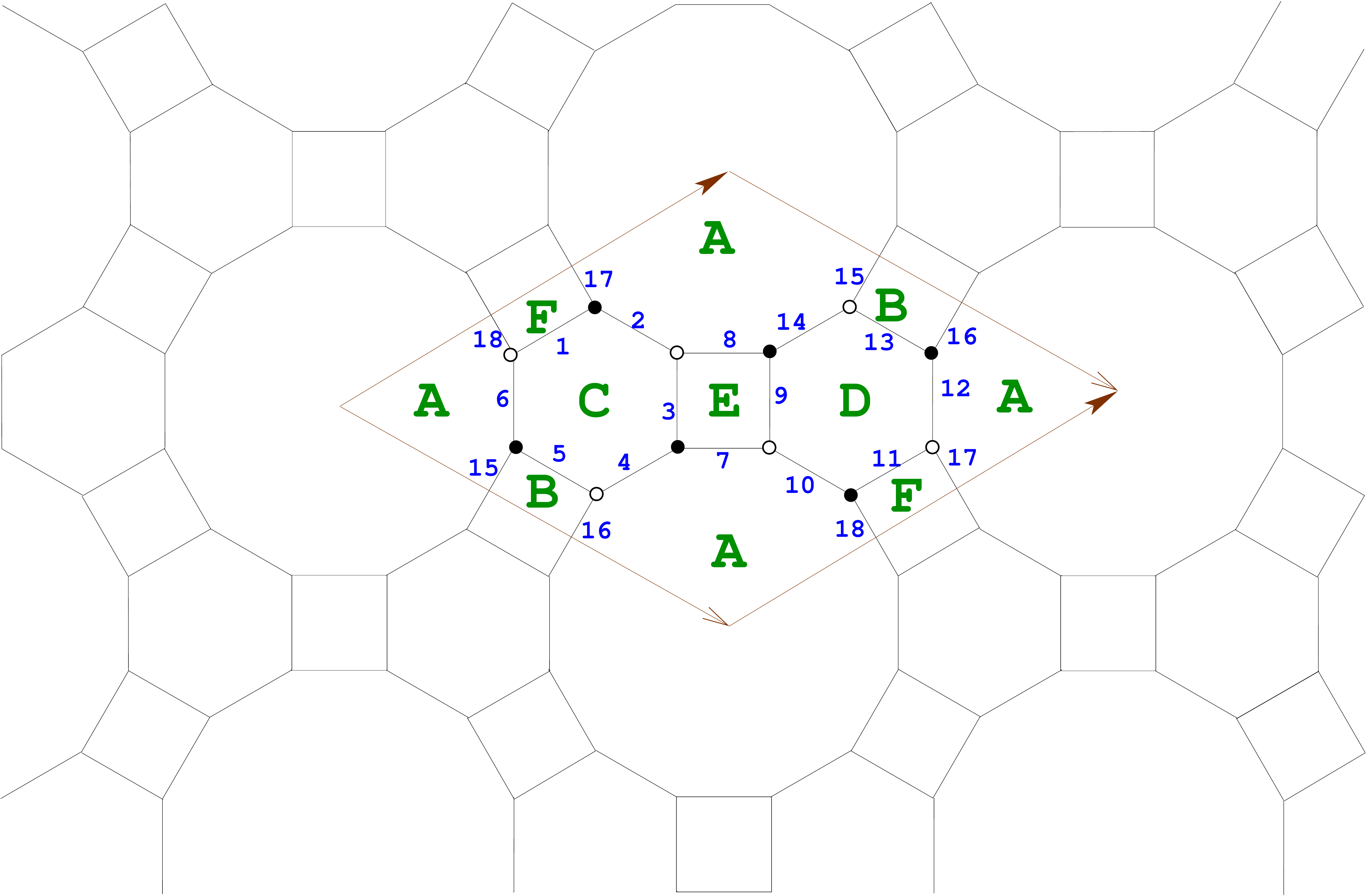}
\end{center}

We have superpotential 
\begin{align} 
W = \tr ( &\Phi_1\Phi_{17}\Phi_{2} + \Phi_{5}\Phi_{15}\Phi_{6} + \Phi_{3}\Phi_{7}\Phi_{4} + \Phi_{8}\Phi_{14}\Phi_{9} + \Phi_{10}\Phi_{11}\Phi_{18} + \Phi_{12}\Phi_{13}\Phi_{16} \nonumber \\
& - \Phi_{1}\Phi_{18}\Phi_{6}- \Phi_{2}\Phi_{3}\Phi_{8}- \Phi_{4}\Phi_{5}\Phi_{16}- \Phi_{7}\Phi_{10}\Phi_{9}- \Phi_{11}\Phi_{17}\Phi_{12}- \Phi_{13}\Phi_{15}\Phi_{14})
\end{align}
and quiver:
\begin{center}
\includegraphics[width=8cm]{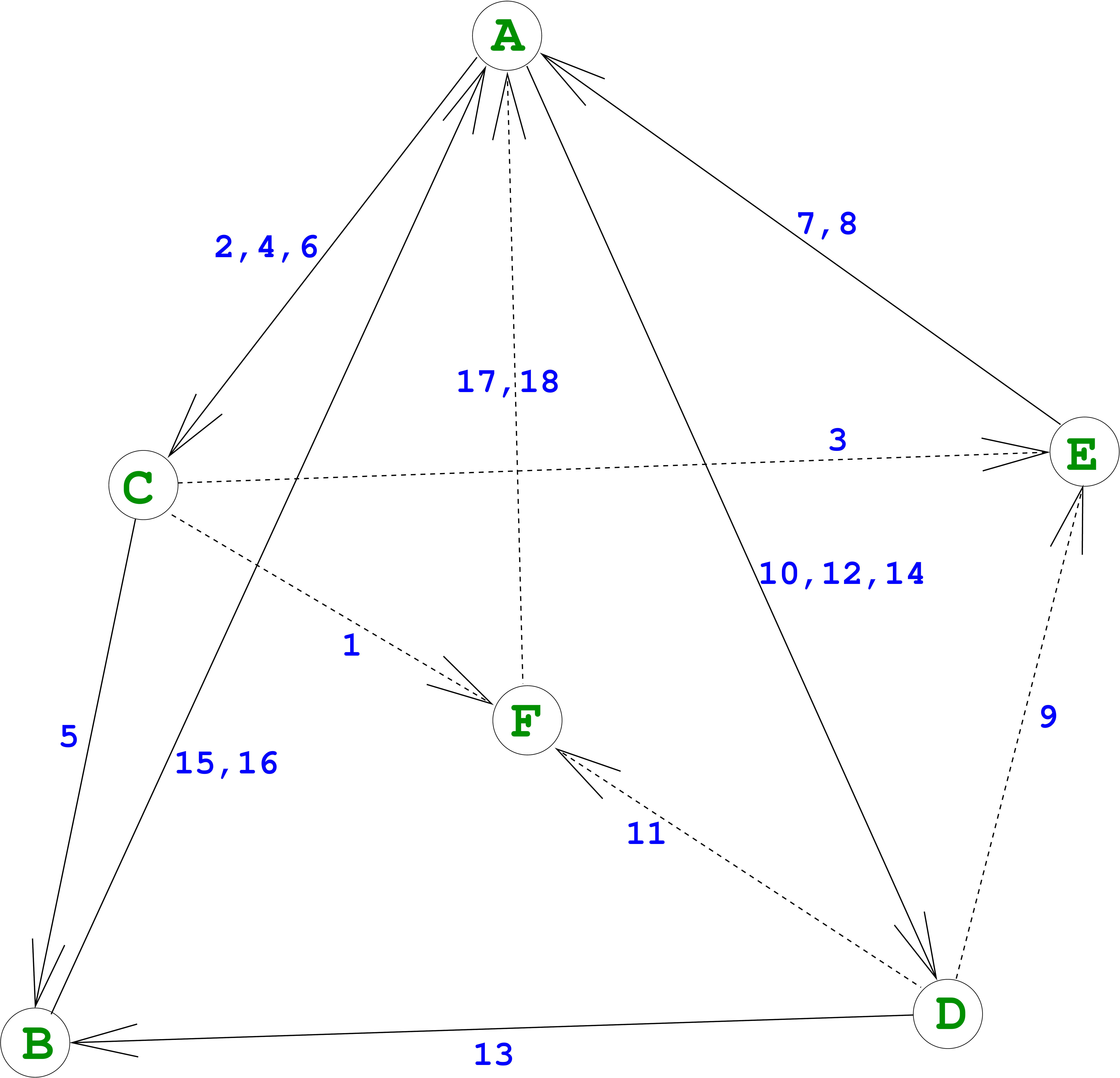}
\end{center}

We note we have 36 gauge invariants of degree 3.

Now if we relabel as follows:
\begin{align}
\Phi_{1} &\rightarrow q_1'  &\Phi_{7}  & \rightarrow p_1'  & \Phi_{13} & \rightarrow X_3   \nonumber \\
\Phi_{2} &\rightarrow W_2   &\Phi_{8}  & \rightarrow p_2'  & \Phi_{14} & \rightarrow W_2   \nonumber \\
\Phi_{3} &\rightarrow p_1   &\Phi_{9}  & \rightarrow p_2   & \Phi_{15} & \rightarrow M_2   \nonumber \\
\Phi_{4} &\rightarrow W_1   &\Phi_{10} & \rightarrow W_1'  & \Phi_{16} & \rightarrow X_{13}\nonumber \\
\Phi_{5} &\rightarrow q_2'  &\Phi_{11} & \rightarrow X_{11}& \Phi_{17} & \rightarrow X_9   \nonumber \\
\Phi_{6} &\rightarrow q_1   &\Phi_{12} & \rightarrow X_8   & \Phi_{18} & \rightarrow M_1 \ .
\end{align}
we get a labelling that gives us the same quiver as phase IV of the del Pezzo surface $dP_3$ in \cite{Feng:2001bn}. 

We also get superpotential:
\begin{align} 
W &= X_3X_{13}X_8 - X_8X_{11}X_9 - W_1q_2'X_{13} - M_2W_2'X_{13} + q_1'X_9W_2 + M_1W_1'X_{11} \nonumber \\
   & - M_1q_1q_1' + M_2q_1q_2' + W_1p_1p_1' - W_2p_1p_2' -W_1'p_2p_1' + W_2'p_2p_2' 
\end{align}
which differs from the superpotential in \cite{Feng:2001bn} in the following terms:
\begin{align}
X_3X_{13}X_8 &\leftrightarrow X_3X_8X_{13} \nonumber\\
-X_8X_{11}X_9&\leftrightarrow -X_8X_9X_{11} \nonumber\\
-M_2W_2'X_{3} &\leftrightarrow -M_2X_3W_2'
\end{align}
Noting that in the case that the $\Phi_i \in \mathbb{C}$ these terms are the same, we see that the master space and vacuum moduli space are the same as phase IV of $dP_3$.

To find the master space, we define the ring $S = \mathbb{C}[\Phi_1,\dots,\Phi_{18}]$ and ideal $I_1 = \left< \partial_i W \right>_{i=1,\dots,18}$ to generate the master space $R = S /I_1$. We find using \cite{mac2} that $I_1$ has dimension 8, degree 92 and Hilbert series
\begin{equation} 
H(t, \cF^{\flat}_{g=1, \ x=(12,6,4)}) = \frac{1+10t + 37t^2+47t^3-15t^4+7t^5+5t^6}{(1-t)^8} \ .
\end{equation}

To find the vacuum moduli space, we consider the ring $R = \mathbb{C} \left[ \Phi_1,\dots\Phi_{18},y_1,\dots,y_{36} \right]$ and ideal $I_2 = \left< \partial_i W,y_j - r_j \right>_{i=1,\dots,18;j=1,\dots,36}$.
We then eliminate all the $\Phi$s and substitute the resulting ideal into ring $R' = \mathbb{C}\left[ y_1,\dots,y_{36}\right]$ to get ideal $V$ representing the vacuum moduli space. Using \cite{mac2}, we see that V has dimension 3, degree 6 and after assigning weights to each $y_j$ equal to the degree of the monomial they represent, we get Hilbert series
\begin{equation} 
H(t, \cM_{g=1, \ x=(12,6,4)}) = \frac{1+4t^3+t^6}{(1-t^3)^3} \ ,
\end{equation}
its palindromic numerator indicating it is a Calabi-Yau 3-fold.

It is nice to see a familiar theory such as the third del Pezzo emerge as one of the semi-regular tessellations.

\subsubsection{Genus 2}

\paragraph*{\fbox{$\mathbf{ x = (12,12,12,4) }$}}
This is a tiling with $V=4,E=8,F=2$.

\begin{center}
\includegraphics[width=15cm]{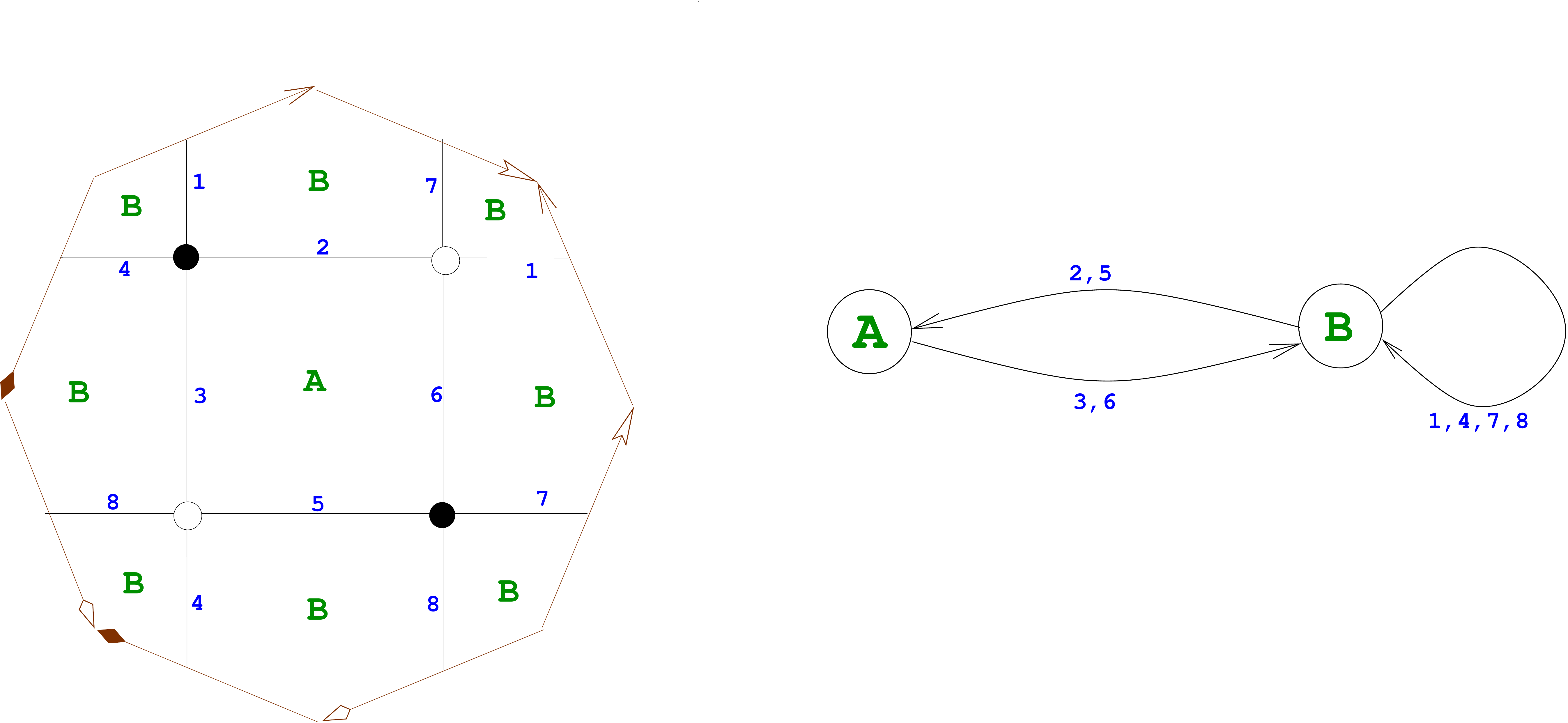}
\end{center}

We have superpotential 
\begin{equation} 
W = \tr (\Phi_1\Phi_2\Phi_3\Phi_4 + \Phi_5\Phi_6\Phi_7\Phi_8 - \Phi_1\Phi_7\Phi_2\Phi_6 - \Phi_3\Phi_8 \Phi_5 \Phi_4)
\end{equation}
and 8 gauge invariants:
\begin{align}
\nonumber
r_1 = \Phi_1 \ , \quad
r_2 = \Phi_4 \ , \quad
r_3 = \Phi_7 \ , \quad
r_4 = \Phi_8 \ , \quad
\\
r_5 = \Phi_2\Phi_3 \ , \quad
r_6 = \Phi_2\Phi_6 \ , \quad
r_7 =\Phi_5\Phi_3 \ , \quad
r_8 = \Phi_5\Phi_6 \ .
\end{align}
To find the master space, we define the ring $S = \mathbb{C}[\Phi_1,\dots,\Phi_8]$ and ideal $I_1 = \left< \partial_i W \right>_{i=1,\dots,8}$ to generate the master space $R = S /I_1$. We find using \cite{mac2} that $I_1$ has dimension 6, degree 4 and Hilbert series
\begin{equation} 
H(t, \cF^{\flat}_{g=2, \ x=(12,12,12,4)}) = \frac{1+2t+3t^2-4t^3+2t^4}{(1-t)^6} \ .
\end{equation}

Using primary decomposition in \cite{mac2}, we get that the curve given by $I_1$ is the union of those given by ideals of which
2 are trivial of dimension 4, degree 1 and have Hilbert series $\frac{1}{(1-t)^4}$.
The \emph{coherent component} is of dimension 6, degree 4 and has Hilbert series 
\begin{equation} 
H(t, \ ^{\text{Irr}}\cF^{\flat}_{g=2, \ x=(12,12,12,4)}) =  \frac{1+2t+t^2}{(1-t)^6} \ .
\end{equation}

To find the vacuum moduli space, we consider the ring $R = \mathbb{C} \left[ \Phi_1,\dots\Phi_8,y_1,\dots,y_8 \right]$ and ideal $I_2 = \left< \partial_i W,y_j - r_j \right>_{i=1,\dots,8;j=1,\dots,8}$.
We then eliminate all the $\Phi$s and substitute the resulting ideal into ring $R' = \mathbb{C}\left[ y_1,\dots,y_8\right]$ to get ideal $V$ representing the vacuum moduli space. Using \cite{mac2}, we see that V has dimension 5, degree 10 and after assigning weights to each $y_j$ equal to the degree of the monomial they represent, we get Hilbert series
\begin{equation} 
H(t, \cM_{g=2, \ x=(12,12,12,4)}) = \frac{1+2t+4t^2+2t^3+t^4}{(1-t^2)^3(1-t)^2} \ ,
\end{equation}
its palindromic numerator indicating it is a Calabi-Yau 5-fold.

We can also write down a permutation triple such that $\sigma_B \sigma_W \sigma_{\infty} = id$: 
\begin{align}
\nonumber
\sigma_B &= (1\;2\;3\;4) (5\;6\;7\;8)\\
\nonumber
\sigma_W &= (1\;6\;2\;7) (3\;5\;4\;8)\\
\sigma_{\infty} &= (1\;5\;4\;8\;2\;7)(3\;6) \ .
\end{align}
So we note we have ramification structure $\{4^2,4^2,2\ 6\}$.

\paragraph*{\fbox{$\mathbf{ x = (8,4,8,4,4) }$}}
This is a tiling with $V=4, E = 10, F=4$.

\begin{center}
\includegraphics[width=15cm]{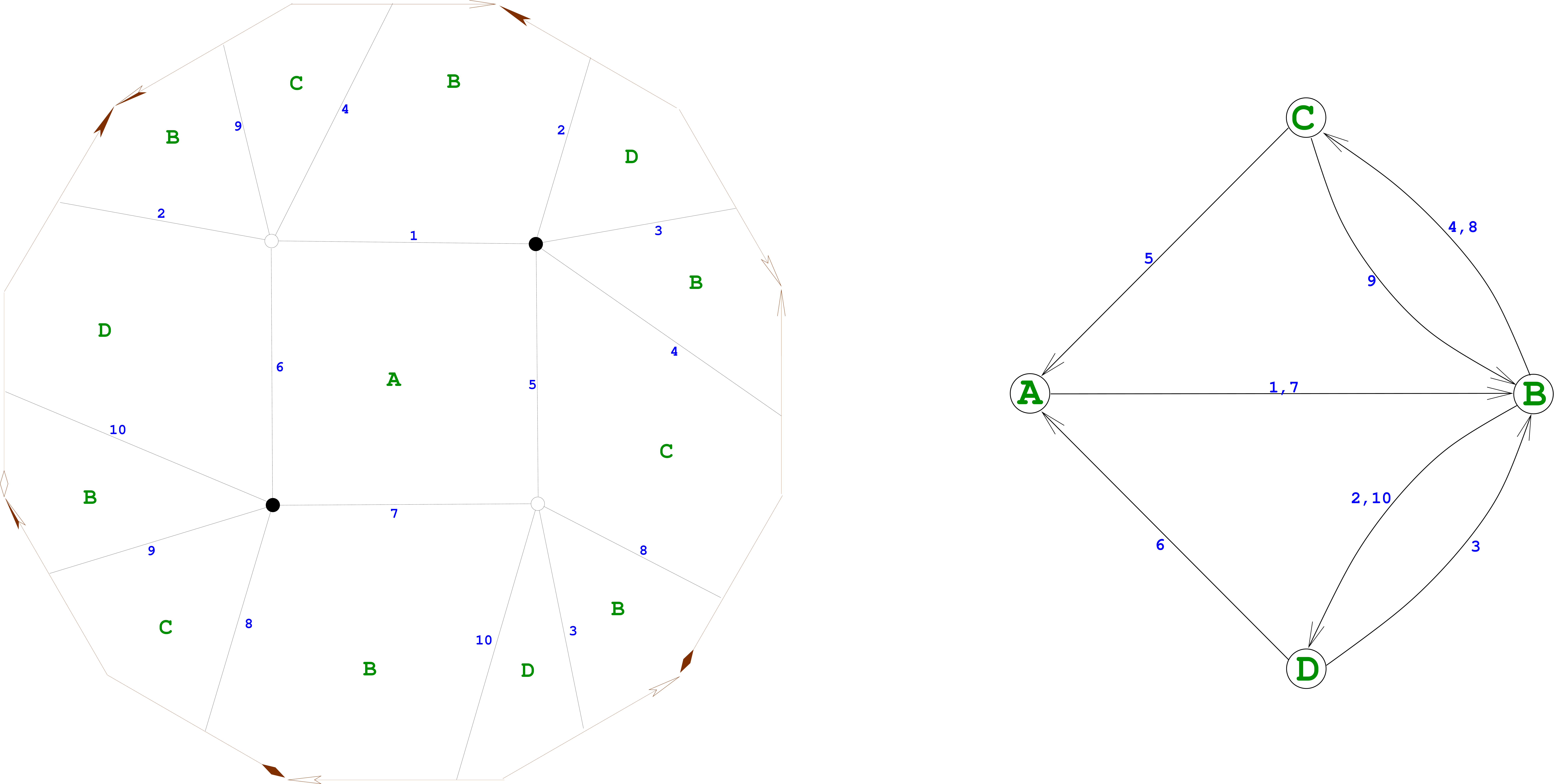}
\end{center}

We have superpotential 
\begin{equation} 
W = \tr (\Phi_1\Phi_2\Phi_3\Phi_4\Phi_5 + \Phi_6\Phi_7\Phi_8\Phi_9\Phi_{10} - \Phi_1\Phi_4\Phi_9\Phi_2\Phi_6 - \Phi_3\Phi_8\Phi_5\Phi_7\Phi_{10}) 
\end{equation}
and 12 gauge invariants:
\begin{align}
\nonumber
&
r_1 = \Phi_1\Phi_4\Phi_5 \ , \quad
r_2 = \Phi_1\Phi_8\Phi_5 \ , \quad
r_3 = \Phi_7\Phi_4\Phi_5 \ , \quad
r_4 = \Phi_7\Phi_8\Phi_5 \ , \quad
\\ \nonumber &
r_5 = \Phi_1\Phi_2\Phi_6 \ , \quad
r_6 = \Phi_1\Phi_{10}\Phi_6 \ , \quad
r_7 = \Phi_7\Phi_2\Phi_6 \ , \quad 
r_8 = \Phi_7\Phi_{10}\Phi_6 \ , \quad 
\\ \nonumber &
r_9 = \Phi_4\Phi_9 \ , \quad
r_{10} = \Phi_8\Phi_9 \ , \quad
r_{11} = \Phi_2\Phi_3 \ , \quad 
r_{12} = \Phi_{10}\Phi_3 \ .
\end{align}
To find the master space, we define the ring $S = \mathbb{C}[\Phi_1,\dots,\Phi_{10}]$ and ideal $I_1 = \left< \partial_i W \right>_{i=1,\dots,10}$ to generate the master space $R = S /I_1$. We find using \cite{mac2} that $I_1$ has dimension 8, degree 6 and Hilbert series
\begin{equation} 
H(t, \cF^{\flat}_{g=2, \ x=(8,4,8,4,4)}) = \frac{1 + 2t + 3t^4 + 4t^4 - 5t^4 - 3t^5 + 4t^6}{(1-t)^8} \ .
\end{equation}

Using primary decomposition in \cite{mac2}, we get that the curve given by $I_1$ is the union of those given by ideals of which
10 are trivial of dimension 6, degree 1 and have Hilbert series $\frac{1}{(1-t)^6}$.
The \emph{coherent component} is of dimension 8, degree 6 and has Hilbert series
\begin{equation} 
H(t, \ ^{\text{Irr}}\cF^{\flat}_{g=2, \ x=(8,4,8,4,4)}) = \frac{1+2t+2t^2+t^3}{(1-t)^8} \ .
\end{equation}

To find the vacuum moduli space, we consider the ring $R = \mathbb{C} \left[ \Phi_1,\dots\Phi_{10},y_1,\dots,y_{12} \right]$ and ideal $I_2 = \left< \partial_i W,y_j - r_j \right>_{i=1,\dots,10;j=1,\dots,12}$.
We then eliminate all the $\Phi$s and substitute the resulting ideal into ring $R' = \mathbb{C}\left[ y_1,\dots,y_{12}\right]$ to get ideal $V$ representing the vacuum moduli space. Using \cite{mac2}, we see that V has dimension 5, degree 197 and after assigning weights to each $y_j$ equal to the degree of the monomial they represent, we get Hilbert series
\begin{equation} 
H(t, \cM_{g=2, \ x=(8,4,8,4,4)}) = \frac{1 + 4t^3 - 13t^5 + 2t^6 + 8t^7 - 8t^8 - 2t^9 + 13t^{10} - 4t^{12} - t^{15}}{(1-t^3)^4(1-t^2)^4} \ .
\end{equation}
Note that while the numerator is not quite palindromic, multiplying both the numerator and denominator by a factor of $(1-t^2)$ would make it palindromic.

We can also write down a permutation triple such that $\sigma_B \sigma_W \sigma_{\infty} = id$: 
\begin{align}
\nonumber
\sigma_B =& (1\;2\;3\;4\;5)(6\;7\;8\;9\;10) \\
\nonumber
\sigma_W =& (1\;6\;2\;9\;4)(3\;10\;7\;5\;8) \\
\sigma_{\infty} =& (1\;3\;7\;9)(2\;10)(4\;8)(6\;5) \ .
\end{align}
Hence we have ramification structure $\{5^2,5^2,2^3\ 4\}$.

\subsubsection{Genus 3}

\paragraph*{\fbox{$\mathbf{ x = (18,18,18,6) }$}}

This is a tiling with $V=6, E=12, F= 2$.

\begin{center}
\includegraphics[width=12cm]{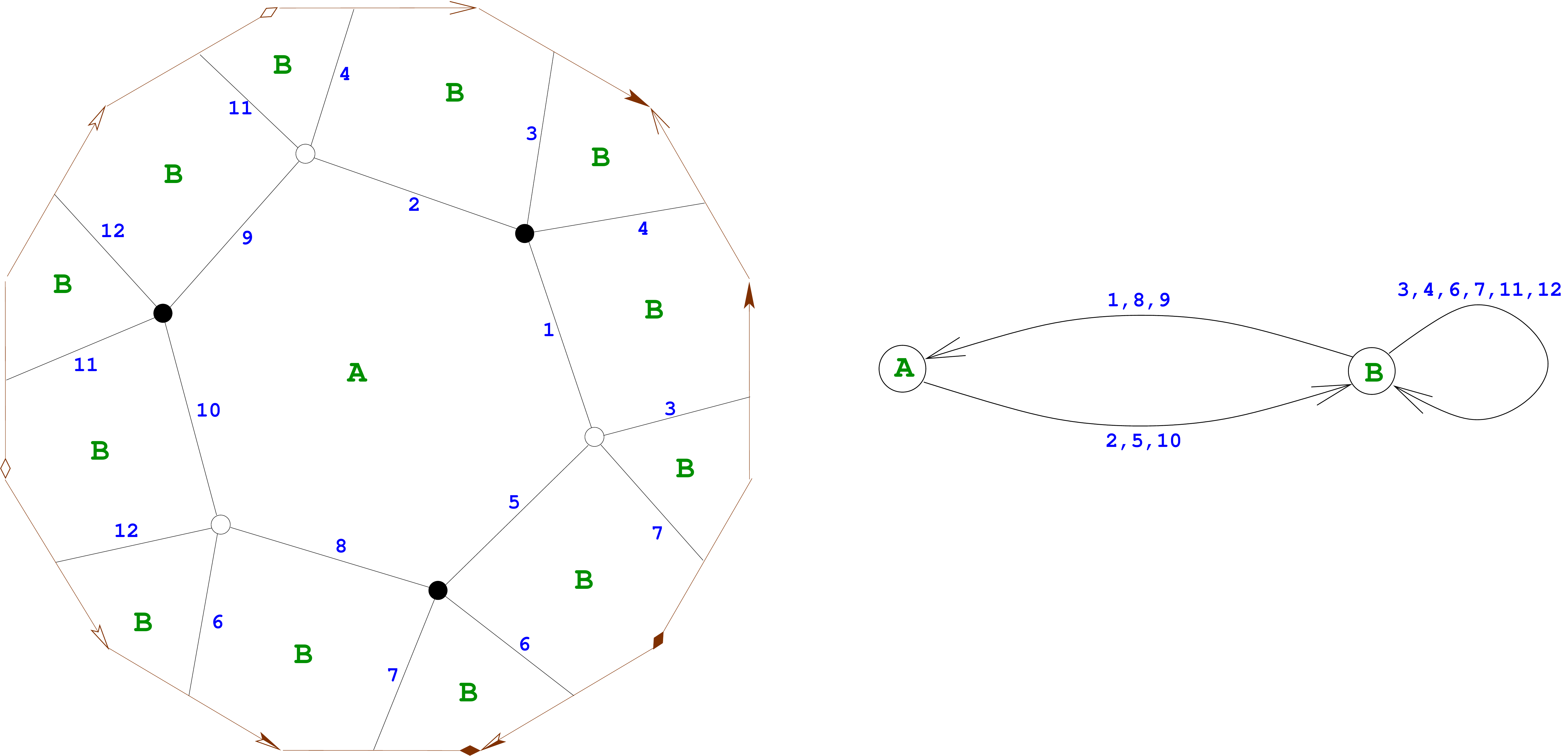}
\end{center}

We have superpotential
\begin{equation}
W = \tr (\Phi_1\Phi_2\Phi_3\Phi_4 + \Phi_5\Phi_6\Phi_7\Phi_8 + \Phi_9\Phi_{10}\Phi_{11}\Phi_{12} - \Phi_1\Phi_5\Phi_7\Phi_3 - \Phi_2\Phi_4\Phi_{11}\Phi_9 - \Phi_6\Phi_8\Phi_{10}\Phi_{12})
\end{equation}
and 15 gauge invariants:
\begin{align}
\nonumber
&
r_1 = \Phi_3\ , \quad 
r_2 = \Phi_4\ , \quad 
r_3 = \Phi_6\ , \quad 
r_4 = \Phi_7\ , \quad 
r_5 = \Phi_{11}\ , \quad 
r_6 = \Phi_{12}\ , \quad 
\nonumber \\ &
r_7 = \Phi_1\Phi_2\ , \quad  
r_8 = \Phi_1\Phi_5\ , \quad  
r_9 = \Phi_1\Phi_{10} \ , \quad  
r_{10} = \Phi_8\Phi_2 \ , \quad  
r_{11} = \Phi_8\Phi_5 \ , \quad  
r_{12} = \Phi_8\Phi_{10} \ , \quad  
\nonumber \\ &
r_{13} = \Phi_9\Phi_2 \ , \quad  
r_{14} = \Phi_9\Phi_5 \ , \quad  
r_{15} = \Phi_9\Phi_{10} \ .
\end{align}

To find the master space, we define the ring $S = \mathbb{C}[\Phi_1,\dots,\Phi_{12}]$ and ideal $I_1 = \left< \partial_i W \right>_{i=1,\dots,12}$ to generate the master space $R = S /I_1$. We find using \cite{mac2} that $I_1$ has dimension 8, degree 16 and Hilbert series
\begin{equation} 
H(t, \cF^{\flat}_{g=3, \ x=(18,18,18,6)}) = \frac{1 + 4t + 10t^2 + 8t^3 - 6t^4 - 20t^5 + 28t^6 - 12t^7 + 3t^8}{(1-t)^8} \ .
\end{equation}

Using primary decomposition in \cite{mac2}, we get that the curve given by $I_1$ is the union of those given by ideals of which:
3 are trivial of dimension 4, degree 1 and have Hilbert series $\frac{1}{(1-t)^4}$;
2 are trivial of dimension 6, degree 1 and have Hilbert series $\frac{1}{(1-t)^6}$;
12 are of dimension 6, degree 2 and have Hilbert series $\frac{1+t}{(1-t)^6}$.
The \emph{coherent component} is of dimension 8, degree 16 and has Hilbert series 
\begin{equation} 
H(t, \ ^{\text{Irr}}\cF^{\flat}_{g=3, \ x=(18,18,18,6)}) = \frac{1+4t+6t^2+4t^3+t^4}{(1-t)^8} \ .
\end{equation}

To find the vacuum moduli space, we consider the ring $R = \mathbb{C} \left[ \Phi_1,\dots\Phi_{12},y_1,\dots,y_{15} \right]$ and ideal $I_2 = \left< \partial_i W,y_j - r_j \right>_{i=1,\dots,12;j=1,\dots,15}$.
We then eliminate all the $\Phi$s and substitute the resulting ideal into ring $R' = \mathbb{C}\left[ y_1,\dots,y_{15}\right]$ to get ideal $V$ representing the vacuum moduli space. Using \cite{mac2}, we see that V has dimension 7, degree 118 and after assigning weights to each $y_j$ equal to the degree of the monomial they represent, we get Hilbert series
\begin{align} 
& H(t, \cM_{g=3, \ x=(18,18,18,6)}) = \nonumber \\
&  \frac{1 + 4t + 14t^2 + 30t^3 + 41t^4 + 18t^5 - 10t^6 - 2t^7 + 25t^8 + 6t^9 -12t^{10} + 3t^{12}}{(1-t^2)^5(1-t)^2}  \ .
\end{align}

We can also write down a permutation triple such that $\sigma_B \sigma_W \sigma_{\infty} = id$: 
\begin{align}
\sigma_B &= (1\;2\;3\;4)(5\;6\;7\;8)(9\;10\;11\;12) \nonumber\\
\sigma_W &= (1\;3\;7\;5)(2\;9\;11\;4)(6\;12\;10\;8) \nonumber\\
\sigma_{\infty} &= (1\;8\;9)(2\;3\;4\;10\;11\;12\;5\;6\;7) \ .
\end{align}
Hence we have ramification structure $\{4^3,4^3,3\ 9\}$

\paragraph*{\fbox{$\mathbf{ x = (12,4,12,4,12,4) }$}}

This is a tiling with $V=4, E=12, F= 4$.

\begin{center}
\includegraphics[width=15cm]{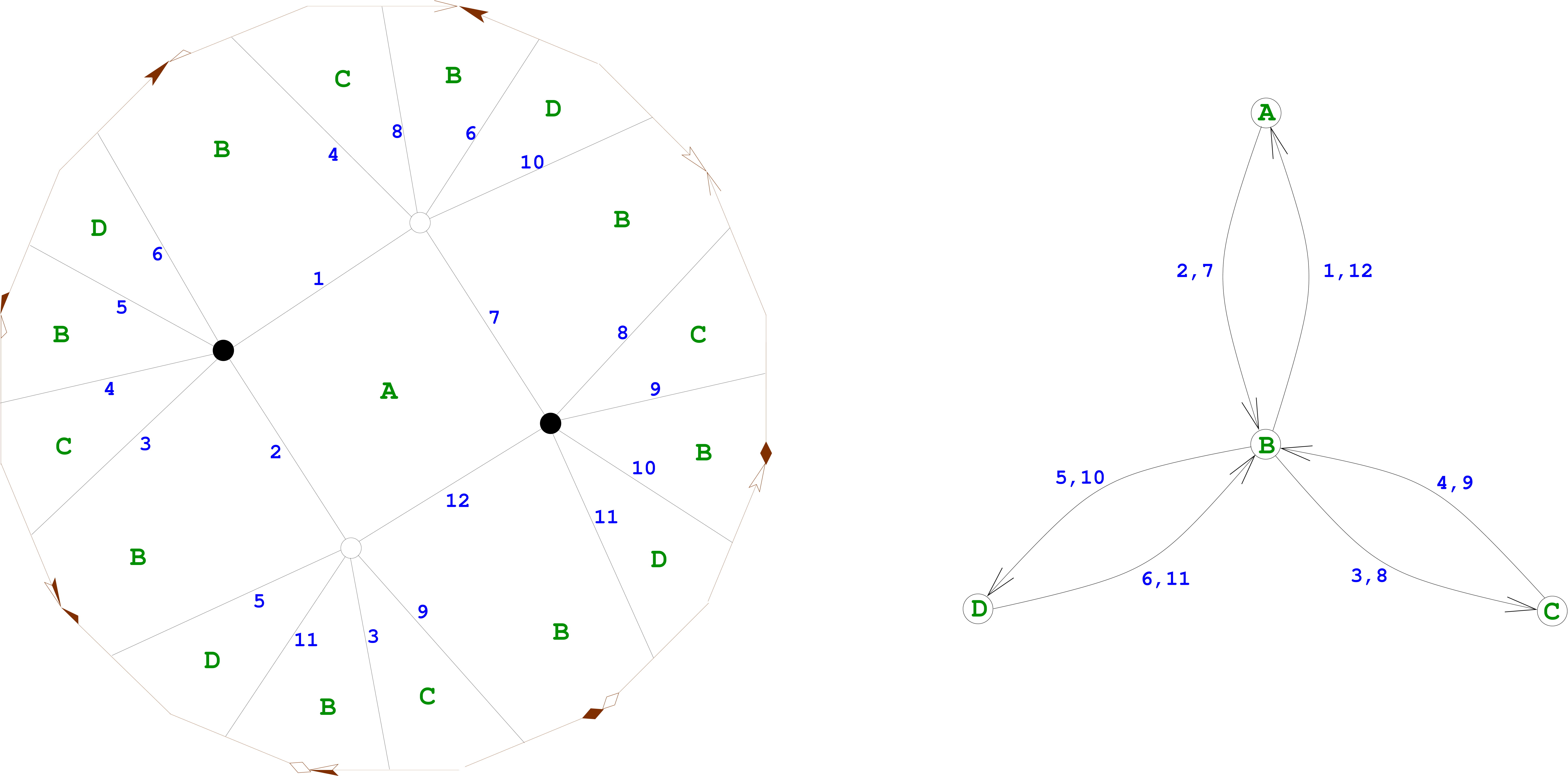}
\end{center}

We have superpotential
\begin{equation}
W = \tr ( \Phi_1\Phi_2\Phi_3\Phi_4\Phi_5\Phi_6 + \Phi_7\Phi_8\Phi_9\Phi_{10}\Phi_{11}\Phi_{12} - \Phi_1\Phi_7\Phi_{10}\Phi_6\Phi_8\Phi_4 - \Phi_2\Phi_5\Phi_{11}\Phi_3\Phi_9\Phi_{12})
\end{equation}
and 12 gauge invariants:
\begin{align}
\nonumber &
r_1 = \Phi_1\Phi_2\ , \quad 	 	
r_2 = \Phi_1\Phi_7\ , \quad 	 	
r_3 = \Phi_{12}\Phi_2\ , \quad  	
r_4 = \Phi_{12}\Phi_7\ , \quad	 	
r_5 = \Phi_3\Phi_4\ , \quad 	 	
r_6 = \Phi_3\Phi_9\ , \quad 	  	
\nonumber \\ &
r_7 	= \Phi_8\Phi_4 \ , \quad
r_8 	= \Phi_8\Phi_9 \ , \quad
r_9 	= \Phi_5\Phi_6 \ , \quad
r_{10}	= \Phi_5\Phi_{11} \ , \quad
r_{11}	= \Phi_{10}\Phi_6 \ , \quad
r_{12}	= \Phi_{10}\Phi_{11} \ .
\end{align}

To find the master space, we define the ring $S = \mathbb{C}[\Phi_1,\dots,\Phi_{12}]$ and ideal $I_1 = \left< \partial_i W \right>_{i=1,\dots,12}$ to generate the master space $R = S /I_1$. We find using \cite{mac2} that $I_1$ has dimension 10, degree 9 and Hilbert series
\begin{equation} 
H(t, \cF^{\flat}_{g=3, \ x=(12,4,12,4,12,4)}) =
\frac{1 + 2t + 3t^2 + 4t^3 + 5t^4 - 6t^5 - 8t^6 + 8t^7}{(1-t)^{10}} \ .
\end{equation}

Using primary decomposition in \cite{mac2}, we get that the curve given by $I_1$ is the union of those given by ideals of which
18 are trivial of dimension 8, degree 1 and have Hilbert series $\frac{1}{(1-t)^8}$;
The \emph{coherent component} is of dimension 10, degree 9 and has Hilbert series 
\begin{equation} 
H(t, \ ^{\text{Irr}}\cF^{\flat}_{g=3, \ x=(12,4,12,4,12,4)}) = \frac{1+ 2t + 3t^2 + 2t^3 + t^4}{(1-t)^{10}} \ .
\end{equation}

To find the vacuum moduli space, we consider the ring $R = \mathbb{C} \left[ \Phi_1,\dots\Phi_{12},y_1,\dots,y_{12} \right]$ and ideal $I_2 = \left< \partial_i W,y_j - r_j \right>_{i=1,\dots,12;j=1,\dots,12}$.
We then eliminate all the $\Phi$s and substitute the resulting ideal into ring $R' = \mathbb{C}\left[ y_1,\dots,y_{12}\right]$ to get ideal $V$ representing the vacuum moduli space. Using \cite{mac2}, we see that V has dimension 7, degree 24 and after assigning weights to each $y_j$ equal to the degree of the monomial they represent, we get Hilbert series
\begin{equation} 
H(t, \cM_{g=3, \ x=(12,4,12,4,12,4)}) = \frac{1 + 5t^2 + 12t^4 + 5t^6 + t^8}{(1-t^2)^7} \ .
\end{equation}  

We can also write down a permutation triple such that $\sigma_B \sigma_W \sigma_{\infty} = id$: 
\begin{align}
\sigma_B &= (1\;2\;3\;4\;5\;6)(7\;8\;9\;10\;11\;12) \nonumber\\
\sigma_W &= (1\;4\;8\;6\;10\;7)(2\;12\;9\;3\;11\;5) \nonumber\\
\sigma_{\infty} &= (1\;12)(2\;4\;6\;7\;9\;11)(3\;8)(5\;10) \ .
\end{align}

Hence we have ramification structure $\{6^2,6^2,2^3\ 6\}$.

\section{Beyond Bipartite Tilings} \label{sec:nonbipartite}\setall

Over the years it has become clear that in supersymmetric gauge theories ``bipartiteness'' (dimer models) and toric moduli spaces are intricately linked \cite{Hanany:2005ve,Franco:2012mm,He:2012js}. \\
While classifying all regular tilings, the natural question arose of the possibility of gauge theories arising from tilings that did not have bipartite structure. That is to say if we relaxed the condition that the superpotential obeys the ``toric condition'' \cite{Feng:2000mi} of having each field appearing only twice with opposite sign, we will lose the bipartite representation of the gauge theory. Consequently, the vacuum moduli space will not necessarily be a toric Calabi-Yau variety, but as far as the physics is concerned, this is no obstruction; we simply move to moduli spaces beyond toric geometry.\\
The story becomes much more complicated and we make a few remarks here.

In our present context of tessellations, we can analyse these in a similar way as described in section \ref{sec:methodology}. As there was no longer a clear distinction between black vertices, which give positive contributions to the superpotential and the negatively contributing white vertices, we can impose arbitrary coefficients to each of the superpotential terms.
The master space analysis is similar as before, now using the computer algebra system {\sf singular} \cite{singular} to do the calculation for a large range of integer and non-integer complex values of the coefficients. Generally in the tilings studied, all choices of coefficients generated the same master space, with the exception of a handful of choices. This is to be expected, as one generally expects there to be a few combinations of coefficients that give ``magical relations" between the partials of the superpotential that exist merely due to the coefficients and not due to the structure of the superpotential. This is ``genericity'' in complex structure \cite{Duncan:2014oja}.

As a toy model, suppose we had a superpotential: 
\begin{equation}
W = \frac{1}{2}\Phi_1^2 + \Phi_1\Phi_2 + \frac{1}{2}\Phi_2^2 
\end{equation}
then it can be easily verified that $W$ satisfies the relation $\partial_1 W = \partial_2 W$, whereas no such a relation exists for the superpotential 
\begin{equation}
\tilde{W} = \pi\Phi_1^2 + \Phi_1\Phi_2 + e^1\Phi_2^2
\end{equation}
even after allowing for multiplication of the partials by some complex number - as we are working with ideals over complex polynomial rings and ideals absorb multiplication by units, i.e. if we consider complex polynomial ring $R$ and ideal $I = \left<a x, b y\right>$ with $x,y \in R$ and $a,b \in \mathbb{C}\setminus \{0\}$ then the ideal $I$ does not depend on the choice of $a$ and $b$.
Most of the ``master spaces" were found to possess non-trivial geometries, despite the arbitrariness of the coefficients.

These master spaces can then be analysed using primary decomposition, and generally there was no Calabi-Yau component (with the exception of some trivial linear pieces), nor was there a unique component (i.e. for any piece of the space, there were usually multiple pieces that had the exact same dimension, degree and Hilbert series), nor was there a top-dimensional component that had the same dimension and degree of the master space. Hence a coherent component does not generally exist. Indeed, we are loosing the nice control provided by toric geoemtry.

Another problem, clearly, is that it is not immediately obvious how to draw an associated (directed) quiver diagram. First, it is no longer possible to use the colour of the vertices to assign a direction to the edges in the quiver. Several proposals, such as assigning an $n$-partite structure, naturally lead to the problem that our arbitrary choice of $n$ changes the whole theory, and in general lead to the problem that, when imposing the structure, our choice which vertex gets which colour also changes the theory, hence not making the structure unique. This does not arise for bipartite structures, as an interchange of black and white vertices does not change the resulting theory.

Furthermore, any tiling that contains a $p$-gon with $p$ an odd number, cannot possibly satisfy the Calabi-Yau condition that states that for every node in the quiver, the number of incoming arrows must equal the number of outgoing arrows, which comes about as an anomaly cancellation condition in 3+1 dimensions \cite{Franco:2005sm, Davey:2009bp}. Though tilings exist with polygons that all have an even number of sides, yet cannot have a bipartite structure imposed on it (such as the $g=3, \{ p,q \} = \{ 6,6 \}$ regular tiling, which is pictured below), we may still have the problem that the quiver will not satisfy this condition.

\includegraphics[width=15cm]{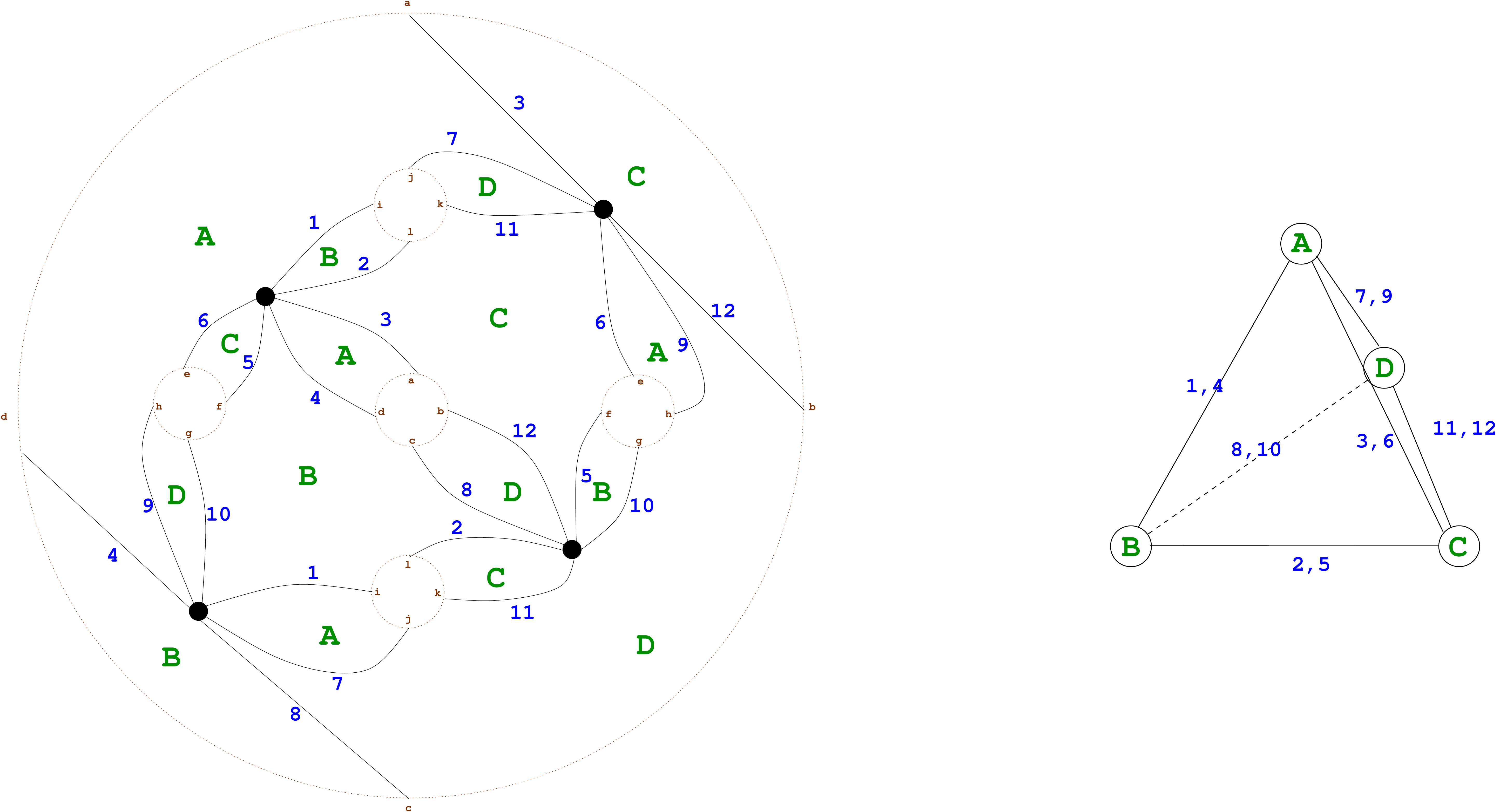}

Taking the $g=3, \{ p,q \} = \{ 6,6 \}$ regular tiling as an example, we see that the arrows between faces $A$ and $B$ both need to point in the same direction, as the edges $1,4$ seperating them connect the same two vertices, and they appear in the same cyclic order around each vertex. Hence by regularity of the tiling, in the quiver diagram, each node would either have 2 more incoming or 2 more outgoing arrows and hence the quiver would not satisfy the Calabi-Yau condition.
Therefore we see that despite the non-trivial geometry of the analysed master spaces, to extract any gauge theory data from non-bipartite remains an interesting challenge.

\section{Conclusions and Outlook} \label{sec:conclusions}\setall

We have studied the properties of field theories arising from regular and semi-regular tilings on Riemann surfaces of different genera. We have classified both the master space and mesonic moduli space arising from these tilings. Some interesting patterns stand out:
\begin{itemize}
\item The master space is always of dimension $F+2g$, where $F$ is the number of faces of the brane tiling (or equivalently the number of nodes in the quiver) and $g$ the genus of the Riemann surface the graph is embedded on. The dimension comes about as a contribution of $F-1$ independent baryonic parameters and $1+2g$ mesonic parameters, thus providing a generalisation of the well-known $F+2$ dimensional master space of toric tilings \cite{Forcella:2008bb,Forcella:2008eh}.
\item The coherent component $^{\text{Irr}}\cF^{\flat}$ is always a Calabi-Yau manifold of the same dimension and degree as the master space, just as in the $g=1$ case.
\item The mesonic moduli space is always of dimension $1+2g$, even when it is not found to be Calabi-Yau. This formula, and that for the dimension in of the master space, agree with the formulae found in \cite{Franco:2012wv} in the case of a Riemann surface without boundary.
\item For a consistent brane tiling, we expect the mesonic moduli space to be Calabi-Yau. Roughly half of the tilings studied, however, generated mesonic moduli spaces that were not Calabi-Yau. \footnote{Some tilings, such as the $g=2, \mathbf{ x = (8,4,8,4,4) }$, have a Hilbert series that does not have a palindromic numerator, but can be transformed into one by simple multiplication (or division) by $\prod_i (1-t^{n_i})$ for appropriate $n_i$. Others though, such as $g=2, \mathbf{\{p,q\} = \{6,4\}}$ have a Hilbert series for which this is not possible, and hence are definitely not Calabi-Yau.} Consistency for $g=1$ brane tilings has been extensively studied \cite{Hanany:2005ss, broomhead} and frequently uses geometric arguments such as zig-zag paths. For example, \cite{broomhead} gives the following consistency check:

\emph{
A dimer model is geometrically consistent if and only if the following conditions hold:
\begin{itemize}
\item No zig-zag flow $\tilde{\eta}$ intersects itself.
\item If $\tilde{\eta}$ and $\tilde{\eta}'$ are zig-zag flows and their homology classes $[\tilde{\eta}], [\tilde{\eta}']$ are linearly independent, then they intersect in precisely one arrow.
\item If $\tilde{\eta}$ and $\tilde{\eta}'$ are zig-zag flows and their homology classes $[\tilde{\eta}], [\tilde{\eta}']$ are linearly dependent, then they do not intersect.
\end{itemize}
}

We found tilings however, that satisfy these conditions yet still do not lead to mesonic moduli spaces that are Calabi-Yau, such as the $g=2$ regular tiling with $\{ p,q \} = \{ 6,4 \} $. Finding consistency checks on dimers, such that they generate Calabi-Yau manifolds as moduli spaces, should be a direction of further research.
\end{itemize}

Another thing to note is that in the case of the torus, we have a 1-1 correspondence between brane tilings and double periodic tilings of the plane (with a fundamental domain imposed). Here it is possible to enlarge the fundamental domain, which corresponds to an orbifold action.\\
However, this property seems to be unique to the torus, as can be seen by the fact that the number of edges - and hence also the number of vertices and faces - of a semi-regular tiling described by $(p_1,\dots,p_q)$ of a genus $g$ Riemann surface is fixed for $g\neq 1$. Hence no operation similar to enlarging the fundamental domain exists for these surfaces. It might be possible to relate different dimers on these surfaces through orbifold actions, but any such relation would be far less obvious geometrically.

Due to the doubly exponential nature of the Buchberger algorithm in its input size, we were unable to fully analyse all the tilings we found. As generally the number of edges in a (semi-)regular tiling goes up with the genus (which can be observed from equation \ref{eq:RegTilingEqn} or \ref{eq:SemiRegTilingEqn}), this soon leads to problems in analysing non-trivial tilings on Riemann surfaces of higher genus. A numerical algorithm to analyse these spaces has been proposed in \cite{Hauenstein:2012xs}, but has not yet been implemented in any program. Using this algorithm, or any other that is able to analyse these spaces, is clearly a vast open direction.


\section*{Acknowledgements}
We would like to thank D. Galloni and R.K. Seong for helpful comments and discussions.

Mark van Loon would like to thank the EPSRC and the Oxford University MPLS division for providing funding to undertake this research, and Merton College, Oxford for their support in providing residence.
He is particularly grateful to Rebecca Dodson for many fruitful discussions and her careful readings of the draft.

Yang-Hui He would like to thank the Science and Technology Facilities Council, UK, for an Advanced Fellowship and for STFC grant ST/J00037X/1, the Chinese Ministry of Education, for a Chang-Jiang Chair Professorship at NanKai University, the city of Tian-Jin for a Qian-Ren Scholarship, the US NSF for grant CCF-1048082, as well as City University, London, the Department of Theoretical Physics and Merton College, Oxford, for their enduring support.

\comment{
\begin{appendix}
\section{Example of used Macaulay2 code} \label{sec:mac2code}
As an example of the code used in Macaulay2, we show here the buffer in Macaulay2 of the calculations for the $g=2, \{p,q\} = \{8,4\}$ regular tiling. \\
Note: input lines correspond to lines with `i', output lines to those with `o'.\\
\\
Master space calculation:

\tiny
\begin{verbatim}
i1 : S=QQ[z1,z2,z3,z4,z5,z6,z7,z8]

o1 = S

o1 : PolynomialRing

i2 : I=ideal(z2*z3*z4 - z5*z3*z6, z1*z3*z4 - z8*z4*z7, z1*z2*z4 - z1*z5*z6, z1*z2*z3 - z2*z8*z7, 
z7*z6*z8 - z1*z3*z6, z5*z7*z8 - z1*z5*z3, z5*z6*z8 - z2*z8*z4, z5*z7*z6 - z2*z4*z7)

o2 = ideal (z2*z3*z4 - z3*z5*z6, z1*z3*z4 - z4*z7*z8, z1*z2*z4 - z1*z5*z6,
     --------------------------------------------------------------------------
     z1*z2*z3 - z2*z7*z8, - z1*z3*z6 + z6*z7*z8, - z1*z3*z5 + z5*z7*z8, -
     --------------------------------------------------------------------------
     z2*z4*z8 + z5*z6*z8, - z2*z4*z7 + z5*z6*z7)

o2 : Ideal of S

i3 : dim I

o3 = 6

i4 : degree I

o4 = 4

i5 : HilbertSeries I

           3      4     5     6
     1 - 8T  + 13T  - 8T  + 2T
o5 = --------------------------
                     8
              (1 - T)

o5 : Expression of class Divide

i6 : reduceHilbert o5

                2     3     4
     1 + 2T + 3T  - 4T  + 2T
o6 = ------------------------
                    6
             (1 - T)

o6 : Expression of class Divide

i7 : primaryDecomposition I

o7 = {ideal (z2*z4 - z5*z6, z1*z3 - z7*z8), ideal (z8, z7, z3, z1), ideal (z6,
     --------------------------------------------------------------------------
     z5, z4, z2)}

o7 : List

i8 : o7 / (i-> dim i)

o8 = {6, 4, 4}

o8 : List

i9 : o7 / (i->degree i)

o9 = {4, 1, 1}

o9 : List

i10 : o7 / (i->HilbertSeries i)

             2    4             2     3    4             2     3    4
       1 - 2T  + T   1 - 4T + 6T  - 4T  + T   1 - 4T + 6T  - 4T  + T
o10 = {------------, -----------------------, -----------------------}
                8                   8                        8
         (1 - T)             (1 - T)                  (1 - T)

o10 : List

i11 : o10 / (i->reduceHilbert i)

                 2
       1 + 2T + T       1         1
o11 = {-----------, --------, --------}
                6          4         4
         (1 - T)    (1 - T)   (1 - T)

o11 : List
\end{verbatim}

\normalsize
Moduli space calculation:

\tiny
\begin{verbatim}
i1 : S=QQ[z1,z2,z3,z4,z5,z6,z7,z8,y1,y2,y3,y4,y5,y6,y7,y8,y9,y10,y11,y12,y13,y14,y15,y16, Degrees => {1,1,1,1,1,1,1,1,2,2,2,2,2,2,2,2,2,2,2,2,2,2,2,2}]

o1 = S

o1 : PolynomialRing

i2 : I=ideal(z2*z3*z4 - z5*z3*z6, z1*z3*z4 - z8*z4*z7, z1*z2*z4 - z1*z5*z6, z1*z2*z3 - z2*z8*z7, z7*z6*z8 - z1*z3*z6, z5*z7*z8 - z1*z5*z3, 
z5*z6*z8 - z2*z8*z4, z5*z7*z6 - z2*z4*z7,      y1 - z1*z2, y2 - z1*z4, y3 - z1*z5, y4 - z1*z6, y5 - z3*z2, y6 - z3*z4, 
y7 - z3*z5, y8 - z3*z6, y9 - z7*z2, y10 - z7*z4, y11 - z7*z5, y12 - z7*z6, y13 - z8*z2, y14 - z8*z4, y15 - z8*z5, y16 - z8*z6)

o2 = ideal (z2*z3*z4 - z3*z5*z6, z1*z3*z4 - z4*z7*z8, z1*z2*z4 - z1*z5*z6,
     --------------------------------------------------------------------------
     z1*z2*z3 - z2*z7*z8, - z1*z3*z6 + z6*z7*z8, - z1*z3*z5 + z5*z7*z8, -
     --------------------------------------------------------------------------
     z2*z4*z8 + z5*z6*z8, - z2*z4*z7 + z5*z6*z7, - z1*z2 + y1, - z1*z4 + y2, -
     --------------------------------------------------------------------------
     z1*z5 + y3, - z1*z6 + y4, - z2*z3 + y5, - z3*z4 + y6, - z3*z5 + y7, -
     --------------------------------------------------------------------------
     z3*z6 + y8, - z2*z7 + y9, - z4*z7 + y10, - z5*z7 + y11, - z6*z7 + y12, -
     --------------------------------------------------------------------------
     z2*z8 + y13, - z4*z8 + y14, - z5*z8 + y15, - z6*z8 + y16)

o2 : Ideal of S

i3 : tmp = eliminate({z1,z2,z3,z4,z5,z6,z7,z8},I)

o3 = ideal (y12*y15 - y11*y16, y8*y15 - y7*y16, y4*y15 - y3*y16, y13*y14 -
     --------------------------------------------------------------------------
     y15*y16, y12*y14 - y10*y16, y11*y14 - y10*y15, y9*y14 - y11*y16, y8*y14 -
     --------------------------------------------------------------------------
     y6*y16, y7*y14 - y6*y15, y5*y14 - y7*y16, y4*y14 - y2*y16, y3*y14 -
     --------------------------------------------------------------------------
     y2*y15, y1*y14 - y3*y16, y12*y13 - y9*y16, y11*y13 - y9*y15, y10*y13 -
     --------------------------------------------------------------------------
     y11*y16, y8*y13 - y5*y16, y7*y13 - y5*y15, y6*y13 - y7*y16, y4*y13 -
     --------------------------------------------------------------------------
     y1*y16, y3*y13 - y1*y15, y2*y13 - y3*y16, y8*y11 - y7*y12, y4*y11 -
     --------------------------------------------------------------------------
     y3*y12, y9*y10 - y11*y12, y8*y10 - y6*y12, y7*y10 - y6*y11, y5*y10 -
     --------------------------------------------------------------------------
     y7*y12, y4*y10 - y2*y12, y3*y10 - y2*y11, y1*y10 - y3*y12, y8*y9 - y5*y12,
     --------------------------------------------------------------------------
     y7*y9 - y5*y11, y6*y9 - y7*y12, y4*y9 - y1*y12, y3*y9 - y1*y11, y2*y9 -
     --------------------------------------------------------------------------
     y3*y12, y4*y8 - y12*y16, y3*y8 - y11*y16, y2*y8 - y10*y16, y1*y8 - y9*y16,
     --------------------------------------------------------------------------
     y4*y7 - y11*y16, y3*y7 - y11*y15, y2*y7 - y10*y15, y1*y7 - y9*y15, y5*y6 -
     --------------------------------------------------------------------------
     y7*y8, y4*y6 - y10*y16, y3*y6 - y10*y15, y2*y6 - y10*y14, y1*y6 - y11*y16,
     --------------------------------------------------------------------------
     y4*y5 - y9*y16, y3*y5 - y9*y15, y2*y5 - y11*y16, y1*y5 - y9*y13, y1*y2 -
     --------------------------------------------------------------------------
     y3*y4)

o3 : Ideal of S

i4 :  R=QQ[y1,y2,y3,y4,y5,y6,y7,y8,y9,y10,y11,y12,y13,y14,y15,y16, Degrees => {2,2,2,2,2,2,2,2,2,2,2,2,2,2,2,2}]

o4 = R

o4 : PolynomialRing

i5 : V = substitute(tmp,R)

o5 = ideal (y12*y15 - y11*y16, y8*y15 - y7*y16, y4*y15 - y3*y16, y13*y14 -
     --------------------------------------------------------------------------
     y15*y16, y12*y14 - y10*y16, y11*y14 - y10*y15, y9*y14 - y11*y16, y8*y14 -
     --------------------------------------------------------------------------
     y6*y16, y7*y14 - y6*y15, y5*y14 - y7*y16, y4*y14 - y2*y16, y3*y14 -
     --------------------------------------------------------------------------
     y2*y15, y1*y14 - y3*y16, y12*y13 - y9*y16, y11*y13 - y9*y15, y10*y13 -
     --------------------------------------------------------------------------
     y11*y16, y8*y13 - y5*y16, y7*y13 - y5*y15, y6*y13 - y7*y16, y4*y13 -
     --------------------------------------------------------------------------
     y1*y16, y3*y13 - y1*y15, y2*y13 - y3*y16, y8*y11 - y7*y12, y4*y11 -
     --------------------------------------------------------------------------
     y3*y12, y9*y10 - y11*y12, y8*y10 - y6*y12, y7*y10 - y6*y11, y5*y10 -
     --------------------------------------------------------------------------
     y7*y12, y4*y10 - y2*y12, y3*y10 - y2*y11, y1*y10 - y3*y12, y8*y9 - y5*y12,
     --------------------------------------------------------------------------
     y7*y9 - y5*y11, y6*y9 - y7*y12, y4*y9 - y1*y12, y3*y9 - y1*y11, y2*y9 -
     --------------------------------------------------------------------------
     y3*y12, y4*y8 - y12*y16, y3*y8 - y11*y16, y2*y8 - y10*y16, y1*y8 - y9*y16,
     --------------------------------------------------------------------------
     y4*y7 - y11*y16, y3*y7 - y11*y15, y2*y7 - y10*y15, y1*y7 - y9*y15, y5*y6 -
     --------------------------------------------------------------------------
     y7*y8, y4*y6 - y10*y16, y3*y6 - y10*y15, y2*y6 - y10*y14, y1*y6 - y11*y16,
     --------------------------------------------------------------------------
     y4*y5 - y9*y16, y3*y5 - y9*y15, y2*y5 - y11*y16, y1*y5 - y9*y13, y1*y2 -
     --------------------------------------------------------------------------
     y3*y4)

o5 : Ideal of R

i6 : dim V

o6 = 5

i7 : HilbertSeries(V)

            4       6       8        10        12        16        18       20       22      24    28
     1 - 55T  + 320T  - 891T  + 1408T   - 1155T   + 1155T   - 1408T   + 891T   - 320T   + 55T   - T
o7 = ------------------------------------------------------------------------------------------------
                                                      2 16
                                                (1 - T )

o7 : Expression of class Divide

i8 : degree(V)

     3
o8 = -
     4

o8 : QQ

i9 : reduceHilbert o7

            2      4    6
     1 + 11T  + 11T  + T
o9 = --------------------
                 2 5
           (1 - T )

o9 : Expression of class Divide
\end{verbatim}
\normalsize

\end{appendix}
}

\end{document}